\documentclass[%
preprint,
 amsmath,amssymb,
aps,
pra,
]{revtex4-2}

\usepackage{graphicx}
\usepackage{dcolumn}
\usepackage{bm}
\usepackage{epstopdf, epsfig}
\usepackage{subfigure}
\usepackage{multirow}
\usepackage{mathrsfs}
\usepackage{ esint }

\begin{document}

\preprint{APS/PRFluids}

\title{Scaling behavior of density gradient accelerated mixing rate in shock bubble interaction}

\author{Bin Yu}
\email{kianyu@sjtu.edu.cn}
\author{Haoyang Liu}
\email{lhy19980810@sjtu.edu.cn}

\author{Hong Liu}
\email{Author to whom correspondence should be addressed: hongliu@sjtu.edu.cn}
\affiliation{School of Aeronautics and Astronautics, Shanghai Jiao Tong University, Shanghai, 200240, China}
\date{\today}

\begin{abstract}
   Variable-density mixing in shock bubble interaction, a canonical flow of Richtermyer-Meshkov instability, is studied by the high-resolution simulation.
   While the dissipation mainly controls the passive scalar mixing rate, an objective definition of variable-density mixing rate characterizing the macroscopic mixing formation is still lacking, and the fundamental behavior of mixing rate evolution is not yet well understood.
   Here, we first show that the variable-density mixing of shock bubble interaction is distinctly different from the previous observations in the passive scalar mixing. The widely-accepted hyperbolic conservation of the first moment of concentration in the scalar mixing, i.e., the conservation of the mean concentration, is violated in variable-density flows.
   We further combine the compositional transport equation and the divergence relation for the miscible flows to provide the evidence that the existence of density gradient accelerated mixing rate, decomposed by the accelerated dissipation term and redistributed diffusion term, contributes to the anomalous decrease or increase of the mean concentration depending on Atwood number.
   Further analyzing a number of simulations for the cylindrical or spherical bubbles under a broad range of shock Mach numbers, Reynolds numbers, and P\'eclet numbers, the density gradient accelerated mixing rate exhibits weak dependent on P\'eclet numbers, and identifies an Atwood number range with high mixing rate, which can be theoretically predicted based on the mode of hyperbolic conservation violation behavior.
\end{abstract}

\maketitle


\section{Introduction}
\label{sec:intro}
Richtmyer-Meshkov (referred to as RM hereafter) instability results from the baroclinic vorticity generation due to the misalignment of pressure gradient and density gradient during shock impact on a density continuity with perturbation \cite{richtmyer1960taylor,meshkov1969instability,brouillette2002richtmyer}. A classical type of RM instability is the shock interacting on a circular bubble of density difference with ambient gas. The high curvature of the density interface bringing strong nonlinear effect impedes the RM linear theory extension to this kind of shock bubble interaction (referred to as SBI hereafter)~\cite{ranjan2011shock}. The resulting interpenetration and mixing between the bubble and ambient gas have the vital application in supernova \cite{klein1994hydrodynamic}, inertial confinement confusion \cite{lindl1992progress}, and supersonic mixing \cite{yang1994model,yu2020two}.
Thus, SBI with relatively simple initial conditions presents ample physical phenomena gaining investigation of this problem ranging from theoretical~\cite{picone1988vorticity,samtaney1994circulation,niederhaus2008computational}, experimental~\cite{ranjan2005experimental,tomkins2008experimental,Zhai2011On}, and numerical~\cite{zabusky1999vortex,Bagabir2001Mach,shankar2011two} perspectives.

SBI, as well as RM instability, defined as the Level-2 mixing by Dimotakis \cite{dimotakis2005turbulent}, shares the same difficulties in mixing study, namely shock compression (specified by Mach number) and variable density (specified by Atwood number) effect, two notoriously challenging problems absent in the Level-1 passive scalar mixing.
The stretching rate of bubble interface after shock impacts is the focus in the study of mixing in SBI. By defining the bubble area, it is found that the mixing leads to the decrease of bubble area \cite{jacobs1992shock}, which reflects the macroscopic mixing behavior due to vortical stretching. Referring to exponent stretching in classic turbulence proposed by Batchelor \cite{batchelor1952effect}, Yang studied the stretching rate of different shock Mach number and density ratio between the bubble and ambient air \cite{yang1993applications}. Different configurations of shock heavy bubble interaction are studied by Kumar \emph{et al.} \cite{kumar2005stretching}, showing that integral measurements like bubble width are insufficient to characterize early time mixing.
Still, integral measurement, such as mixing zone width, is one crucial indicator of mixing performance among the studied in RM instability due to the small perturbation of density discontinuity \cite{Mikaelian1998Analytic}.
Although integral geometric parameters such as mixing width can reflect the general mixing status, it is molecular diffusion combining with the stretching or growth rate that controls chemical reactions \cite{marble1985growth} and dilution of peak concentration \cite{villermaux2019mixing}. Thus advection/diffusion characteristic is vital to the Level-2 variable-density flows.

Advection of multi-component species is converted into a density evolution equation, describing the mixing of two incompressible fluids with different densities, $\rho_1$ and $\rho_2$, known as buoyancy-driven Rayleigh-Taylor (referred to as RT hereafter) instability \cite{livescu2007buoyancy,livescu2008variable}.
Through introducing the advection of density mass flux and mole fraction mixing rate, the mixing width growth rate in RT instability is successfully built by Cook \emph{et al.} \cite{cook2004mixing}.
Recently, mean mass fraction and mean molecular fraction is theoretically predicted in RT instability based on the asymptotic analysis of the mass fraction advection equation \cite{ruan2020density}.
As for RM instability flows, from the conservation equation for the mass fraction of diffusive multi-species component \cite{besnard1992turbulence}, the evolution of density self-correlation (DSC) of turbulent mixing in RM flows is investigated \cite{tomkins2013evolution}, suggesting a form of equilibrium of DSC as the onset of mixing transition.
Recently, Nobel \cite{noble2020high} applied the normalized scalar advection-diffusion equation to propose a model that predicts the growth rate of a shocked mixing width.
The view of the transport equation of mass fraction and density offers a new perspective in the research of RT/RM-type variable-density mixing.

It is worth noting that the advection-diffusion equation of a passive scalar has been studied for decades, which can be described as \cite{villermaux2019mixing}:
\begin{equation}\label{eq: PS-ADE-1}
    \left(\frac{\partial}{\partial t}+u_j\frac{\partial}{\partial x_j}-\mathscr{D}\frac{\partial^2}{\partial x_j^2}\right)\varphi=0,
\end{equation}
where $\varphi(x,t)$ is the scalar concentration that shows the conservation characteristic, leading to the time derivative of mean concentration zero. To distinguish the mixing structure of conservative scalar $\varphi(x,t)$, scalar energy  $\frac{1}{2}\varphi(x,t)^2$ is defined, and its evolution follows \cite{buch1996experimental}:
\begin{equation}\label{eq: PS-ADE-2}
  \left(\frac{\partial}{\partial t}+u_j\frac{\partial}{\partial x_j}-\mathscr{D}\frac{\partial^2}{\partial x_j^2}\right)
  \frac{\varphi^2}{2}=-\mathscr{D}\frac{\partial\varphi}{\partial x_j}\frac{\partial\varphi}{\partial x_j},
\end{equation}
where the term on the right side is the well-known scalar dissipation or scalar mixing rate, $\chi=\mathscr{D}\frac{\partial\varphi}{\partial x_j}\frac{\partial\varphi}{\partial x_j}$, which is strictly positive to dissipate scalar. Thus lots of studies pay attention to the degree of mixing that reflects the macroscopic mixing increase from the local scalar dissipation rate. Cetegen and Mohamad \cite{cetegen1993experiments} experimentally studied the passive scalar mixing in shear flows. By defining the mixedness, $f=4\varphi(1-\varphi)$, ranging from 0 to 1, the time evolution of $f$ is controlled by scalar dissipation $\chi$ by connecting the diffusivity of scalar $\mathscr{D}$ due to the hyperbolic conservation of mean concentration (i.e., $\mathrm{D}\left<\varphi\right>/\mathrm{D}t=0$ where $\left<\cdot\right>$ is spatial averaging). Theoretically analyzing the advection-diffusion equation in the form of a vortical flow, a passive scalar's mixing time follows the dependence of a 1/3 scaling law on P\'eclet number \cite{meunier2003vortices,wonhas2001mixing}. The mixedness $f$ and scalar dissipation rate $\chi$ suggests the quantification of the mixing behavior in all kinds of flows.

In RM-type flows, few pioneering studies applied scalar dissipation $\chi$, to investigate mixing. Tomkins \emph{et al.} \cite{tomkins2008experimental} found that the scalar dissipation rate is mainly connected to the large-scale strain field of the non-turbulent region in shock accelerated heavy bubble.  Several mixing indicators, one of which is scalar dissipation, are studied in shock accelerated gas curtain \cite{orlicz2013incident}. The result shows that the mixing rate decays faster in higher shock Mach number due to the higher degree of stirring. Scalar dissipation rate can be enhanced in the RM instability with reshock \cite{wong2019high}.
The idea of scalar energy and dissipation rate has also been extended to the other forms of variable-density mixing, such as in the RT convection of porous media~\cite{jha2011fluid,de2019rayleigh} and combustion flows, such as in explaining the local flame extinction~\cite{peters1983local}.
It can be concluded that scalar dissipation not only displays the mixing rate of different flow structures in RM flows more clearly, but connects the similarities in mixing behavior with passive scalar and discerns the differences from the fundamental nature of the variable-density effect and shock Mach number effect in RM flows \cite{soulard2018permanence}. However, it is noteworthy that the passive scalar definition is still strictly applied in the above variable-density flows.

Since the mass fraction of specific species in variable-density flows no longer follows the advection-diffusion equation [Eq.~(\ref{eq: PS-ADE-1})] but transport equation of mass fraction $Y$ obeying Fickian's law \cite{weber2012turbulent}:
\begin{equation}\label{eq: trans for mass}
  \frac{\partial\left(\rho Y\right)}{\partial t}+\frac{\partial}{\partial x_j}\left(\rho Y u_j\right)=\frac{\partial}{\partial x_j}\left(\mathscr{D}\rho\frac{\partial Y}{\partial x_j}\right).
\end{equation}
This leads to the mass fraction dissipation different from the scalar energy function in Eq.~(\ref{eq: PS-ADE-2}), in which density effect can not be neglected \cite{su2010quantitative}. Knowing the evolution of mass fraction and its energy evolution is vital for modeling the reaction rate \cite{bilger1979turbulent} and extinction in non-premixed combustion \cite{peters1983local}. Thus one of the inherent difficulties for further analyzing mixing in the form of the mass fraction is to define correctly a mixing rate that controls mass fraction and its energy evolution, which is still lacking and urgently needed. Through high-resolution numerical simulation, this paper investigates the mixing rate of mass fraction and its energy in variable-density mixing of SBI. The density gradient brings the accelerated dissipation and redistributed diffusion terms for mixedness linear growth before an asymptotic limit of mixing is reached.
The time-averaged density gradient accelerated mixing rate shows the nontrivial time-dependent behavior and a weak dependence on Pe number under a broad range of systematic parameters.
Moreover, the growth rate of the mean mass fraction and its energy determines a density ratio range with a high mixing rate pattern, which can be theoretically predicted based on the local and global mode of hyperbolic conservation violation behavior.
The accelerated dissipation of variable-density mixing found in this paper implies the new standpoint for auto-ignition \cite{mastorakos2009ignition} and extinction in non-premixed combustion \cite{peters1983local} in the extensive variable-density problems.

\section{Methodology and cases presentation}
\label{sec:method}
The governing equations for compressible flows comprised of different miscible species, which are controlled by Navier-Stokes equations (referred to as NS equations hereafter), in the Cartesian frame of reference are:
\begin{equation}\label{eq: masscon}
  \frac{\partial\widetilde{\rho}}{\partial \widetilde t}+\frac{\partial (\tilde{\rho}\widetilde{u}_i)}{\partial \widetilde x_i}
  +\vartheta\frac{\tilde{\rho}\widetilde{u}_r}{\widetilde{x}_r}=0,
\end{equation}
\begin{equation}\label{eq: momcon}
  \frac{\partial (\tilde{\rho}\widetilde{u}_i)}{\partial \widetilde t}+\frac{\partial(\tilde{\rho}\widetilde{u}_i\widetilde{u}_j)}{\partial\widetilde x_j}
  +\vartheta\frac{\tilde{\rho}\widetilde{u}_i\widetilde{u}_r}{\widetilde x_r}
  =\left(\frac{\partial \widetilde x_r}{\widetilde x_r}\right)^{\vartheta(j-1)}
  \frac{\partial\widetilde{\sigma}_{ij}}{\partial\widetilde x_j}
  -\frac{\partial\widetilde{p}}{\partial\widetilde x_i},
\end{equation}
\begin{equation}\label{eq: enercon}
  \frac{\partial(\tilde{\rho}\widetilde{E})}{\partial\widetilde t}+\frac{\partial(\tilde{\rho}\widetilde{u}_i\widetilde{H})}{\partial \widetilde x_i}
  +\vartheta\frac{\tilde{\rho}\widetilde{u}_r\widetilde{H}}{\widetilde x_r}
  =\left(\frac{\partial \widetilde x_r}{\widetilde x_r}\right)^{\vartheta(i-1)}\cdot
  \left[\frac{\partial\left(\widetilde{u}_j\widetilde{\sigma}_{ij}\right)}{\partial\widetilde x_i}-\frac{\partial \widetilde{q}_i}{\partial\widetilde x_i}\right],
\end{equation}
\begin{equation}\label{eq: mfcon}
  \frac{\partial(\tilde{\rho}\widetilde{Y}_m)}{\partial\widetilde t}+\frac{\partial(\tilde{\rho}\widetilde{u}_i\widetilde{Y}_m)}{\partial\widetilde x_i}
  +\vartheta\frac{\tilde{\rho}\widetilde{u}_r\widetilde{Y}_m}{\widetilde x_r}
  =\left(\frac{\partial \widetilde x_r}{\widetilde x_r}\right)^{\vartheta(i-1)}\cdot
  \left[\frac{\partial}{\partial \widetilde x_i}\left(\widetilde{\rho}\mathscr{D}\frac{\partial\widetilde{Y}_m}{\partial\widetilde x_i}\right)\right],
  \quad m=1,2,\cdots,s-1.
\end{equation}
Here, $\widetilde{\rho}$, $\widetilde{p}$, $\widetilde{u}_i$, $\widetilde{E}$ and $\widetilde{H}$ are density, pressure, velocity, energy, and enthalpy respectively. The mass fraction of species $m$ is denoted as $\widetilde{Y}_m$. There are $s$ components in total.
Parameter $\vartheta$ determines the axisymmetric coordinate ($\vartheta=1$) or symmetric coordinate ($\vartheta=0$). Moreover, subscript $r$ does not conduct Einstein's summation, and if $x$ coordinate is set as the axis of symmetry,  $r=2$ \cite{houim2013ghost,murugan2013study}.

$\widetilde{\sigma}_{ij}=\mu[\partial\tilde{u}_i/\partial\tilde x_j+\partial\tilde{u}_j/\partial \tilde x_i-2/3\delta_{ij}\partial\tilde{u}_k/\partial\tilde x_k]$ is the viscous stress tensor in which $\mu$ is the constant dynamics viscosity. $\widetilde{q}_i=-\lambda\partial\widetilde{T}/\partial\tilde x_i$ is the heat flux calculated as $\lambda=C_p\mu/\mathrm{Pr}$ where $C_p$ is constant-pressure specific heat~\cite{Kee1991CHEMKIN}, and Prandtl number is chosen as $\mathrm{Pr}=0.72$~\cite{houim2011low}.
$\mathscr{D}$ is Fickian diffusivity set to be constant in all cases. Then kinetic viscosity $\nu$ can be estimated as $\nu=\mu/\overline{\rho}$, where $\overline{\rho}=[(\rho_1^*)'+(\rho_2^*)']/2$ is the average of post-shock light bubble density $(\rho_2^*)'$, and post-shock heavy ambient air density $(\rho_1^*)'$ obtained from one-dimensional shock dynamics~\cite{Cook2001Transition}.

In this paper, the NS equations are solved using our in-house high-resolution code \emph{ParNS3D}~\cite{wang2018scaling,liang2019hidden,liu2020optimal} to study the mixing process of SBI. Three-order TVD Runge-Kutta method~\cite{gottlieb1998total} is applied for time marching, and convection terms are discretized by the fifth-order WENO scheme~\cite{liu1994weighted} while the discretion of viscous terms is dealt with the central difference method.

\begin{figure}
  \centering
  \includegraphics[clip=true,trim=0 0 0 0,width=.99\textwidth]{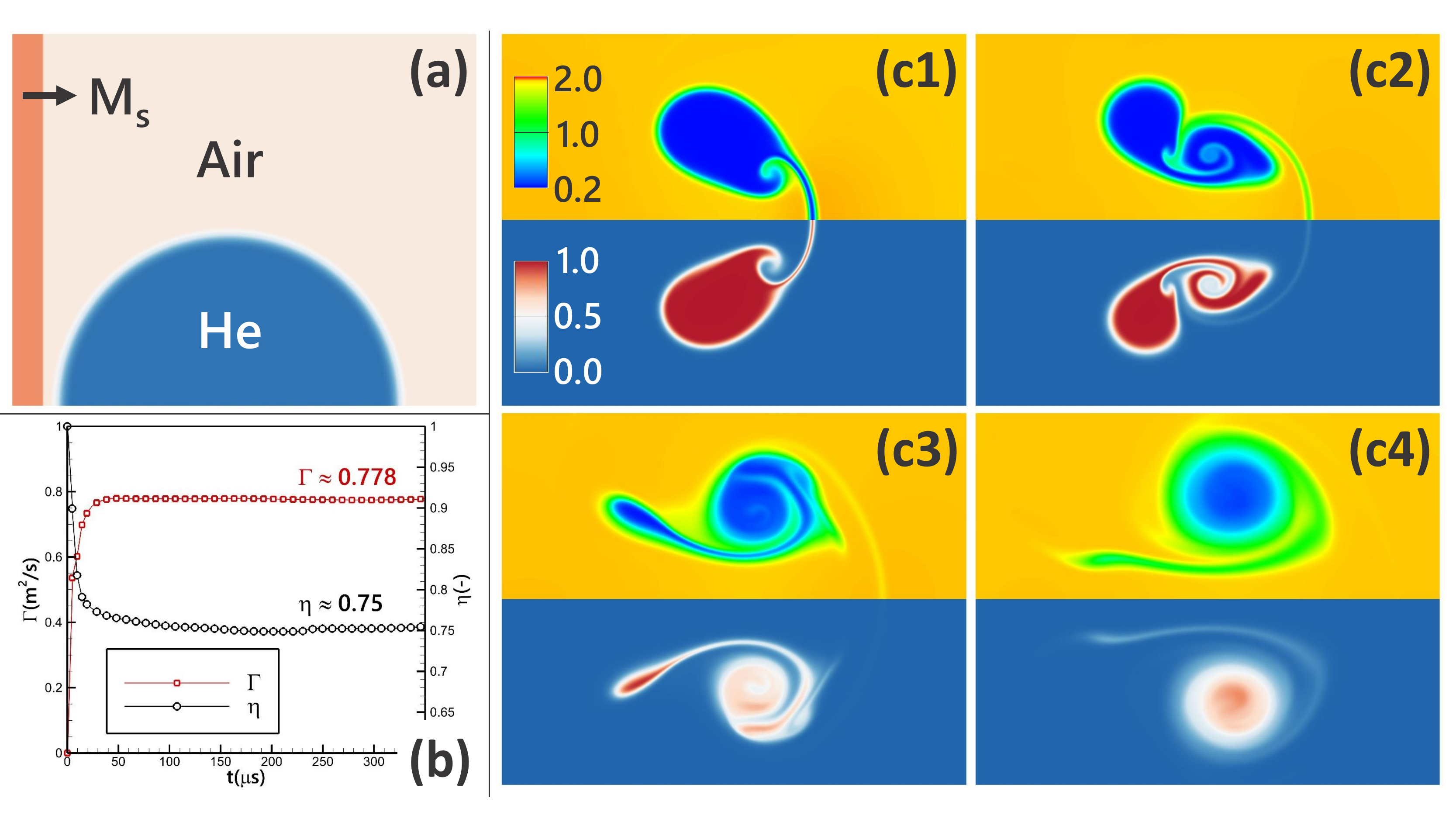}\\
  \caption{(a) Initial conditions of shock cylindrical bubble interaction. (b) Time history of circulation and compression rate. (c) Up: density contour, bottom: mass fraction contour; (c1) $\widetilde{t}=57.6\mu$s ($t=1.65$); (c2) $\widetilde{t}=96\mu$s ($t=2.74$); (c3) $\widetilde{t}=187.2\mu$s ($t=5.35$); (c4) $\widetilde{t}=321.6\mu$s ($t=9.2$). }\label{fig1}
\end{figure}
The initial conditions for a two-dimensional shock strength of Ma=1.22 (only half cylindrical bubble is shown) are plotted in Fig.~\ref{fig1}(a). The bubble is contained full of light gas helium with ambient air around before impacted by shock. The post-shock parameters are determined by the Rankine-Hugoniot equation~\cite{diegelmann2016pressure}. The bubble boundary is set as a diffusive layer to avoid spurious vorticity production from the grid step~\cite{niederhaus2008computational}. The distribution of the diffusive layer is the same as the one reported in Ref.~\cite{wang2018scaling}. Boundary conditions are applied as fourth-order extra-interpolation to avoid pseudo-pressure reflection wave interference with flow structures and classical symmetry conditions at the bubble axis. The constant diffusivity $\mathscr{D}$ is set as $142\times10^{-6}\textrm{ m}^2/\textrm{s}$, and dynamics viscosity $\mu$ is set as $125\times10^{-6}\textrm{ Pa}\cdot\textrm{s}$ for the cases concerned.

After shock passages, the time evolution of bubble deformation is depicted in Figs.~\ref{fig1}(c1)-(c4) at four specific time instants. Due to the baroclinic vorticity deposited along the bubble boundary formed from the misalignment of the pressure gradient of shock and density gradient of bubble, the bubble's roll-up is gradually growing with time. A bridge structure links the upper part and the lower part of the bubble, which forms the typical kidney shape of SBI at an early time \cite{ranjan2008shock}. The main vortex is entraining the bubble lobe through the connector between them, as shown in Fig.~\ref{fig1}(c2). During the entrainment process presented in Fig.~\ref{fig1}(c3), mixing happens mainly in the vortex region and partly along the bubble's edge because of concentration gradient diffusion. Finally, the main vortex becomes stable after absorbing the major baroclinic vorticity and maintains pure diffusion with a low degree of mixing, as shown in Fig.~\ref{fig1}(c4). Three points can be summarized: First, the general pattern of density and mass fraction is similar. Second, mixing happens during the growth of the main vortex. Third, particular mixing structures such as bridge decrease its region, which we will show that this decrease is caused by accelerated dissipation in variable-density flows.

For the qualitative value, the circulation, $\Gamma=\iiint_\mathcal{V}\omega(x,t)d\mathcal{V}$, is obtained from the area integration of the vorticity inside the bubble region. Fig.~\ref{fig1}(b) shows the near-constant value of circulation, which controls the mixing from stirring.
Once we get the controlling system parameters, variables considered are made dimensionless as follows in the following study:
\begin{equation}\label{eq: dimensionless}
  x_i=\frac{\widetilde{x}_i}{D} ~,~~ u_i=\frac{\widetilde{u}_i}{u^*} ~,~~ t=\frac{\widetilde{t}}{t^*} ~,~~
  \rho=\frac{\widetilde{\rho}}{\rho^*_1} ~,~~ p=\frac{\widetilde{p}}{p^*_1} ~,~~
\end{equation}
where $D=5.2$ mm is the diameter of bubble. $u^*=\Gamma/D$ and $t^*=D/u^*$ in which $\Gamma$ is the main circulation of the bubble after shock impacts. $\rho^*_1$, $p^*_1$ are density and pressure ahead of shock, respectively. Then we can define Re number \cite{glezer1988formation} and Pe number \cite{meunier2003vortices}:
\begin{equation}\label{eq: Re-Pe}
   \mathrm{Re}\equiv\frac{\Gamma}{\nu} ~,~~ \mathrm{Pe}\equiv\frac{\Gamma}{\mathscr{D}}=\mathrm{Re}\cdot\mathrm{Sc}
   ~,~~ \eta\equiv\frac{\iiint_\mathcal{V}\mathcal{X}(x,t\rightarrow\infty)d\mathcal{V}}{\mathrm{V}_0} ~,~~
\end{equation}
where Sc$=\nu/\mathscr{D}$ is Schmidt number. The volume fraction is $\mathcal{X}(x,t)=(\rho Y_2/M_2)/(\sum_{m=1}^{s}\rho Y_m/M_m)$ (subscript 2 is denoted as light helium gas concerned). Then, the initial volume of the bubble is calculated as $\mathrm{V}_0=\iiint_\mathcal{V}\mathcal{X}(x,t=0)d\mathcal{V}$.
Here, the compression rate, $\eta$, can be defined referring to Ref.~\cite{giordano2006richtmyer}, which is one fundamental dimensionless parameter reflecting the main shock compression.
As shown in Fig.~\ref{fig1}(b), the near-constant compression rate is found immediately after shock passages, which shows the apparent compression of bubble volume, as illustrated in Fig.~\ref{fig1}(c1). This compression volume maintains until the late time evolution.
We will show that the compression rate controls the asymptotic scaling behavior of mixing in general.

\section{Hyperbolic conservation violation of the mean mass fraction in shock bubble interaction}
\label{sec:hyer}
\begin{figure}
  \centering
  \includegraphics[clip=true,trim=0 0 300 0,width=.8\textwidth]{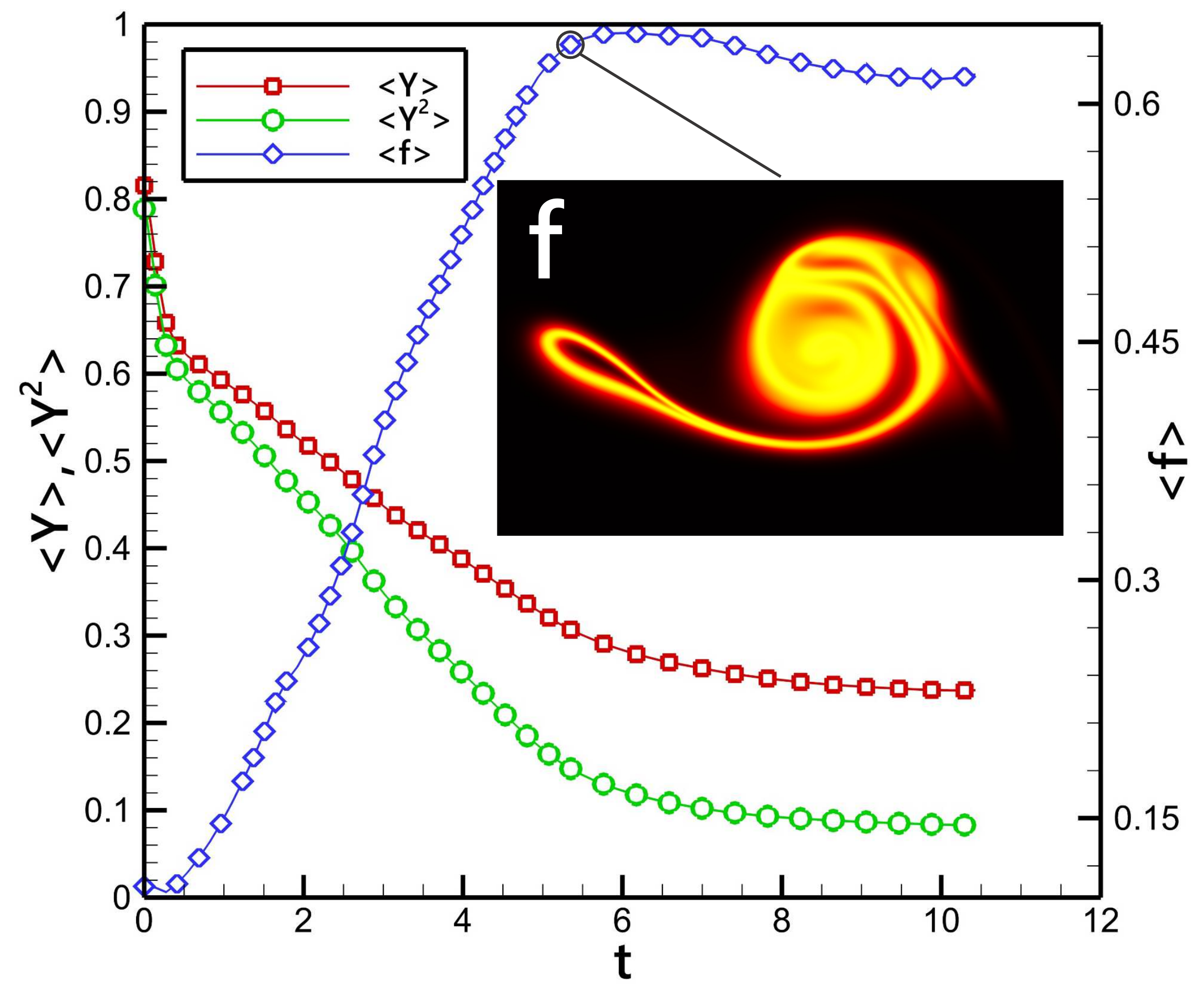}\\
  \caption{Time history of the first moment of mass fraction $\left<Y\right>$, the second moment of mass fraction $\left<Y^2\right>$ and mixedness $\left<f\right>$. Insert: mixedness contour at $t=5.35$.}\label{fig2}
\end{figure}
One important mixing descriptor is the mean concentration of scalar and scalar energy \cite{buch1996experimental}. Here, we study the first and second moment of mass fraction, i.e., mean mass fraction and mean mass fraction energy, based on volume integration defined as:
\begin{equation}\label{eq: meanYI}
  \left< Y \right>(t)=\iiint_\mathcal{V} Y(x,t) d\mathcal{V},
\end{equation}
\begin{equation}\label{eq: meanYIsqu}
  \left<Y^2\right>(t)=\iiint_\mathcal{V} Y(x,t)^2 d\mathcal{V}.
\end{equation}
Then, the mixedness can be defined locally as \cite{cetegen1993experiments}:
\begin{equation}\label{eq: mixed-local}
  f(x,t)=4Y(x,t)(1-Y(x,t)).
\end{equation}
The bulk-integrated mixedness with time has a direct relationship with mean mass fraction $\left<Y\right>$ and mass fraction energy $\left<Y^2\right>$:
\begin{equation}\label{eq: mixed}
  \left<f\right>(t)=4\times\left(\left<Y\right>(t)-\left<Y^2\right>(t)\right).
\end{equation}
Figure~\ref{fig2} illustrates the time evolution of volumetric mean mass fraction $\left<Y\right>$, mass fraction energy $\left<Y^2\right>$, and mixedness $\left<f\right>$. The fundamental observation is the decay of both mean mass fraction and mass fraction energy. The decrease in the mean mass fraction indicates $\mathrm{D}\left<Y\right>/\mathrm{D}t\ne0$. This phenomenon violates the widely-accepted concept of hyperbolic conservation of passive scalar obeying Eq.~(\ref{eq: PS-ADE-1}), which can derive $\left<\varphi\right>$=const. The faster decay of mean mass fraction energy is the inherent characteristic of mixing, leading to an increase of mixedness profile, as shown in Fig.~\ref{fig2}. After $t\approx6$, the mixing indicator turns into a steady status that means the well-mixed state is obtained. Thus, the source of decay of mean mass fraction and mass fraction energy is the key to understanding the mixing enhancement behavior in the variable-density vortical flows. Obviously, the scalar dissipation rate defined from the advection-diffusion equation [Eq.~(\ref{eq: PS-ADE-1})] can not explain the mixing behavior in such RM-type variable-density flows.

\section{Density gradient accelerated dissipation and redistributed diffusion mechanism}
\label{sec:DGAD}
\begin{figure}
    \centering
    \subfigure[]{
    \label{fig3: K1} 
    \includegraphics[clip=true,trim=5 5 5 5,width=.48\textwidth]{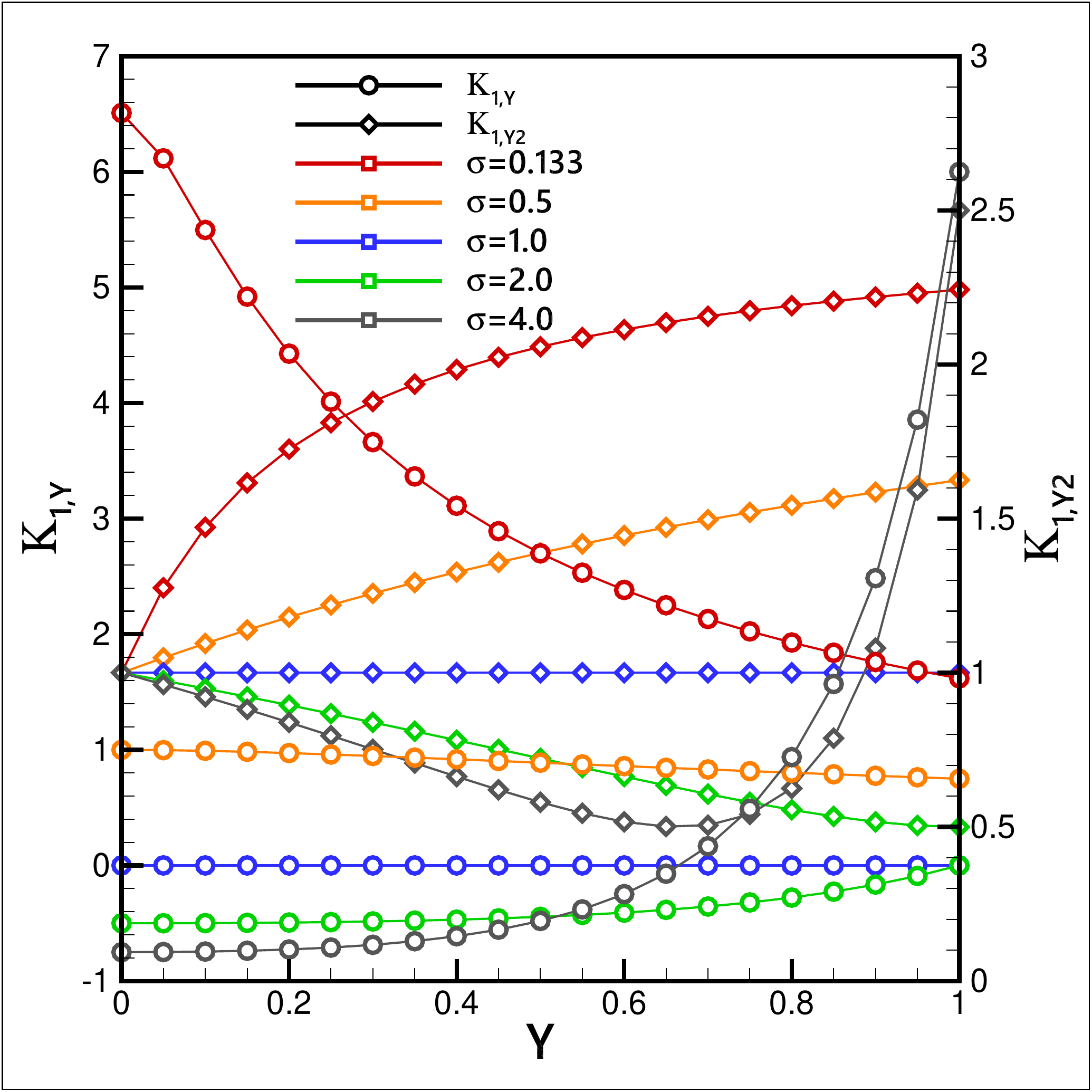}}
    \subfigure[]{
    \label{fig3: K2} 
    \includegraphics[clip=true,trim=5 5 5 5,width=.48\textwidth]{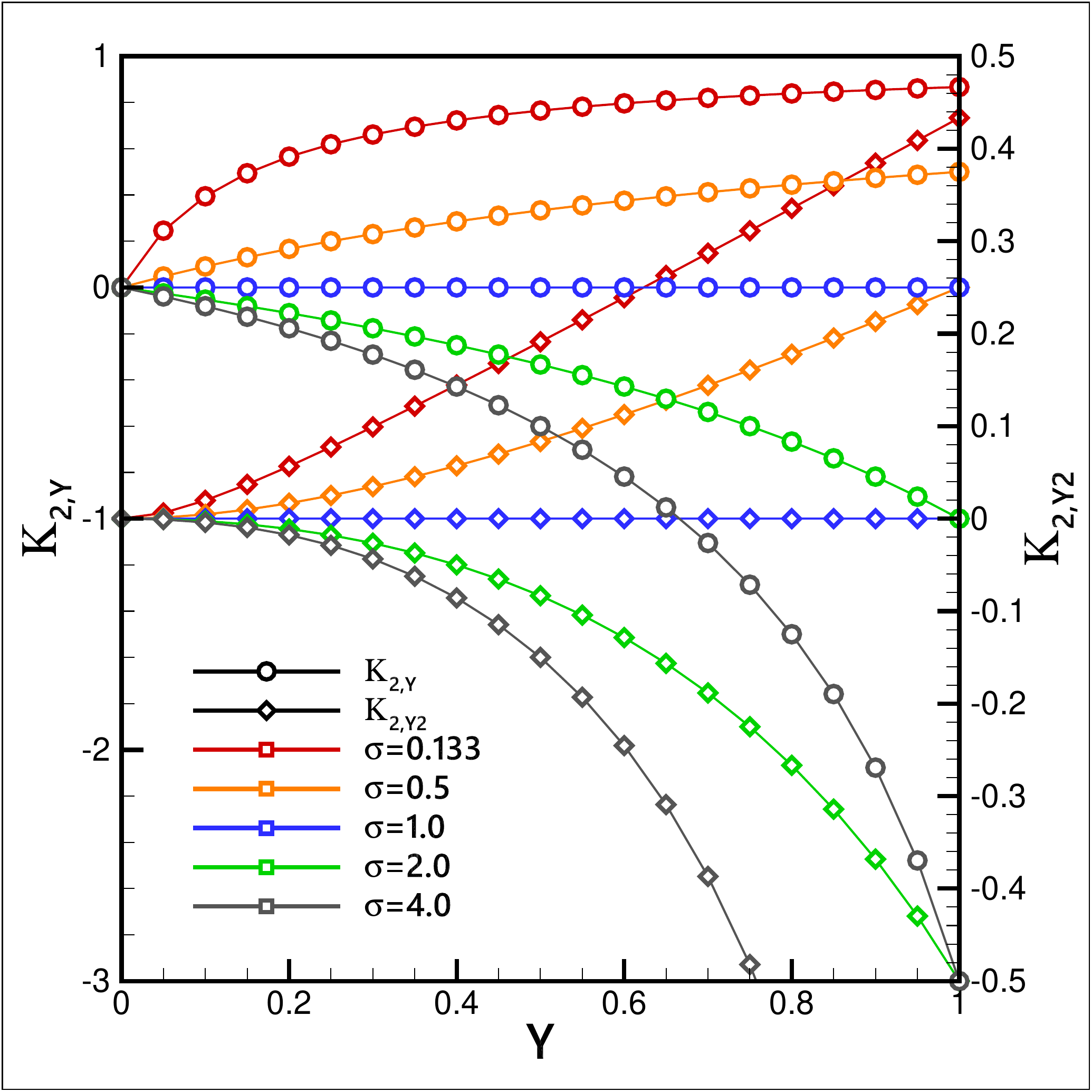}}
    \caption{
    Coefficients of density gradient accelerated dissipation, $\mathscr{K}_{1,Y}$ for the mass fraction $Y$ and $\mathscr{K}_{1,Y^2}$ for the mass fraction energy $Y^2/2$ (a).
    Coefficients of density gradient redistributed diffusion, $\mathscr{K}_{2,Y}$ for the mass fraction $Y$ and $\mathscr{K}_{2,Y^2}$ for the mass fraction energy $Y^2/2$ (b).
    The coefficients are shown with the variation of mass fraction and post-shock density ratio $\sigma$ ranging from 0.133 to 4.0. \label{fig3: K} }
\end{figure}
\subsection{Mixing rate of mass fraction in variable-density flows}
\label{subsec:mixingrate}
Here, we sort to reveal the mechanism that causes the hyperbolic conservation violation in the compressible variable-density mixing flows. We start from the fundamental behavior of material derivative of time in a non-zero divergence of the velocity field.
The first thing we should obtain is the time derivative expression of the mass fraction concerned in this paper. For arbitrary scalar field $\phi(x,t)$, its time derivative of volumetric mean value can be decomposed as:
\begin{equation}\label{eq: derivative}
  \frac{\mathrm{D}  \left(\iiint_{\mathcal{V}}\phi d\mathcal{V}\right)  }{\mathrm{D}t}=\iiint_{\mathcal{V}} \frac{\mathrm{D}\phi}{\mathrm{D}t}d\mathcal{V} + \iiint_{\mathcal{V}} \phi\frac{\mathrm{D}\left(d\mathcal{V}\right) }{\mathrm{D}t}.
\end{equation}
For the first term on the right of Eq.~(\ref{eq: derivative}), it reads the local rate of change of scalar field $\phi$. Due to the conservative characteristic of passive scalar, this term is zero in limits of large Pe number~\cite{jha2013synergetic,cetegen1993experiments}. However, we will show that this term is the leading source for decreasing the mean mass fraction of RM variable-density mixing.
For the second term on the right of Eq.~(\ref{eq: derivative}), it reflects the rate of change in the volume occupied by the scalar field.
In the compressible flows, the material derivative of a finite volume is controlled by the divergence of velocity \cite{anderson2010fundamentals}:
\begin{equation}\label{eq: divergence}
  \frac{\mathrm{D}\left(d\mathcal{V}\right) }{\mathrm{D}t}=\left(\nabla\cdot \textbf{V}\right) d\mathcal{V}.
\end{equation}
Although this term is usually not modeled in conservative passive scalar mixing of incompressible mixing, which leads $\mathrm{D}\left<\varphi\right>/\mathrm{D}t=\left<\mathrm{D}\varphi/\mathrm{D}t\right>=0$, velocity divergence will occur in compressible passive scalar mixing that makes the mixing area of passive scalar $\varphi$ decrease or increase due to either compression or expansion of local flow element \cite{pan2010mixing}. Nevertheless, the divergence-free assumption is accepted by most studies since if $\nabla\cdot\textrm{V}\ne0$, the concentration of scalar $\varphi$ will take values larger than 1 or take negative values in the form of advection-diffusion equations with the source of Fisher-Kolmogorov-Petrovskii-Piskunov reaction rate~\cite{branco2007numerical}. For most variable-density flows, velocity divergence exists even in incompressible flows (\cite{sandoval1995dynamics} and see Appendix~\ref{sec:app5}). Once the complete source of the time derivative of the scalar is known, problems remained are the exact expression that reveals the physical mechanism leading to the anomalous decreasing of mean concentration.

For mass fraction $Y$, the time derivative of its volumetric mean can be expressed as:
\begin{equation}\label{eq: dYIdt}
  \frac{\mathrm{D}\left<Y\right>}{\mathrm{D}t}=\left<\frac{\mathrm{D}Y}{\mathrm{D}t}\right>
  +\left<Y\left(\nabla\cdot \textbf{V}\right)\right>,
\end{equation}
and one of mass fraction energy is expressed as:
\begin{equation}\label{eq: dYIsqudt}
  \frac{\mathrm{D}\left<Y^2/2\right>}{\mathrm{D}t}=\left<\frac{\mathrm{D}(Y^2/2)}{\mathrm{D}t}\right>
  +\left<Y^2/2\left(\nabla\cdot \textbf{V}\right)\right>.
\end{equation}

We firstly model the first term on the right hand of Eqs.~(\ref{eq: dYIdt}) and (\ref{eq: dYIsqudt}).
By using the canonical correlation between mass fraction and density in multi-species miscible flows \cite{tomkins2013evolution}:
\begin{equation}\label{eq: Y-rho}
  \frac{1}{\rho}=\frac{Y}{\rho'_2}+\frac{1-Y}{\rho'_1},
\end{equation}
and introducing $\sigma=\rho'_2/\rho'_1$ as the post-shock density ratio and $\rho'_2=\rho_2/\eta$ (see more details in Appendix~\ref{sec:app3}).
From the dimensionless transport equation of species as Eq.~(\ref{eq: trans for mass}), we can obtain:
\begin{equation}\label{eq: dYdt-SDR}
  \left(\frac{\partial}{\partial t}+\textbf{V}\cdot\nabla-\frac{1}{\mathrm{Pe}}\nabla^2\right)Y=-\frac{1}{\mathrm{Pe}}\frac{1-\sigma}{(1-\sigma)Y+\sigma}\nabla Y\cdot\nabla Y.
\end{equation}
Due to $\left<\frac{1}{\mathrm{Pe}}\nabla^2Y\right>=0$ (proof and discussion are shown in Appendix~\ref{sec:app4}), then we get:
\begin{equation}\label{eq: dYdt-SDR-int}
  \left<\frac{\mathrm{D}Y}{\mathrm{D}t}\right>=-\left<\frac{1}{\mathrm{Pe}}\frac{1-\sigma}{(1-\sigma)Y+\sigma}\nabla Y\cdot\nabla Y\right>.
\end{equation}
From Eq.~(\ref{eq: dYdt-SDR}), one can obtain the convection-diffusion equation for mass fraction energy:
\begin{equation}\label{eq: dY2dt-SDR}
  \left(\frac{\partial}{\partial t}+\textbf{V}\cdot\nabla-\frac{1}{\mathrm{Pe}}\nabla^2\right)\frac{1}{2}Y^2=-\frac{1}{\mathrm{Pe}}\left(2-\frac{\sigma}{(1-\sigma)Y+\sigma}\right)\nabla Y\cdot\nabla Y.
\end{equation}
Due to $\left<\frac{1}{\mathrm{Pe}}\nabla^2(Y^2/2)\right>=0$, then one obtains:
\begin{equation}\label{eq: dY2dt-SDR-int}
  \left<\frac{\mathrm{D}(Y^2/2)}{\mathrm{D}t}\right>=-\left<\frac{1}{\mathrm{Pe}}\left(2-\frac{\sigma}{(1-\sigma)Y+\sigma}\right)\nabla Y\cdot\nabla Y\right>.
\end{equation}
More details of the above derivation are shown in Appendix~\ref{sec:app2}. Here we can find a strictly negative term for advection equation of mass fraction in Eq.~(\ref{eq: dYdt-SDR}), partly explaining the decrease of mean volumetric mass fraction observed. Moreover, this term takes a similar form of scalar dissipation and converges to zero as $\sigma=1$, i.e., the constant-density passive scalar mixing scenario.

For the second term of on the right hand of Eqs.~(\ref{eq: dYIdt}) and (\ref{eq: dYIsqudt}), velocity divergence exists even for incompressible flows in the variable-density mixing. Besides, due to the first shock impact brings the velocity divergence embedded in shock, we can express the velocity divergence term as:
\begin{equation}\label{eq: div}
  \nabla\cdot\textbf{V}=-\nabla\cdot\left(\frac{1}{\mathrm{Pe}}\frac{\nabla\rho}{\rho}\right)+\left(\nabla\cdot\textbf{V}\right)_S.
\end{equation}
The second part of divergence becomes small immediately after shock impacts (see discussion in Appendix~\ref{sec:app5}). By using Eq.~(\ref{eq: Y-rho}), then one can obtain the complete expression for the right term of Eq.~(\ref{eq: dYIdt}) in the form of mass fraction:
\begin{eqnarray}\label{eq: dYdt-SDR-com}
  \frac{\mathrm{D}\left<Y\right>}{\mathrm{D}t} & = & \underbrace{\left<-\frac{1}{\mathrm{Pe}}\mathscr{K}_{1,Y}(\sigma,Y)\nabla Y\cdot\nabla Y\right>}_{\text{DG Accelerated Dissipation}}+\underbrace{\left<\frac{1}{\mathrm{Pe}}\mathscr{K}_{2,Y}(\sigma,Y)\nabla^2 Y\right>}_{\text{DG Redistributed Diffusion}} \\
  & = & \left<-\chi_{ad,Y}\right>+\left<-\chi_{rd,Y}\right>,
\end{eqnarray}
with coefficients $\mathscr{K}_{1,Y}$ on the density gradient accelerated dissipation term (DGAD for short) and $\mathscr{K}_{2,Y}$ on redistributed diffusion term (DGRD for short):
\begin{equation}\label{eq: dYIdt-K}
\left\{ \begin{array}{l}
  \mathscr{K}_{1,Y}(\sigma,Y)=\Psi\cdot\left(1+\Psi Y\right), \\
  \mathscr{K}_{2,Y}(\sigma,Y)=\Psi Y, \\
  \Psi=(1-\sigma)/\left((1-\sigma)Y+\sigma\right),
\end{array} \right.
\end{equation}
and complete expression for the decay rate of mass fraction energy [the right term of Eq.~(\ref{eq: dYIsqudt})] in the form of mass fraction:
\begin{eqnarray}\label{eq: dY2dt-SDR-com}
  \frac{\mathrm{D}\left<Y^2/2\right>}{\mathrm{D}t} & = & \underbrace{\left<-\frac{1}{\mathrm{Pe}}\mathscr{K}_{1,Y^2}(\sigma,Y)\nabla Y\cdot\nabla Y\right>}_{\text{DG Accelerated Dissipation}}+\underbrace{\left<\frac{1}{\mathrm{Pe}}\mathscr{K}_{2,Y^2}(\sigma,Y)\nabla^2 Y\right>}_{\text{DG Redistributed Diffusion}} \\
  & = & \left<-\chi_{ad,Y^2}\right>+\left<-\chi_{rd,Y^2}\right>,
\end{eqnarray}
also with the coefficient on the accelerated dissipation term and redistributed diffusion term:
\begin{equation}\label{eq: dY2dt-K}
\left\{ \begin{array}{l}
  \mathscr{K}_{1,Y^2}(\sigma,Y)=\left(1+\Psi Y\right)^2/2+1/2, \\
  \mathscr{K}_{2,Y^2}(\sigma,Y)=\Psi Y^2/2.
\end{array} \right.
\end{equation}

To gain the effect of the density ratio $\sigma$ on these coefficients, we plot the coefficients of accelerated dissipation and redistribution diffusion with the variation of mass fraction $Y$, as shown in Fig.~\ref{fig3: K}. The first observation is that when $\sigma=1$, the coefficient degenerated to the constant-density passive scalar mixing \cite{buch1996experimental}, that is $\mathscr{K}_{1,Y}(\sigma,Y)=\mathscr{K}_{2,Y}(\sigma,Y)=\mathscr{K}_{2,Y^2}(\sigma,Y)=0$ and $\mathscr{K}_{1,Y^2}(\sigma,Y)=1$.
which shows the generalization of mixing rate expression under a wide range density difference.

Secondly, for the density gradient acceleration term $\mathscr{K}_{1,Y}(\sigma,Y)$ and $\mathscr{K}_{1,Y^2}(\sigma,Y)$, the coefficient is much larger than passive scalar mixing when $\sigma<1$ for the light gas case, as shown in Fig.~\ref{fig3: K1}. When the mass fraction is low ($\sigma=0.133$), the coefficient $\mathscr{K}_{1,Y}(\sigma,Y)$ is near 7, which means that the mixing will decay faster due to the existence of density gradient at lower value of the mass fraction~\cite{liu2020mixing}. Thus, we call this term density gradient accelerated dissipation because the density gradient amplifies dissipation of the scalar mass fraction.
As far as heavy gas ($\sigma>1$) is concerned, $\mathscr{K}_{1,Y}(\sigma,Y)$ for mass fraction is negative, leading to the increase of mass fraction, which also occurs in the variable-density cases studied in the later section.

Thirdly, as for the redistributed diffusion term $\mathscr{K}_{2,Y}(\sigma,Y)$ or $\mathscr{K}_{2,Y^2}(\sigma,Y)$, as shown in Fig.~\ref{fig3: K2}, it is generally lower than the dissipation term when $\sigma<1$. Monotonous growth with mass fraction shows that more diffusion will gain when the mass fraction concentration is higher.
When $\sigma>1$, the redistributed diffusion term's coefficients become negative, whose absolute value is higher when density difference is higher. In this scenario, the redistributed diffusion term will take a dominant role in the mass fraction or its energy growth rate.
One thing needs to note that the diffusion term $\nabla^2Y$ is not strictly negative, meaning that the diffusion term redistributes the local growth of the mass fraction and mass fraction energy when variable-density mixing happens.
The behavior of density gradient accelerated dissipation and redistributed diffusion in the case of SBI will be introduced in the next section.

\subsection{Behavior of density gradient accelerated dissipation and redistributed diffusion}
\label{subsec:DGAD}
\begin{figure}
    \centering
    \subfigure[]{
    \label{fig4: dYdt} 
    \includegraphics[clip=true,trim=20 20 25 60,width=.48\textwidth]{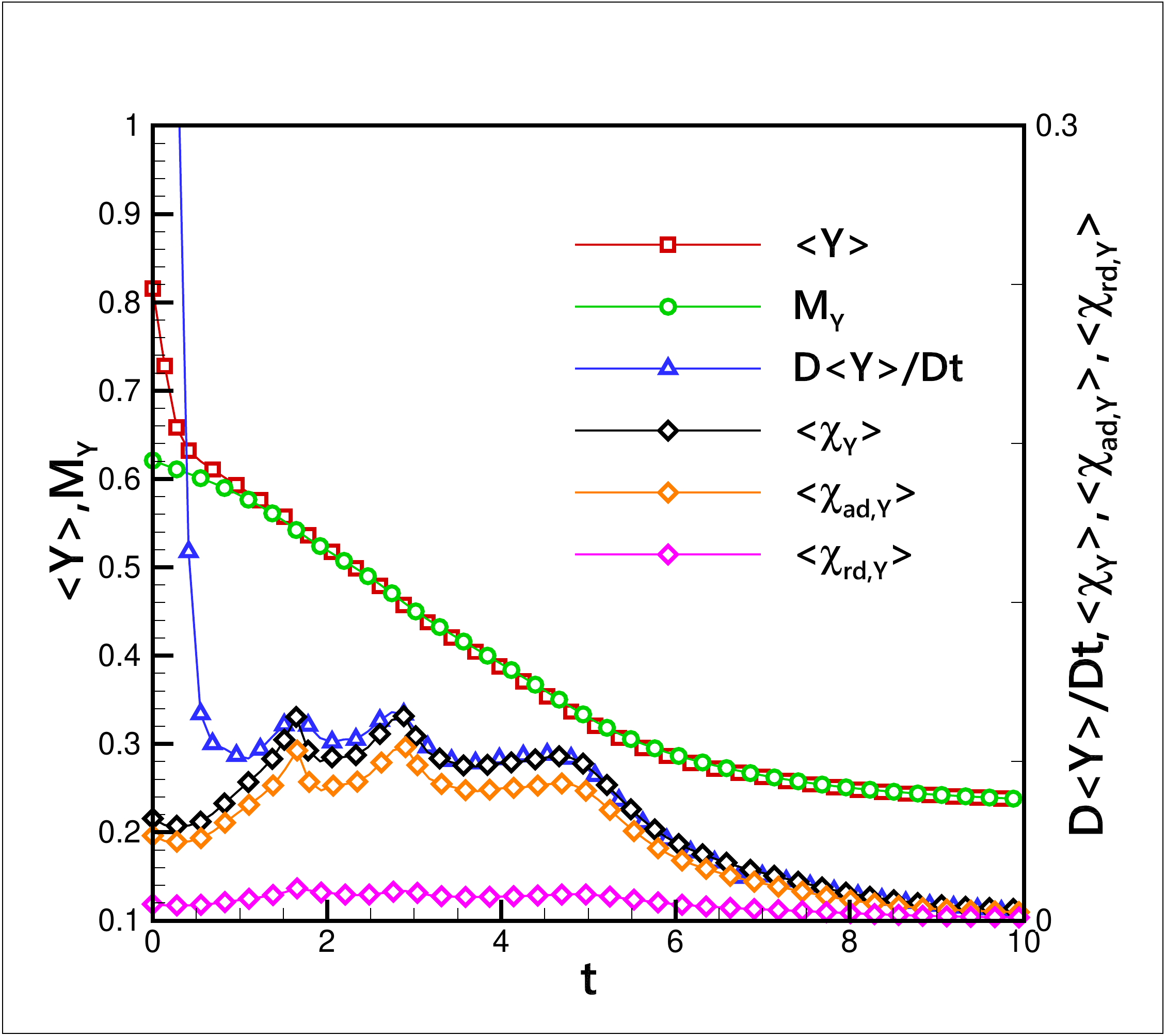}}
    \subfigure[]{
    \label{fig4: dY2dt} 
    \includegraphics[clip=true,trim=20 20 25 60,width=.48\textwidth]{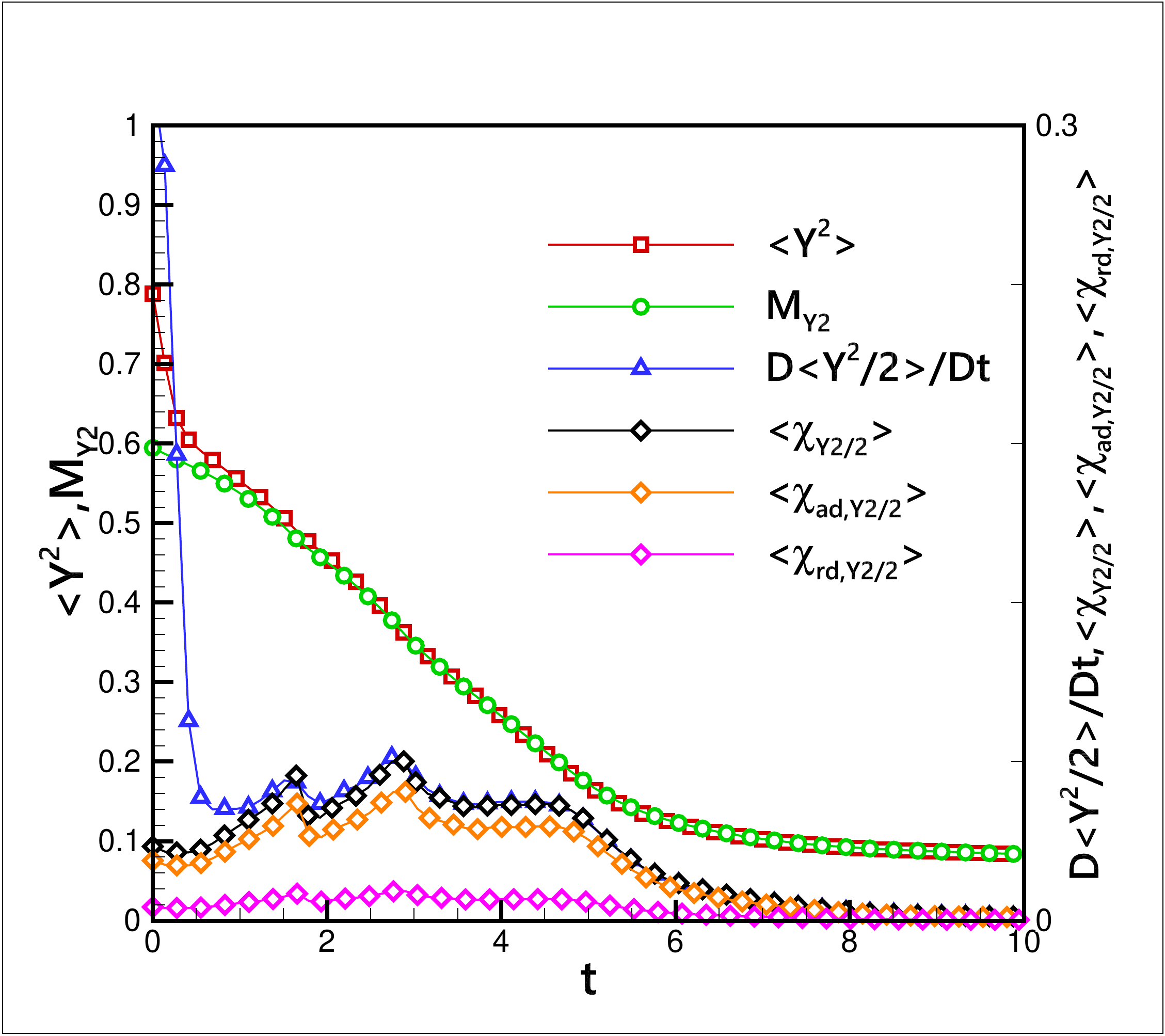}}
    \caption{(a) Mixing rate of $\left<Y\right>$ composed by DGAD and DGRD term, validating Eq.~(\ref{eq: dYdt-SDR-com}). (b) Mixing rate of $\left<Y^2\right>$ composed by DGAD and DGRD term, validating Eq.~(\ref{eq: dY2dt-SDR-com}). Both mean mass fraction $\left<Y\right>$ and mass fraction energy $\left<Y^2\right>$ decreases monotonically with time. The integration of mixing rate is defined in Eqs.~(\ref{eq: int_chi_Y}) and (\ref{eq: int_chi_Y2}). The general agreement between local mixing rate and time derivative of macroscopic mixing can be found.\label{fig4} }
\end{figure}
Now, we pay attention to the DGAD and DGRD behavior in SBI.
In order to validate the local mixing rate and global mean mixing descriptor, the time integral of DGAD and DGRD is defined and compared with the first moment of mass fraction:
\begin{equation}\label{eq: int_chi_Y}
  \mathcal{M}_Y(t)=\mathcal{M}_{Y,0}-\int_0^t\left(\left<\chi_{ad,Y}\right>(t')+\left<\chi_{rd,Y}\right>(t')\right)\mathrm{d}t' ,
\end{equation}
and with the second moment of mass fraction:
\begin{equation}\label{eq: int_chi_Y2}
  \mathcal{M}_{Y^2}(t)=\mathcal{M}_{Y^2,0}-
       2\int_0^t\left(\left<\chi_{ad,Y^2}\right>(t')+\left<\chi_{rd,Y^2}\right>(t')\right)\mathrm{d}t'.
\end{equation}
The sudden decrease of mean value from the first compression from shock is eliminated by introducing the initial integration of $\mathcal{M}_{Y,0}$ and $\mathcal{M}_{Y^2,0}$.
Figure~\ref{fig4} shows the comparison between mean concentration $\left<Y\right>$ ($\left<Y^2\right>$) and mixing rate integral $\mathcal{M}_{Y}$ ($\mathcal{M}_{Y^2}$) in shock helium cylindrical bubble interaction. General agreement is observed, validating that both the density gradient accelerated dissipation and redistributed diffusion contributes to the decrease of mass fraction in a variable-density problem.
In accordance with the analysis on the coefficient of DGAD and DGRD, the accelerated dissipation contributes much larger than the redistributed diffusion to the decrease of mean mass fraction when $\sigma$ is small, as depicted in Fig.~\ref{fig4}.
Moreover, the time derivative of mean concentration and volumetric integration of the mixing rate composed by DGAD and DGRD also collapse with good agreement.

\begin{figure}
    \centering
    \subfigure[]{
    \label{fig5-1} 
    \includegraphics[clip=true,trim=0 100 0 0,width=.87\textwidth]{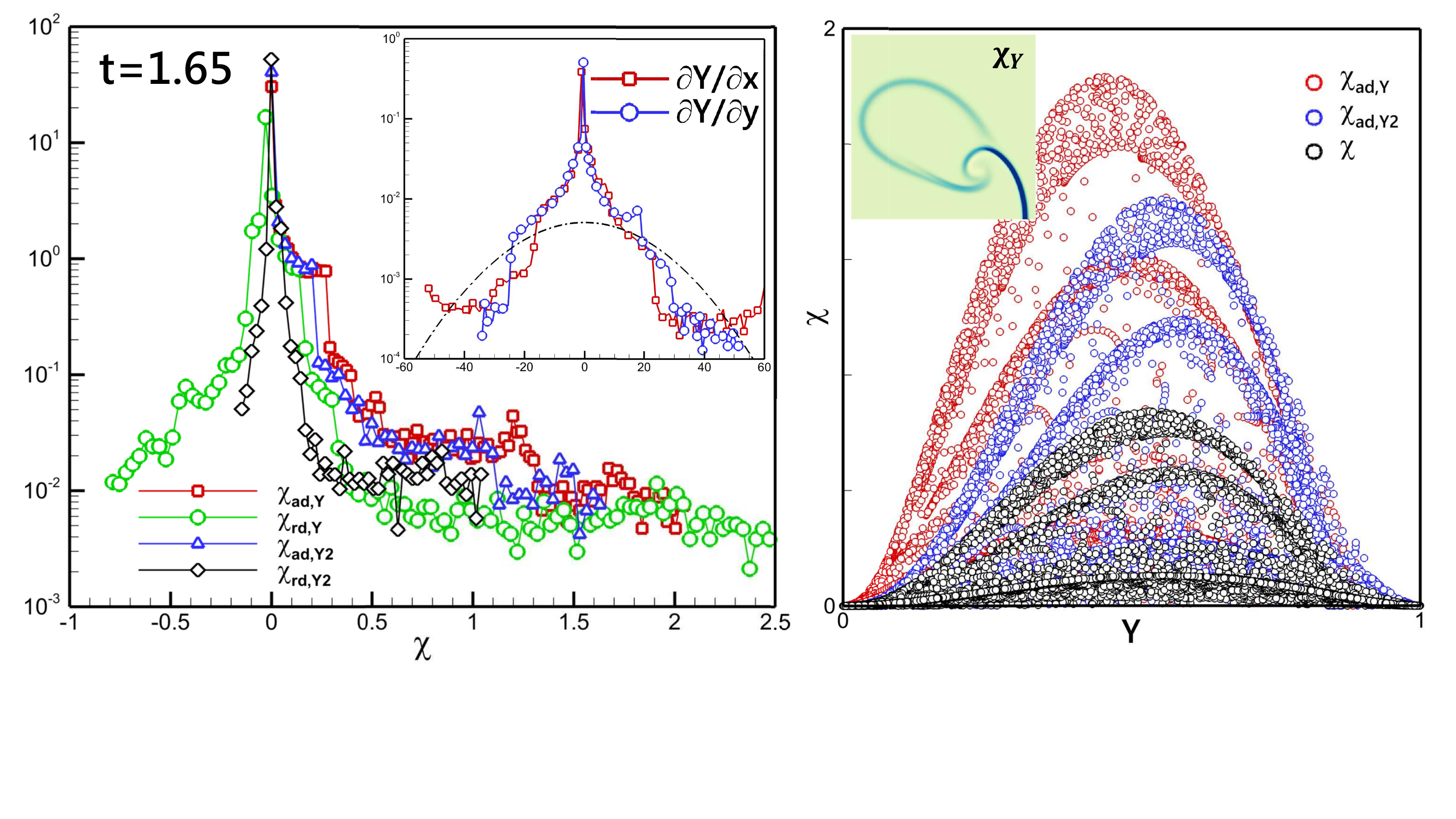}} \\
    \subfigure[]{
    \label{fig5-2} 
    \includegraphics[clip=true,trim=0 100 0 0,width=.87\textwidth]{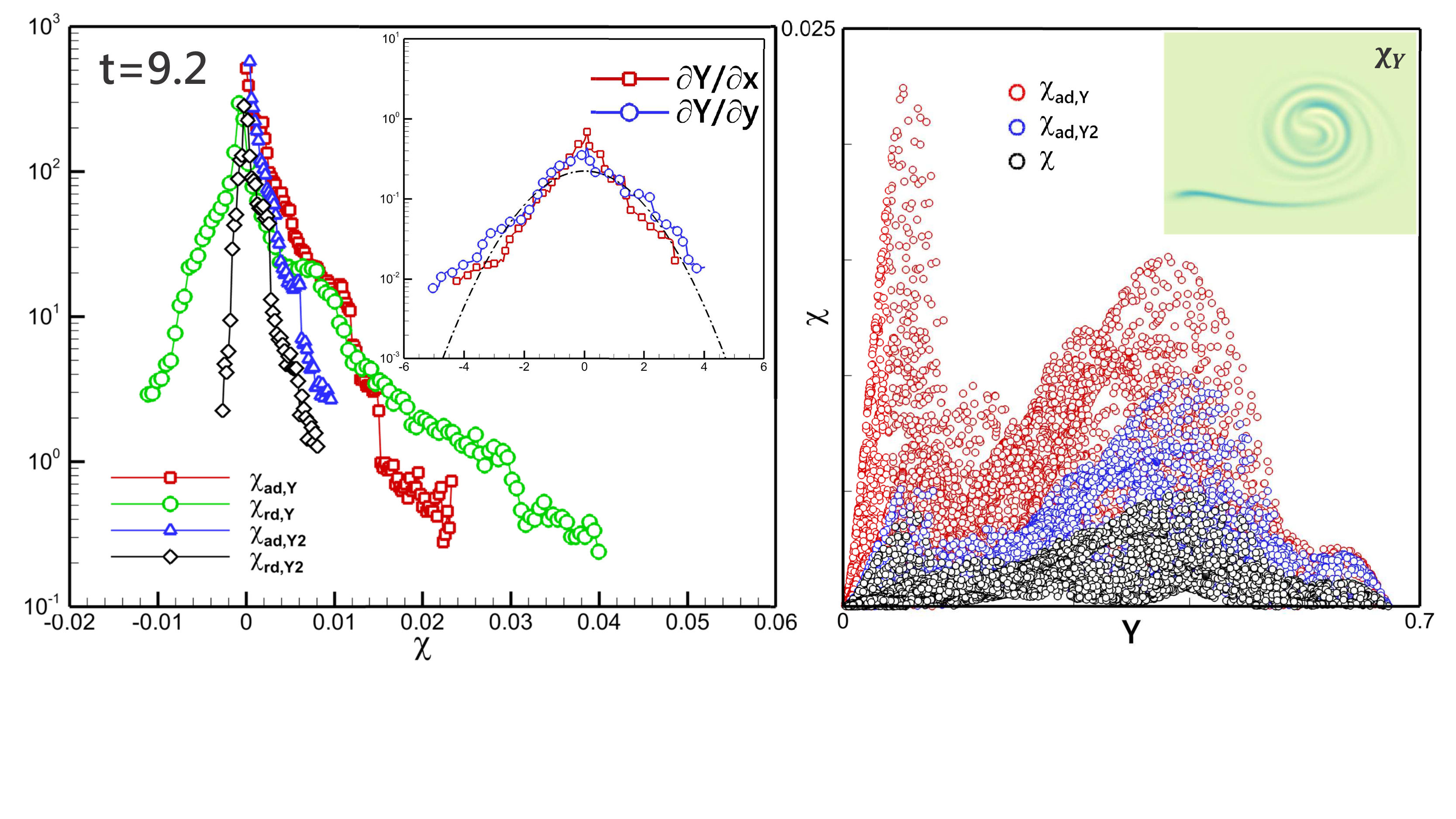}}
    \caption{ Left: probability density function (PDF) of DGAD, DGRD, and derivative of the mass fraction (inserted figure). Right: the dissipative structure of density gradient accelerated dissipation with mass fraction comparing with the constant-density scalar dissipation $\chi=1/\mathrm{Pe}\nabla Y\cdot\nabla Y$. The contour of the mass fraction mixing rate is inserted. At an early time of $t=1.65$ (a), $\partial Y/\partial x$ shows the characteristic exponential tails while it tends toward a Gaussian behavior (dashed dot line) at late time $t=9.2$ (b), meaning the homogeneity of the flow. \label{fig5} }
\end{figure}
Further, probability density function (PDF for short) offers the dissipative structure of DGAD and DGRD, as shown in Fig.~\ref{fig5}. At an early time of $t=1.65$, the DGAD term's PDF shows the steep distribution of both mass fraction and its energy, indicating the small amount of high mixing rate, as shown in the left part of Fig.~\ref{fig5-1}. As for the DGRD term, considerable counts show the opposite sign. However, nearly the same amount counts of DGRD term is positive, which off-set the negative part. Moreover, the DGRD term with high-value points is much less than DGAD, meaning the minor effect of DGRD term on the mixing rate. Thus, the scatter points of the DGAD term are plotted in the right part of Fig.~\ref{fig5-1}. It can be found that at about $Y=0.5$, the dissipation rate is the highest. The density gradient accelerated dissipation of mass fraction, and its energy is more extensive than the value obtained from the passive scalar dissipation, showing the inherent nature of faster decay in variable-density mixing. Also, the figure inserted is the mixing rate of the mass fraction. It offers the information that a high mixing rate concentrates on the bridge structure. At a later time of $t=9.2$, both DGAD and DGRD values are much lower than those early, indicating a steady mixing state. Still, DGRD becomes dominant to homogenize the mass fraction. This homogenization is also validated by the Gaussian distribution of the mass fraction gradient \cite{li2019gaussian}, as shown in the inserted figure in the left part of Fig.~\ref{fig5-2}.

\begin{figure}
  \centering
  \includegraphics[clip=true,trim=0 75 90 0,width=.99\textwidth]{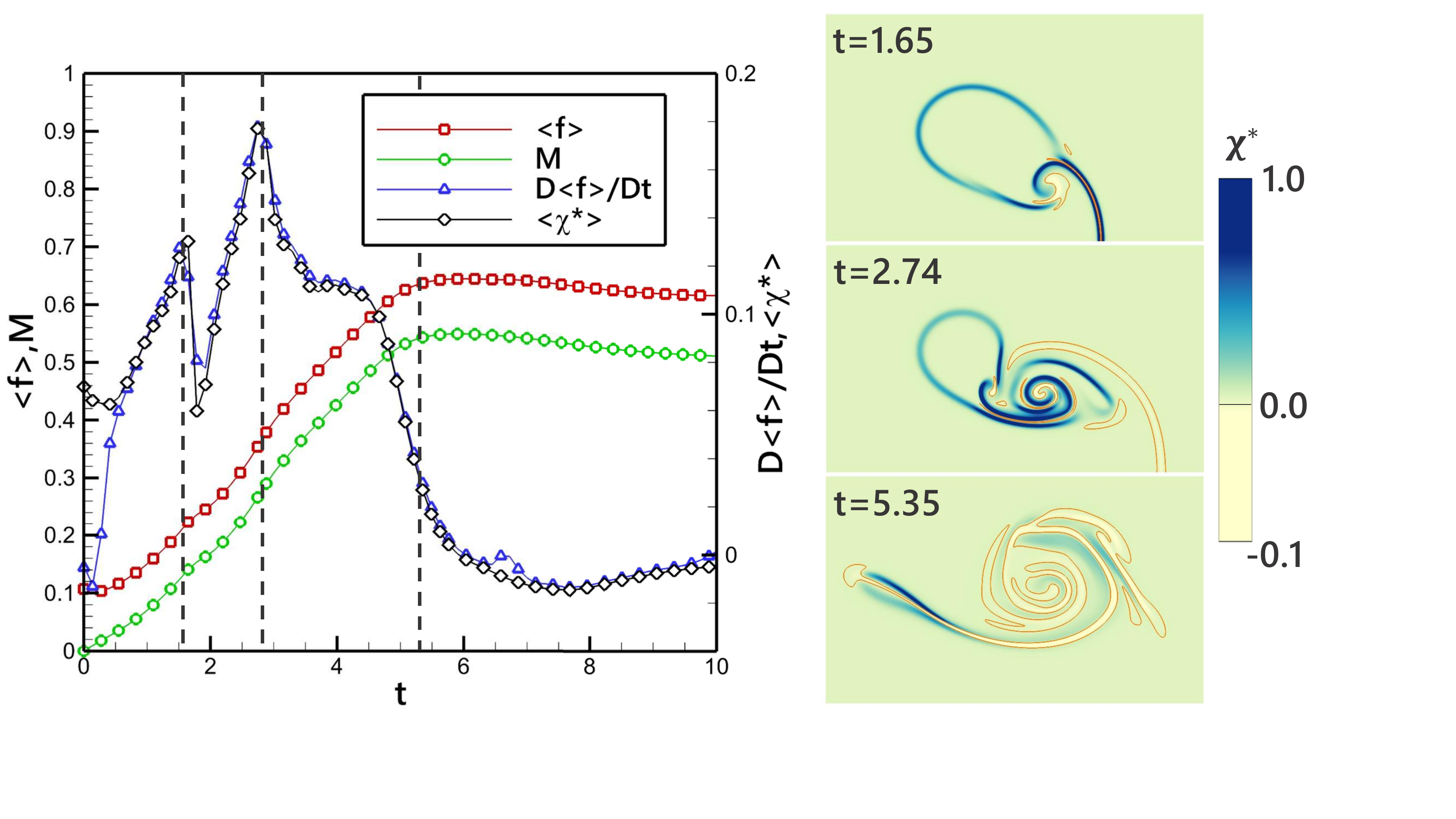}\\
  \caption{Left: Time evolution of mixedness and its time derivative comparing with $\left<\chi^*\right>$. Right: The flow structures, rendered by contour maps of density gradient accelerated mixing rate $\chi^*$ are also explicitly given at three specific time instants. The isoline of $\chi^*=0$ is also plotted to show the region with the negative mixing rate.}\label{fig6}
\end{figure}
\subsection{Mixedness formation}
Once the dissipation rate of mass fraction and its energy are obtained, the dissipation rate of mixedness can be easily derived based on the definition of mixedness Eq.~(\ref{eq: mixed}):
\begin{equation}\label{eq: dfdt}
  \frac{\mathrm{D}\left<f\right>}{\mathrm{D}t}=\frac{\mathrm{D}\left<4Y\right>}{\mathrm{D}t}
                                                             -\frac{\mathrm{D}\left<4Y^2\right>}{\mathrm{D}t}=\left<\chi^*\right>.
\end{equation}
A new dissipation rate $\chi^*$ for mixedness in variable-density mixing can be expressed as:
\begin{equation}\label{eq: dfdt-SDR}
  \chi^*=\frac{4}{\mathrm{Pe}}\mathscr{K}_{1,f}(\sigma,Y)\nabla Y\cdot\nabla Y+
               \frac{4}{\mathrm{Pe}}\mathscr{K}_{2,f}(\sigma,Y)\nabla^2 Y,
\end{equation}
also with the coefficient of density gradient accelerated dissipation term and redistributed diffusion term:
\begin{equation}\label{eq: dfdt-K}
\left\{ \begin{array}{l}
  \mathscr{K}_{1,f}(\sigma,Y)=5-\frac{2(\sigma+1)}{(1-\sigma)Y+\sigma}+\frac{\sigma}{\left((1-\sigma)Y+\sigma\right)^2}, \\
  \mathscr{K}_{2,f}(\sigma,Y)=\frac{Y(1-Y)(1-\sigma)}{(1-\sigma)Y+\sigma}.
\end{array} \right.
\end{equation}
A time integral of mixing rate $\left<\chi^*\right>$ is defined to compare with the mixedness profile.
\begin{equation}\label{eq: int_SDR}
  \mathcal{M}(t)=\int_0^t\left<\chi^*\right>(t')\mathrm{d}t'.
\end{equation}

Figure~\ref{fig6} shows the time evolution of mixedness and relative variables.
Except for a little discrepancy observed at the early time due to the first shock compression, the general agreement is found between the density accelerated mixing rate and the time derivative of mixedness.
The time history of mixing rate temporal integral shows the remarkable similarity with mixedness except that the initial mixedness from the diffusive layer is not considered in Eq.~(\ref{eq: int_SDR}).
Three specific time instant of mixing rate $\chi^*$ are given on the right side of Fig.~\ref{fig6}. A high mixing rate occurs at the bridge structure and connector of the vortex and lobe from the observation. This causes the local peak of the volumetric integrated mixing rate. At later times, the mixing rate becomes negative due to the redistributed diffusion phenomenon, which will not occur in passive scalar mixing. From the mixedness profile and time integral of the mixing rate, two mixing stages can be determined. The first stage is the mixing growth stage, mainly due to the stretching of the vortex. The second stage is the steady mixing stage, in which redistributed diffusion dominates even decreases the degree of mixing. This two-stage mixing rate shows the stirring effect from the baroclinic vortex and equilibrium diffusion at the late time, which implies a vital scaling behavior of the mixing rate.

\section{Scaling behavior of mixing rate and mixedness}
\label{sec:scaling}
\subsection{Scaling behavior of mixing rate on Pe number and Re number}
\label{subsec:scaling1}
\begin{figure}
  \centering
  \includegraphics[clip=true,trim=0 0 300 0,width=.6\textwidth]{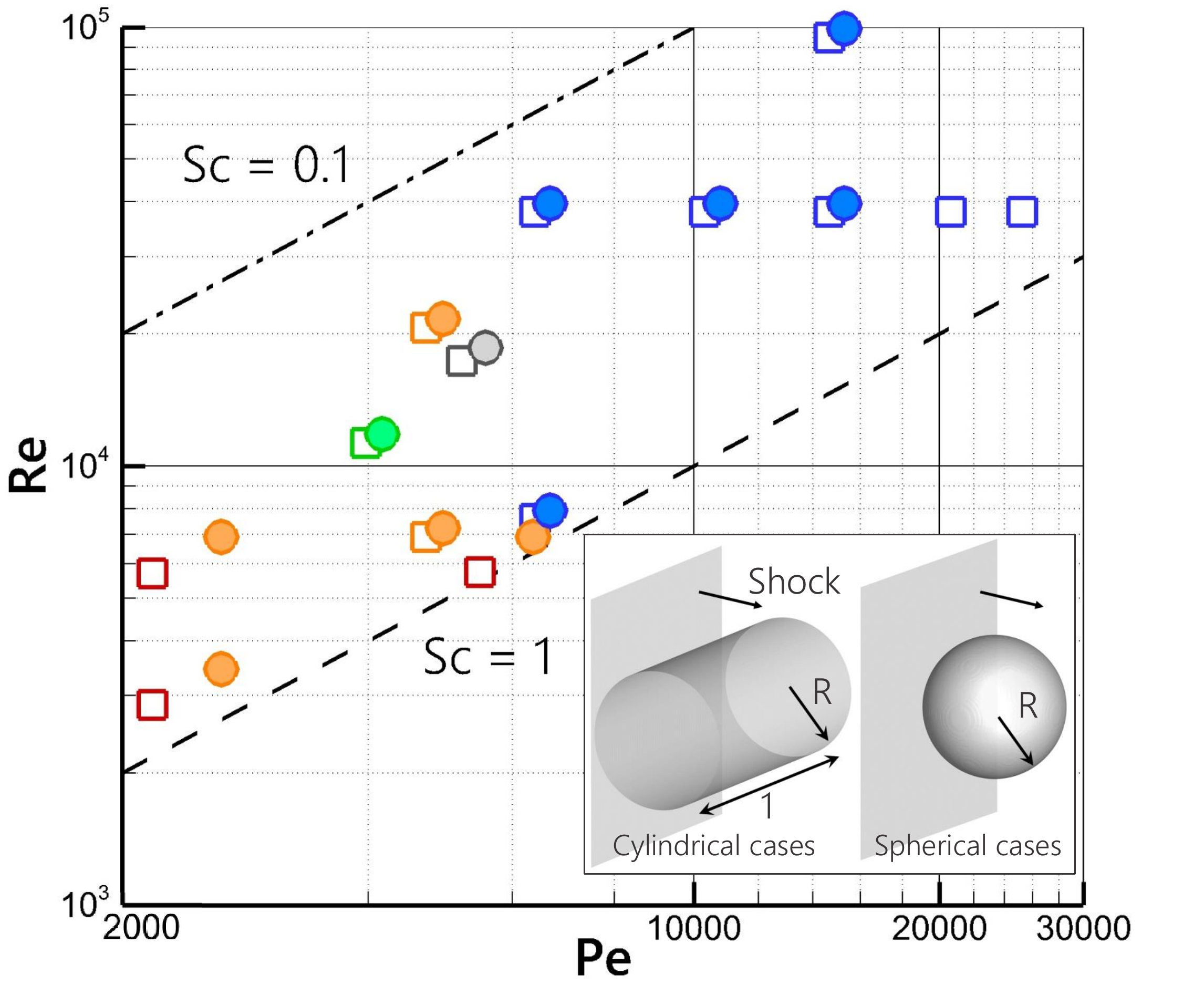}\\
  \caption{All helium bubble cases concerned range from Pe~$\approx2500\sim25000$ and Re~$\approx3\times10^3\sim1\times10^5$. Square dots represent cylindrical cases, and circular dots represent spherical cases. Red: Ma = 1.22; Orange: Ma = 1.8; Blue: Ma = 2.4; Green: Ma = 3; Black: Ma~=~4. Isoline of Sc = 0.1 and 1 are plotted as dashed-dot line and dashed line respectively. The inserted figure illustrates the volumetric difference between cylindrical cases and spherical cases.  }\label{Figres1-cases}
\end{figure}
Although the time-dependent mixing rate exhibits the ups and downs during the mixing growth due to specific mixing structures such as bridge, the overall mixedness growth shows the quasi-linear behavior, indicating the constant average mixing rate.
Here we examine shock interacting with cylindrical and spherical bubbles with a wide range of Ma, Re, and Pe numbers (Sc numbers within a range of 0.1$\sim$1.0, typical in the gaseous mixture \cite{ranjan2011shock,wong2019high}), as shown in Fig.~\ref{Figres1-cases}.
Detailed controlling parameters can be found in Supplementary Material \cite{supp}.
The spherical bubble is simulated by two-dimensional axisymmetric boundary conditions as introduced in Sec.~\ref{sec:method}.
It is noteworthy that the integration of an axisymmetric variable $\phi$ is $\iint_\mathcal{V} 2\pi y\phi(x,y,t) dydx$, where $y$ is the distance to axis. In order to compare the spherical cases with cylindrical cases in the same level of magnitude, we revise the integration of the non-dimensional axisymmetric variables, such as mean mass fraction or mixedness, by a coefficient $\mathrm{V}_c/\mathrm{V}_s=3/(2D)$, where $\mathrm{V}_c$ and $\mathrm{V}_s$ are the volumes of a cylinder and a sphere with diameter $D$, as illustrated in Fig.~\ref{Figres1-cases}.

\begin{figure}
  \centering
  \includegraphics[clip=true,trim=5 2 5 10,width=.99\textwidth]{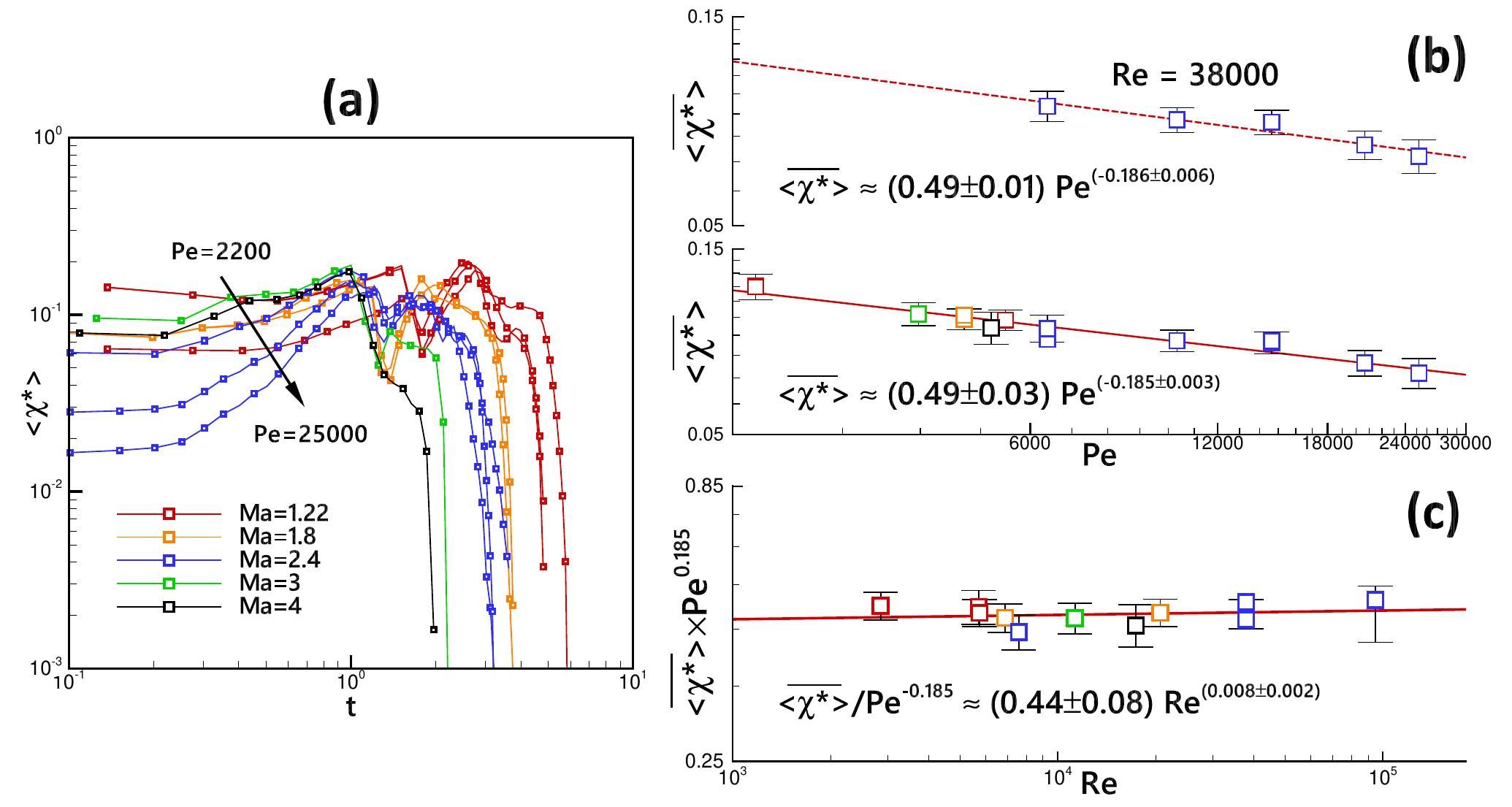}\\
  \caption{(a) Time evolution of density gradient accelerated mixing rate $\left<\chi^*\right>$ for cylindrical helium bubble cases.
  For clarity, Ma = 2.4 curves presented are the ones at Re = 38000, Pe = 6400; Pe = 15000 and Pe = 25000 respectively.
  (b) and (c) show the time-averaged accelerated mixing rate calculated as $\overline{\left<\chi^*\right>}=\mathcal{M}(T)/T$, where $T$ is a time window before the rapid decay of second stage mixing for averaging (at time intervals of $\delta t=0.1$). (b) Upper: the scaling of time-averaged mixing rate $\overline{\left<\chi^*\right>}$ on Pe number at constant Re = 38000 and Ma = 2.4. Lower: the scaling of $\overline{\left<\chi^*\right>}$ on Pe number for all cases. (c) The scaling of time-averaged mixing rate independent of Pe number, $\overline{\left<\chi^*\right>}\cdot\textrm{Pe}^{0.185}$, on Re number. We indicate, with a dashed or solid line for each figure, the best power-law fit over the concerned data, which shows the weak dependence of mixing rate on Pe number or Re number. }\label{Figres2-cylinscale}
\end{figure}
To outline the influence of Re and Pe numbers on mixing rate, we first analyze the time history of the mixing rate $\left<\chi^*\right>$ of all cylindrical cases as depicted in Fig.~\ref{Figres2-cylinscale}(a). It can be found that a similar magnitude of mixing rate in all cases is obtained. The mixing rate slightly decreases with the increase of Pe number. Still, for higher Mach number, the steady mixing state is earlier than low Mach number cases due to the strong compression leading to a smaller quantity of mass fraction, as analyzed in the following part.
To compare more precisely the mixing rate of different cases, we introduce a time-average dissipation rate during the first stage of mixing growth, defined as $\overline{\left<\chi^*\right>}=\mathcal{M}(T)/T$. The integration time $T$ is longer for lower Mach number and shorter for higher Mach number. Since log coordinate is used in Fig.~\ref{Figres2-cylinscale}(a), the second stage of mixing rate with negative value is invisible. Thus, we choose the integration time window that reaches $\left<\chi^*\right>=0.01$, which can be deemed the end of the first stage mixing growth. The integration time independence is studied to set the upper integration bound of $\left<\chi^*\right>=0.015$ and lower integration bound $\left<\chi^*\right>=0.005$, shown as up error bar and down error bar.

By controlling Re = 38000 of all Ma = 2.4 cases, we first examine scaling dependence of the time average mixing rate on Pe number, as depicted in the upper half of Fig.~\ref{Figres2-cylinscale}(b).
Considering the fluctuations of mean mixing rate, we fit a power law to $\overline{\left<\chi^*\right>}$ obtained from high-resolution simulations, as a function of Pe number. The results yield a best fit $\overline{\left<\chi^*\right>}\approx(0.49\pm0.01)\mathrm{Pe}^{-0.186\pm0.006}$ for Pe number dependence.
If all cases are taken into account, as shown in the bottom half of Fig.~\ref{Figres2-cylinscale}(b), the Pe number scaling shows a similar exponent as $-0.185$, which suggests that the weak scaling is robust for the cases concerned in the present paper.
Further analyzing the Re number effect, it can be found that if the Pe number dependence is removed, an independent behavior of mixing rate on Re number appears, $\overline{\left<\chi^*\right>}/\mathrm{Pe}^{-0.185}\approx(0.44\pm0.08)\mathrm{Re}^{0.008\pm0.002}$, as shown in Fig.~\ref{Figres2-cylinscale}(c).

\begin{figure}
  \centering
  \includegraphics[clip=true,trim=5 2 5 10,width=.99\textwidth]{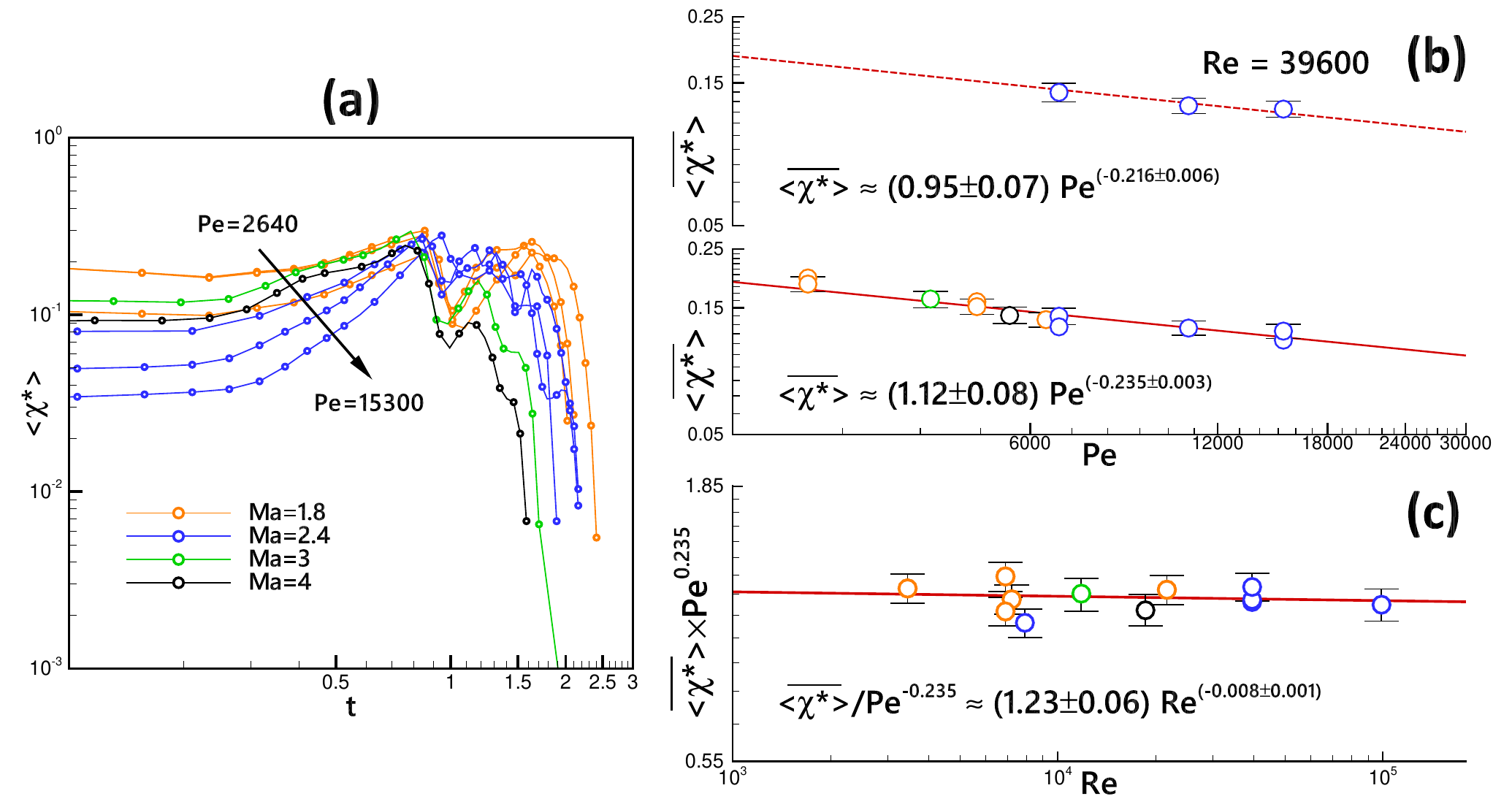}\\
  \caption{(a) Time evolution of density gradient accelerated mixing rate $\left<\chi^*\right>$ for spherical helium bubble cases.
  For clarity, Ma = 1.8 curves presented are the ones at Re = 3500, Pe = 2640; Re = 6900, Pe = 2640 and Re = 7200 Pe = 4900 respectively.
  Ma = 2.4 curves presented are the ones at Re = 39600, Pe = 6700; Pe = 10000 and Pe = 15000 respectively.
  (b) Upper: the scaling of time-averaged mixing rate $\overline{\left<\chi^*\right>}$ on Pe number at constant Re = 39600 and Ma = 2.4. Lower: the scaling of $\overline{\left<\chi^*\right>}$ on Pe number for all cases. (c) The scaling of time-averaged mixing rate independent of Pe number, $\overline{\left<\chi^*\right>}\cdot\textrm{Pe}^{0.235}$, on Re number. A dashed or solid line for each figure shows the best power-law fit over the concerned data. }\label{Figres3-sphscale}
\end{figure}
The effect of Pe and Re number on the density gradient accelerated mixing rate of a spherical bubble is examined in the same way as the cylindrical cases.
Figure~\ref{Figres3-sphscale}(a) shows the mixing rate temporal evolution. Interestingly, by introducing the coefficient $\mathrm{V}_c/\mathrm{V}_s$ into spherical cases, we observe a similar magnitude of mixing rate between spherical bubble cases and cylindrical bubble cases. Moreover, the scaling dependence of mixing rate on Pe and Re number, as shown in Fig.~\ref{Figres3-sphscale}(b) and (c), illustrates a similar trend as the cylindrical cases.
While mixing rate dependence for even lower Reynolds number (such as reported in Ref.~\cite{liu2020contribution}) or higher P\'eclet number (i.e., higher Sc number) deserves further validation, the scaling provides the conclusive evidence that the density gradient accelerated mixing rate in RM-type mixing with large density variation predicts, in the regime of high P\'eclet number and Reynolds number concerned, a weak dependence of mixing rate on Pe number by a scaling exponent $-0.185$ for cylindrical bubble and $-0.235$ for spherical bubble, and near independence on Re number.

\subsection{Scaling behavior of mixedness evolution}
\label{subsec:scaling2}
\begin{figure}
    \centering
    \subfigure[]{
    \label{fig4cylin-1} 
    \includegraphics[clip=true,trim=28 28 70 60,width=.32\textwidth]{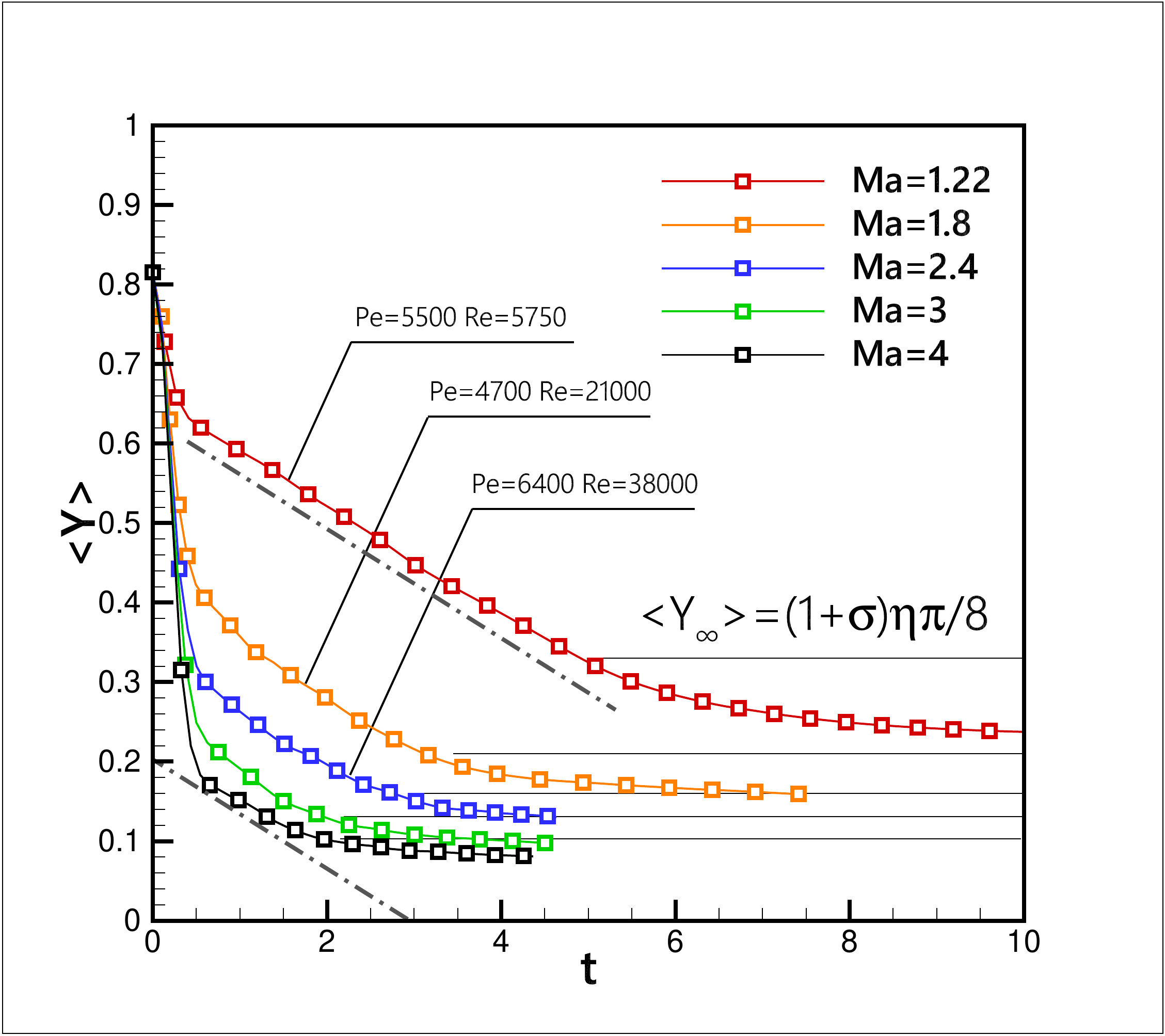}}
    \subfigure[]{
    \label{fig4cylin-2} 
    \includegraphics[clip=true,trim=28 28 70 60,width=.32\textwidth]{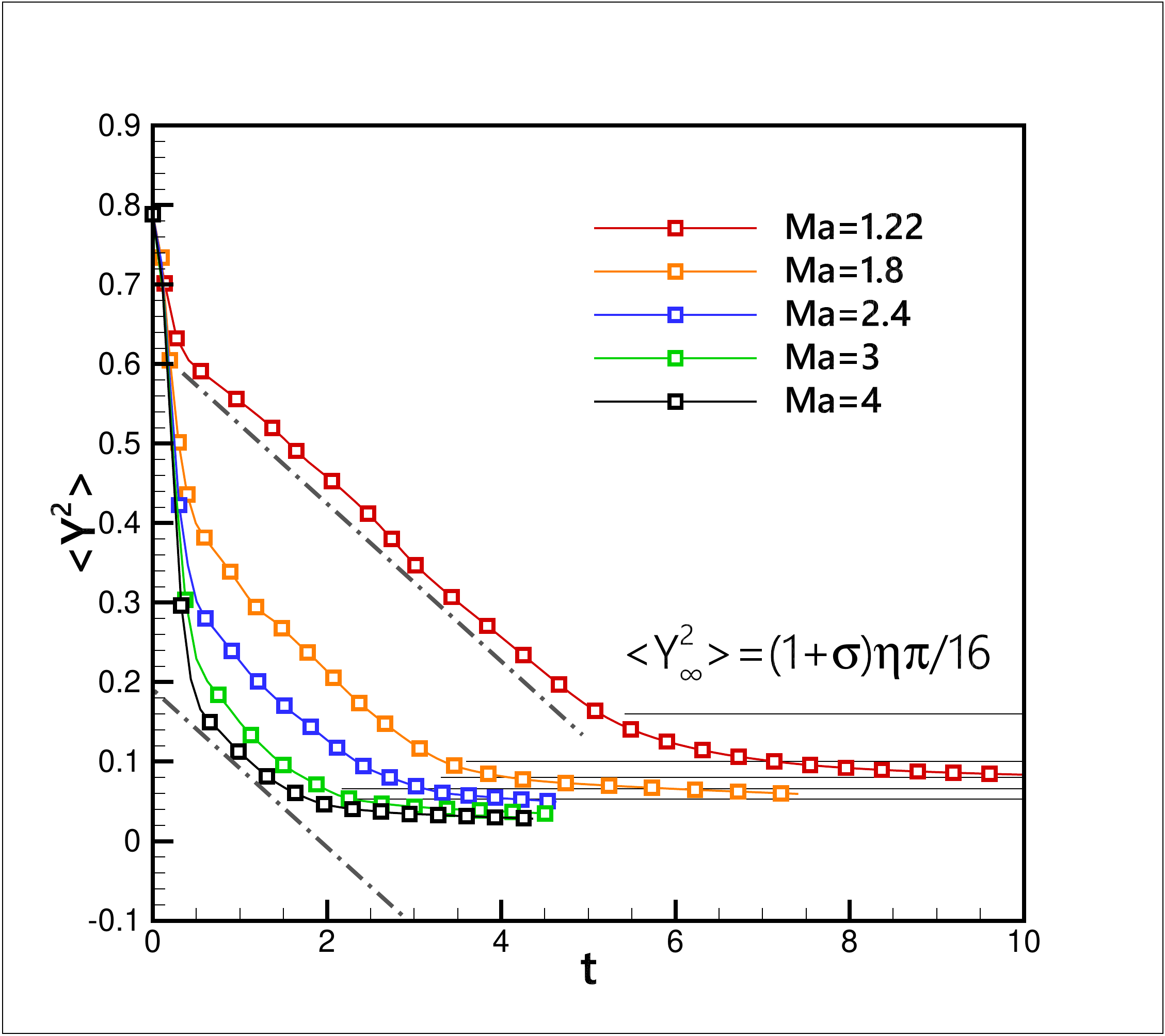}}
    \subfigure[]{
    \label{fig4cylin-3} 
    \includegraphics[clip=true,trim=28 28 70 60,width=.32\textwidth]{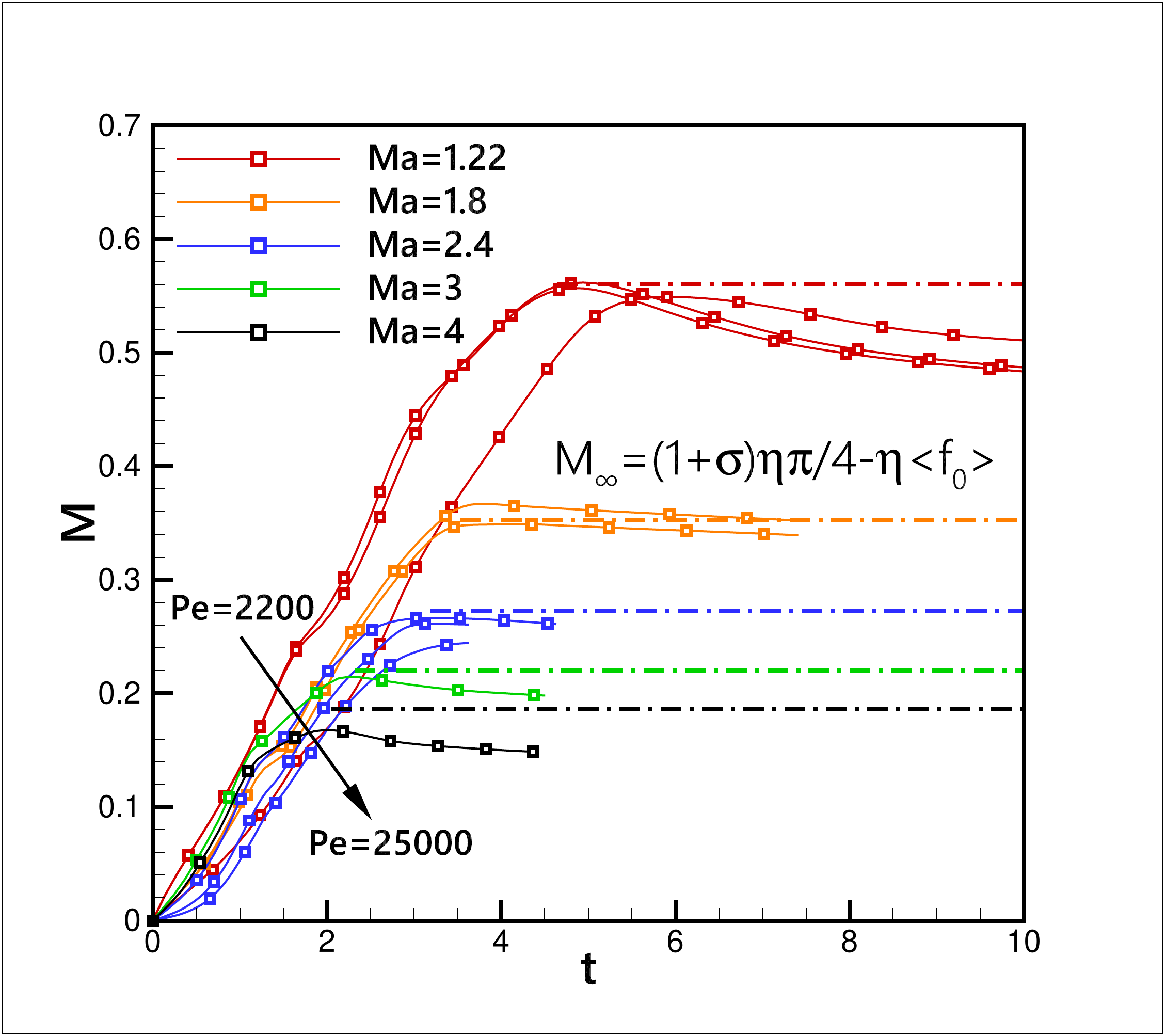}}\\
    \subfigure[]{
    \label{fig4sphe-1} 
    \includegraphics[clip=true,trim=28 28 70 60,width=.32\textwidth]{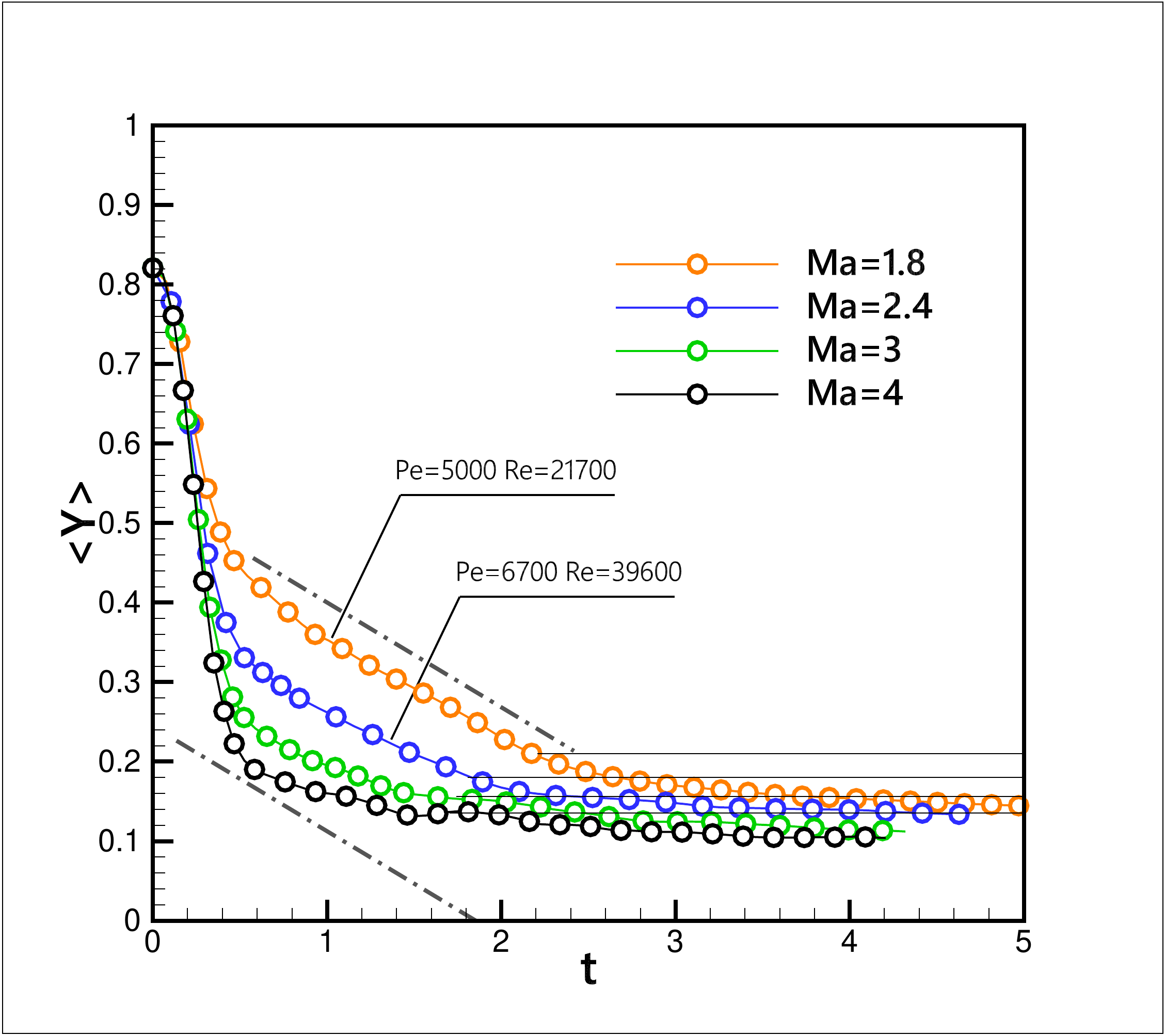}}
    \subfigure[]{
    \label{fig4sphe-2} 
    \includegraphics[clip=true,trim=28 28 70 60,width=.32\textwidth]{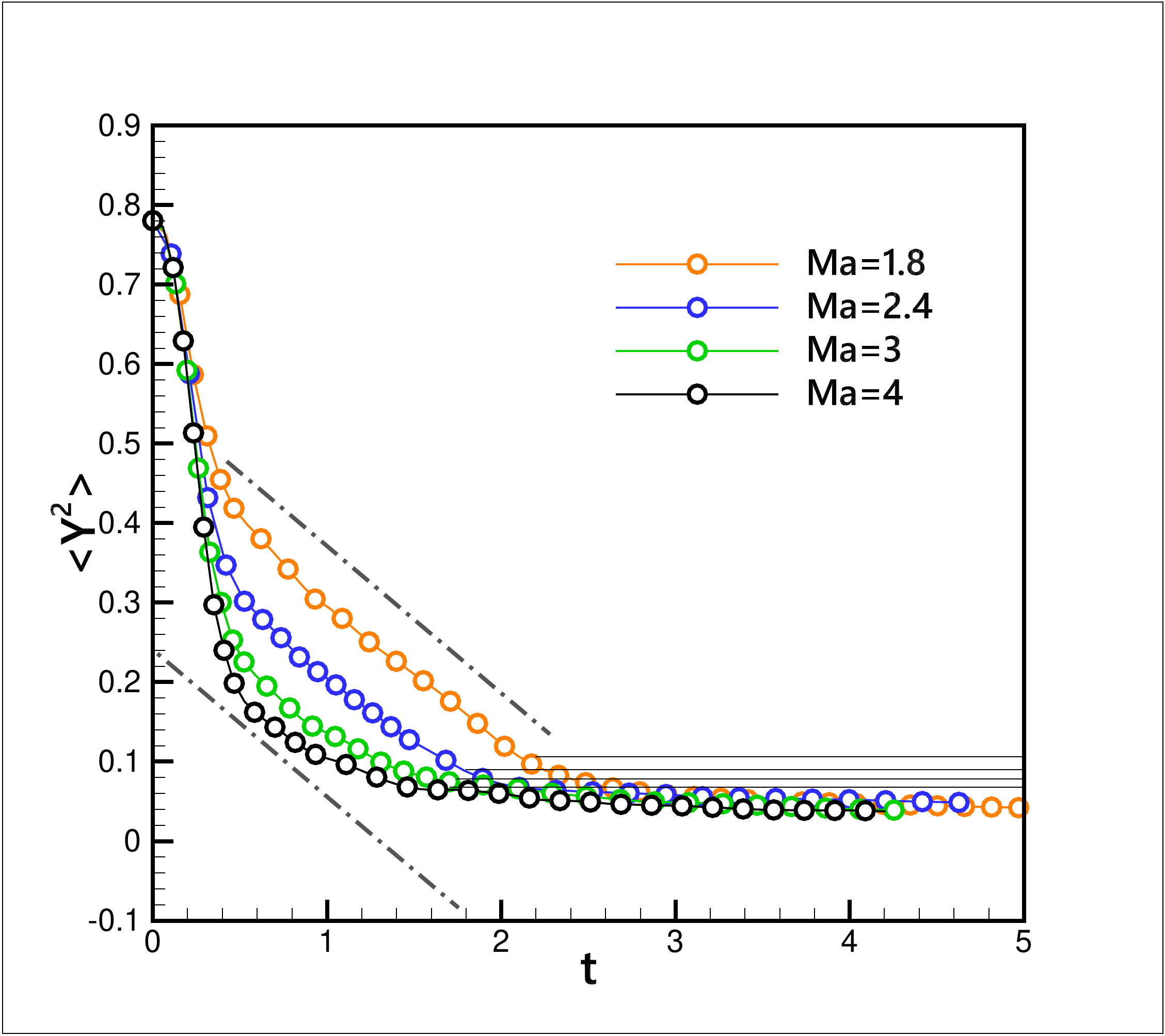}}
    \subfigure[]{
    \label{fig4sphe-3} 
    \includegraphics[clip=true,trim=28 28 70 60,width=.32\textwidth]{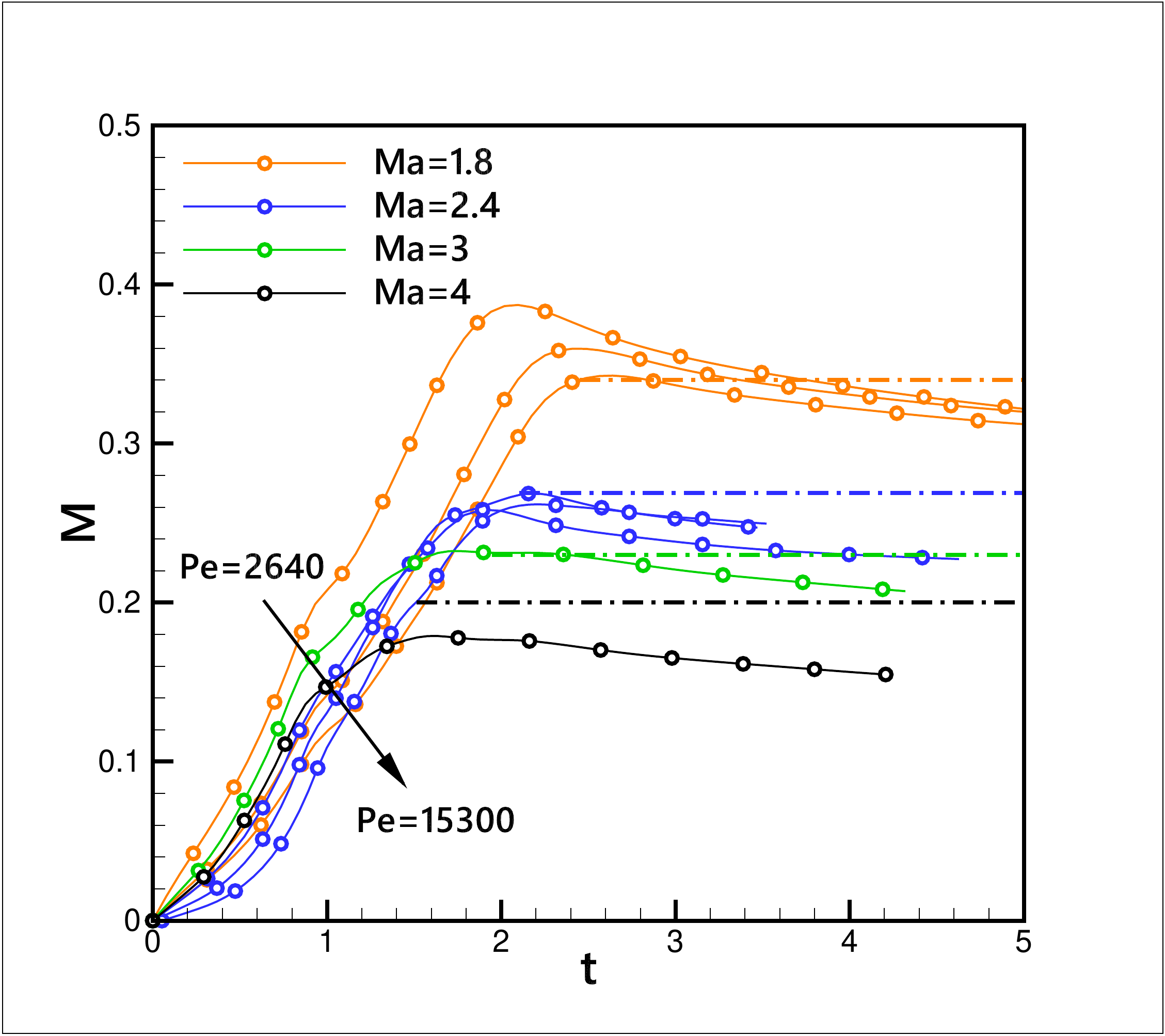}}
    \caption{The robust linear growth of $\left<Y\right>$ (a), $\left<Y^2\right>$ (b), and $\mathcal{M}$ (c) of cylindrical bubble cases; $\left<Y\right>$ (d), $\left<Y^2\right>$ (e), and $\mathcal{M}$ (f) of spherical bubble cases.
    The curves presented in cylindrical cases in (c) and spherical cases in (f) are under the same Re and Pe number as the ones in Fig.~\ref{Figres2-cylinscale}(a) and Fig.~\ref{Figres3-sphscale}(a) respectively.
    \label{Figres4-mix} }
\end{figure}
From the characteristic of weak dependence on Pe and Re number, we can further find a robust scaling that controls mixedness evolution in SBI.
Figures~\ref{Figres4-mix}(a-c) show the time history of the mean mass fraction, its energy and the time integral of the mixing rate for several typical cylindrical cases.  Figures~\ref{Figres4-mix}(d-f) show the same variables of spherical cases as the cylindrical bubble. The hyperbolic conservation violation of mean mass fraction is observed in all cases.
As for mixedness representative, $\mathcal{M}$, our results indicate the scaling law $\mathcal{M}(t)\sim t^1$, which characterizes the mixing regime dominated by the convective stirring of a vortex.
Furthermore, the linear slope of mixing indicator $\mathcal{M}$ varies with Pe number, following Pe number scaling on mixing rate.

Interestingly, $\mathcal{M}(t)$ has the same asymptotic limit at a later time, which is insensitive to Pe and Re numbers and can be predicted upon the integration of Eq.~(\ref{eq: dfdt}) from $t=0$ to $t\rightarrow \infty$:
\begin{equation}\label{eq: int_SDR_infty}
  \mathcal{M}_\infty=4\left[\left(\left<Y_\infty\right>-\left<Y^2_\infty\right>\right)-\varepsilon_0\right].
\end{equation}
From the initial conditions, the mass fraction inside the bubble area is 1, $Y(x,t=0)=1$, thus $\left<Y_0\right>\approx\left<Y^2_0\right>\approx\pi/4$.
For the well-mixed state, it can be found that the equilibrium state of final mixing is composed of a vortex pair containing the well-distributed mass fraction. From Fig.~\ref{fig1}(c4), it is reasonable to assume the upper half and lower half of the vortex pair at a later time as two bubbles with the same radius $a$. Inside the two bubbles, the mass fraction is $Y(x,t\rightarrow\infty)=1/2$ due to homogenous mixing. Then, we obtain $\left<Y_\infty\right>=\pi a^2/D^2$ and $\left<Y^2_\infty\right>=\pi a^2/(2D^2)$. Since the volume fraction can be expressed in form of the mass fraction as $\mathcal{X}(x,t\rightarrow\infty)=Y/[(1-\sigma)Y+\sigma]=1/(1+\sigma)$ (see Appendix~\ref{sec:app7}), the mean volume fraction $\left<\mathcal{X}_\infty\right>=2\pi a^2/[(1+\sigma)D^2]$. Thus, the compression rate can be expressed as $\eta=\left<\mathcal{X}_\infty\right>/\left<\mathcal{X}_0\right>=8a^2/[(1+\sigma)D^2]$, and we further model the mean mass fraction and its energy as:
\begin{equation}\label{eq: meanmass}
  \left<Y_\infty\right>=\frac{(1+\sigma)\eta\pi}{8}, \quad \left<Y^2_\infty\right>=\frac{(1+\sigma)\eta\pi}{16}.
\end{equation}
Although the spatial distribution of mass fraction at equilibrium state, such as vortex pair radius $a$, is still unknown, the mean mass fraction $\left<Y_\infty\right>$ and its energy $\left<Y^2_\infty\right>$ can be modeled through using the conservative characteristic of mean volume fraction, behaving as the near-constant compression rate $\eta$.

As for $\varepsilon_0$ in Eq.~(\ref{eq: int_SDR_infty}), it is the initial mixedness (due to the diffusion layer at the initial condition) after the shock compression, which can be estimated by $\varepsilon_0\approx\eta\left(\left<Y_0\right>-\left<Y^2_0\right>\right)=\eta\left<f_0\right>/4$. This means that the integration of Eq.~(\ref{eq: dfdt}) starts after the shock compression. Rearranging Eq.~(\ref{eq: int_SDR_infty}) and using Eq.~(\ref{eq: meanmass}), we obtain:
\begin{equation}\label{eq: finalmixed}
  \mathcal{M}_\infty= \frac{(1+\sigma)\eta\pi}{4}-\eta\left<f_0\right>.
\end{equation}
Mixedness of initial state $\left<f_0\right>$ can be theoretically integrated if the diffusion layer distribution is known, and it is the same about 0.14 in all cases. Such prediction, which is explicitly presented for the different Ma number cases by the solid black lines in Fig.~\ref{Figres4-mix}, represents the asymptotic limits of mixing evolution that is mainly affected by the compression rate and density ratio. This behavior also implies that the shock effect on the mixing rate is relatively weak but stirring from vortex dominates the mixing rate growth after initial shock passages.

\begin{figure}
    \centering
    \subfigure[]{
    \label{fig4sec-1} 
    \includegraphics[clip=true,trim=28 28 70 60,width=.4\textwidth]{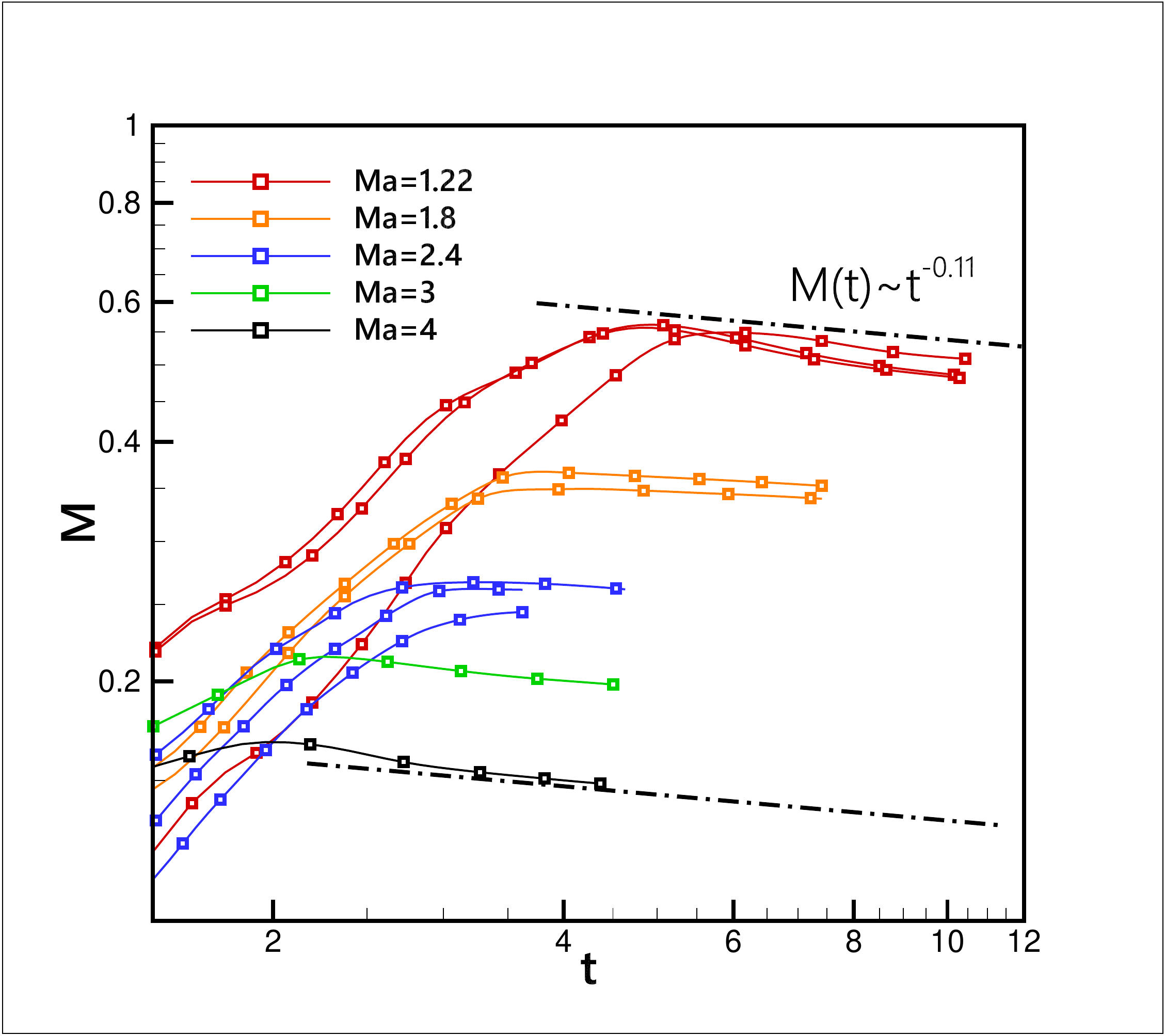}} \hspace{12mm}
    \subfigure[]{
    \label{fig4sec-2} 
    \includegraphics[clip=true,trim=28 28 70 60,width=.4\textwidth]{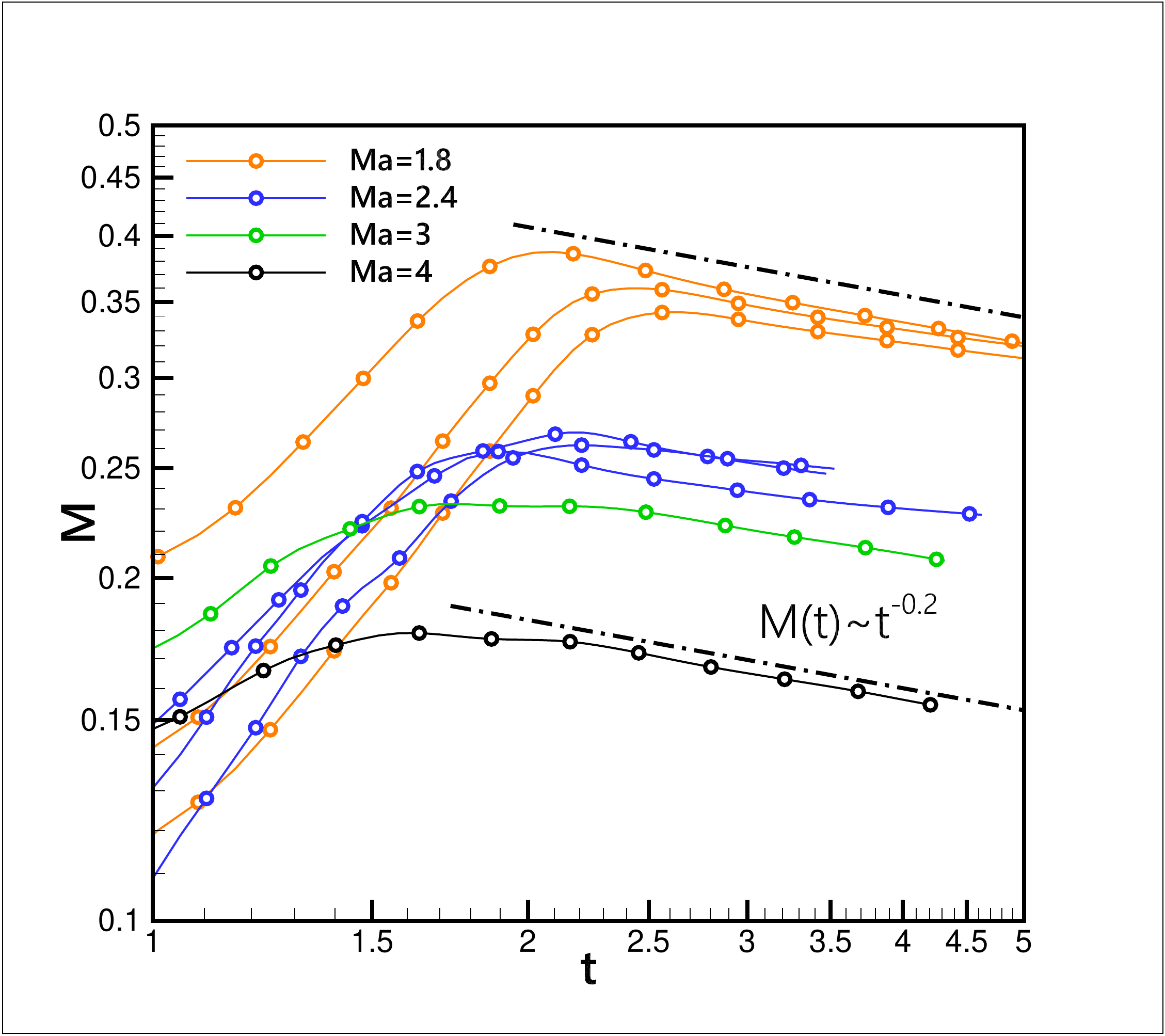}}
    \caption{The second stage scaling of mixing rate integral $\mathcal{M}$ with time for cylindrical bubble cases (a) and spherical bubble cases (b). \label{Figres4-sec} }
\end{figure}
The scaling of mixing in the second stage with time for both spherical and cylindrical cases is plotted in Fig.~\ref{Figres4-sec} through the log chart. It suggests that the mixedness decreases after reaching the equilibrium state under the scaling law $\mathcal{M}(t)\sim t^\alpha$, where exponent $\alpha=-0.11$ for cylindrical cases and $\alpha=-0.2$ for spherical bubble cases are determined via exponential fitting of $\mathcal{M}(t)$. Considering that $\left<\chi^*\right>(t)=\partial_t\mathcal{M}(t)$, we can infer the scaling law of mixing rate with time as $\left<\chi^*\right>(t)\sim t^{-1.11}$ for a cylinder and $\sim t^{-1.2}$ for a sphere. It is noteworthy that the mixedness always increases in the passive scalar mixing due to the hyperbolic conservation \cite{cetegen1993experiments}, while the cause of the decrease of mixedness in variable-density flows is the production of DGRD with a negative value as analyzed in Sec.~\ref{subsec:DGAD}.


\subsection{Effect of density ratio on mixing rate and mixedness}
\label{subsec:scaling3}
\begin{figure}
    \centering
    \subfigure[]{
    \label{Figres5-1} 
    \includegraphics[clip=true,trim=5 5 10 15,width=.9\textwidth]{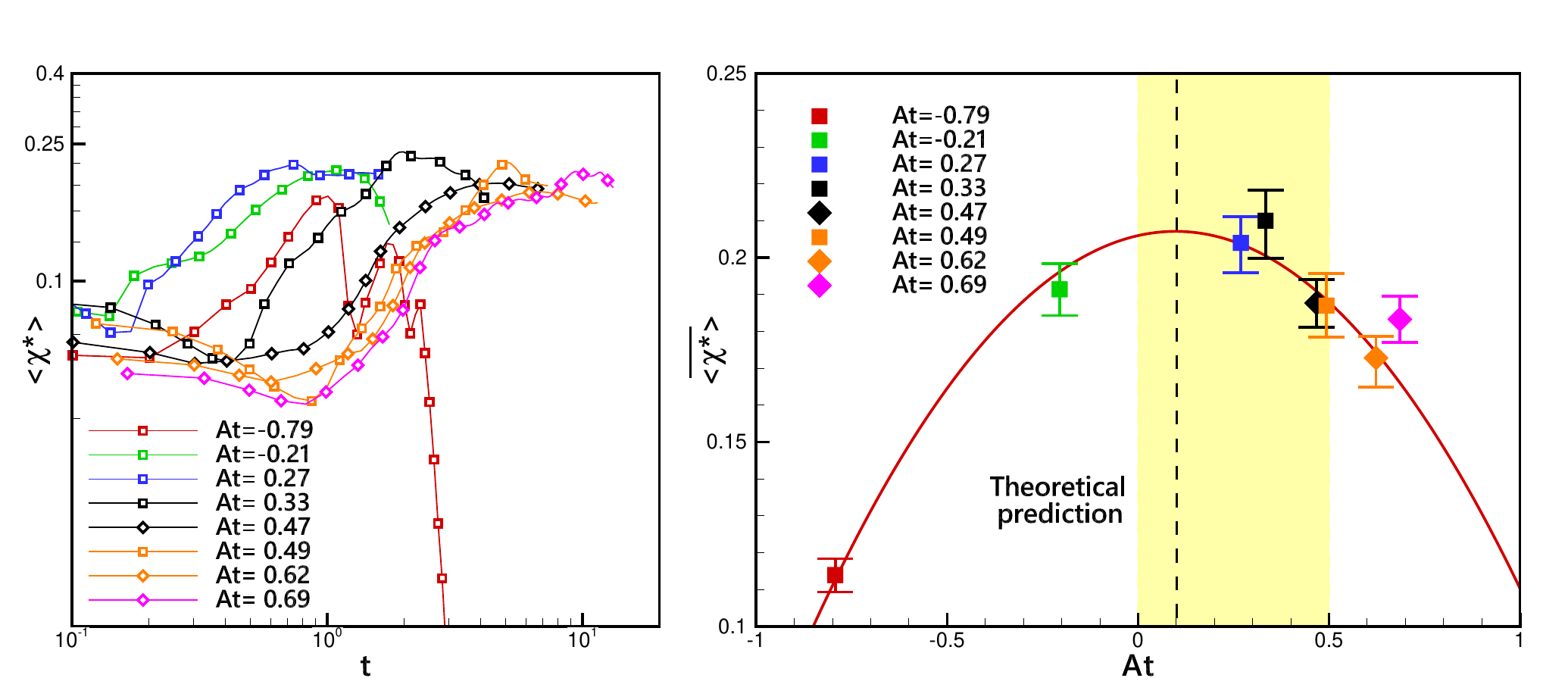}}
    \subfigure[]{
    \label{Figres5-2} 
    \includegraphics[clip=true,trim=0 230 0 0,width=.9\textwidth]{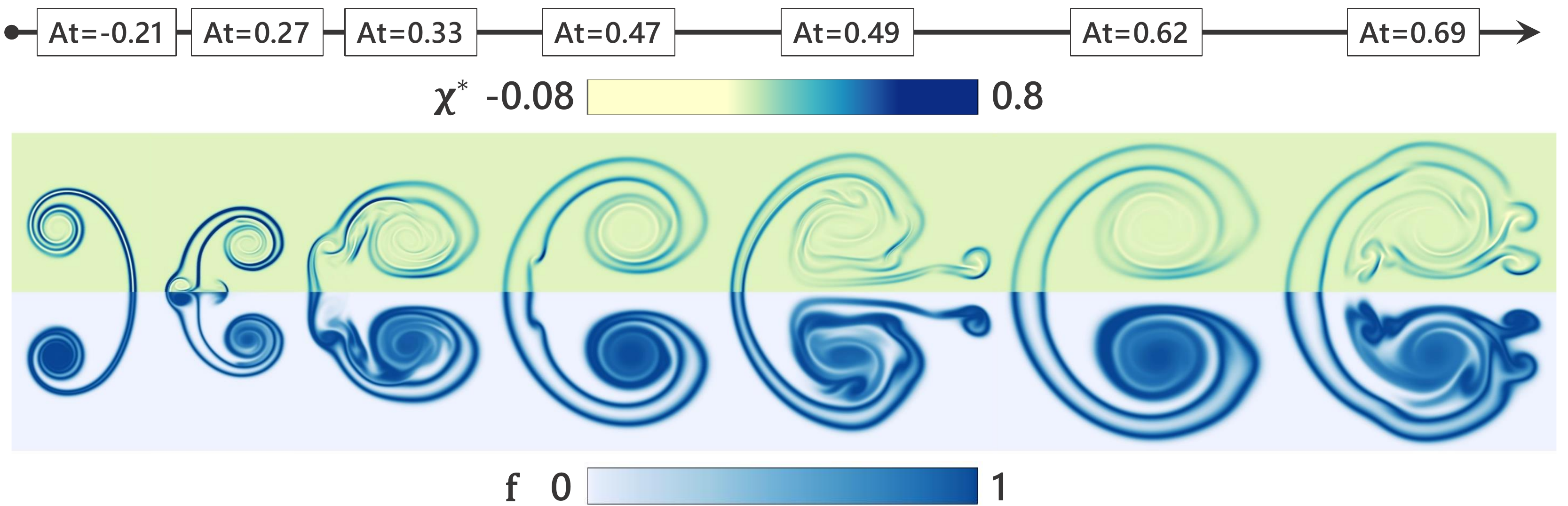}}
    \caption{
    (a) Left: Time evolution of density gradient accelerated mixing rate $\left<\chi^*\right>$ of cylindrical bubble ranging from At = $-$0.79 to At = 0.69, where Atwood number is defined as At $=(\sigma-1)/(\sigma+1)$. Right: Variation of time-averaged accelerated mixing rate $\overline{\left<\chi^*\right>}$ with Atwood numbers. Square dots: Ma = 2.4; Diamond dots: Ma = 1.22. Red: He; Green: CH$_4$; Blue: CO$_2$; Black: Kr; Orange: Xe; Pink: SF$_6$.
    (b) Characteristic flow structure of different Atwood number cases. Up: density gradient accelerated mixing rate; bottom: mixedness. For each case, At = $-$0.21 at $t = 1.75$; At = 0.27 at $t = 1.59$; At = 0.33 at $t = 4.19$; At = 0.47 at $t = 5.97$; At = 0.49 at $t = 7.33$; At = 0.62 at $t = 9.66$; At = 0.69 at $t = 10.24$.
        \label{Figres5-heavyscale} }
\end{figure}
From Eqs.~(\ref{eq: meanmass}) and (\ref{eq: finalmixed}), it appears the importance of density ratio on mixing indicators. Based on the same Pe and Re number through controlling diffusivity $\mathscr{D}$ and dynamics viscosity $\mu$, we change the bubble component from helium to methane (CH$_4$), carbon dioxide (CO$_2$), krypton (Kr), xenon (Xe), and sulfur hexafluoride (SF$_6$) to analyze the influence of density difference on mixing (see Supplementary Material \cite{supp} for the set-up details). The density ratio among these gases ranges from $\sigma=0.12\sim5.4$ or in the form of Atwood number At $=(\sigma-1)/(\sigma+1)=-0.79\sim0.69$. The left part of Fig.~\ref{Figres5-1} shows the time history of mixing rate $\left<\chi^*\right>$ of all At numbers. Unlike the helium bubble case, the mixing rate of other cases still maintains a high level during the whole computing time and shows a declining trend at late time.
The mixing status at the end of simulation for all At number cases can be explored in Fig.~\ref{Figres5-2}, showing that the bridge structure maintains a long time and contributes to continuous mixing \cite{tomkins2008experimental}.

To obtain the appropriate time-averaged mixing rate, we set the integration time window $T$ as $\left<\chi^*\right>>0.5\left<\chi^*\right>_{max}$ when the mixing rate is high, characterizing the first stage mixing growth. The integration time independence is therefore investigated by setting the upper and lower integration bound $\left<\chi^*\right>=0.5\left<\chi^*\right>_{max}\pm\left<\Delta\chi^*\right>(5\delta t)$. The variation of time-averaged mixing rate illustrates a rise and decline with At number increasing, as plotted in the right part of Fig.~\ref{Figres5-1}.
Based on a best quadratic fitting, it predicts a higher mixing rate around At~$\approx0.1$.
Since the mixing rate is composed by the time derivative of mean mass fraction and mass fraction energy from Eq.~(\ref{eq: dfdt}), we further examine the differences of the mean mass fraction and its energy evolution between three typical At numbers to seek the reason that causes the high mixing rate behavior with density ratio.

\begin{figure}
    \centering
    \subfigure[]{
    \label{figres6-1} 
    \includegraphics[clip=true,trim=20 20 23 60,width=.32\textwidth]{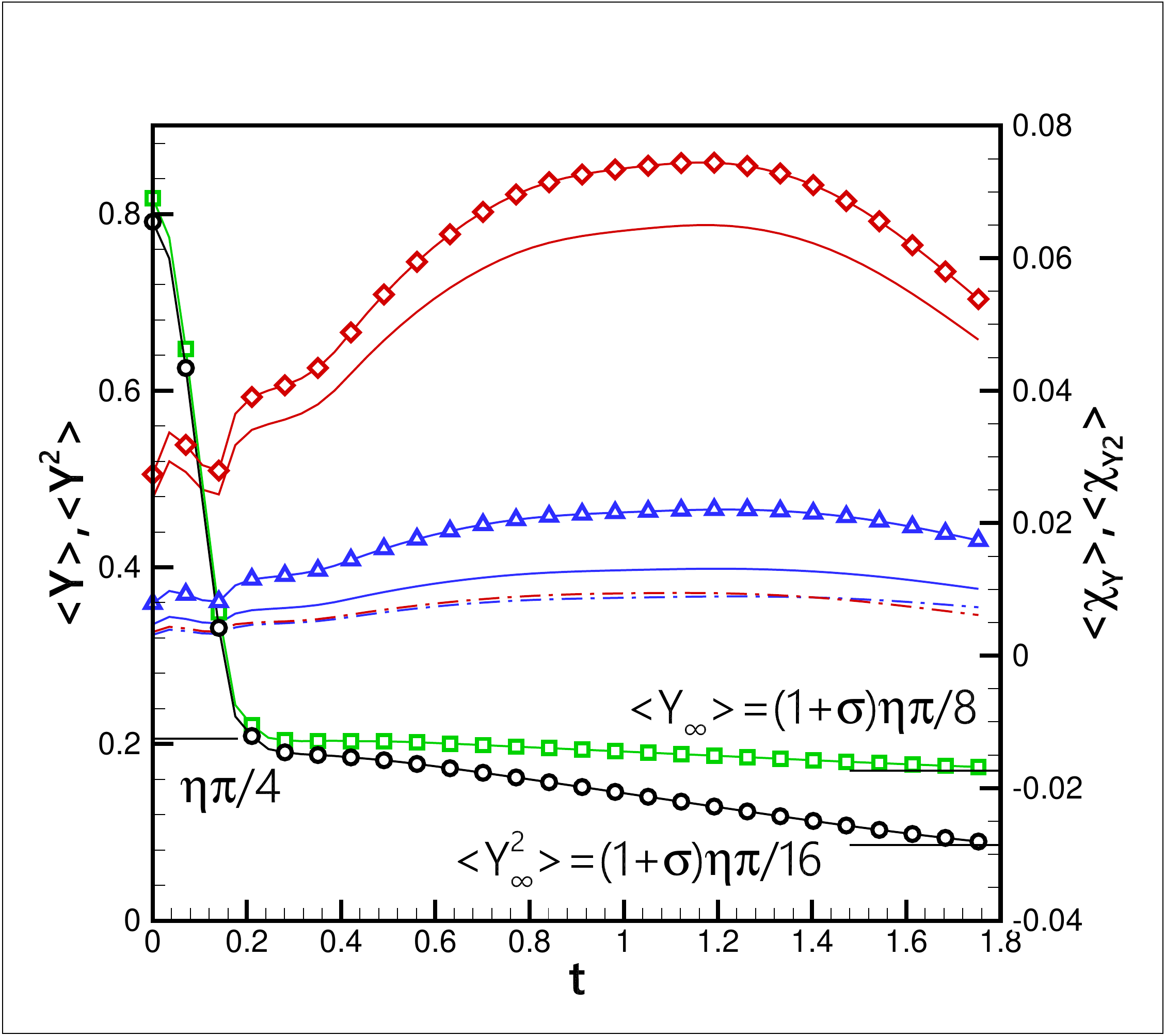}}
    \subfigure[]{
    \label{figres6-2} 
    \includegraphics[clip=true,trim=20 20 23 60,width=.32\textwidth]{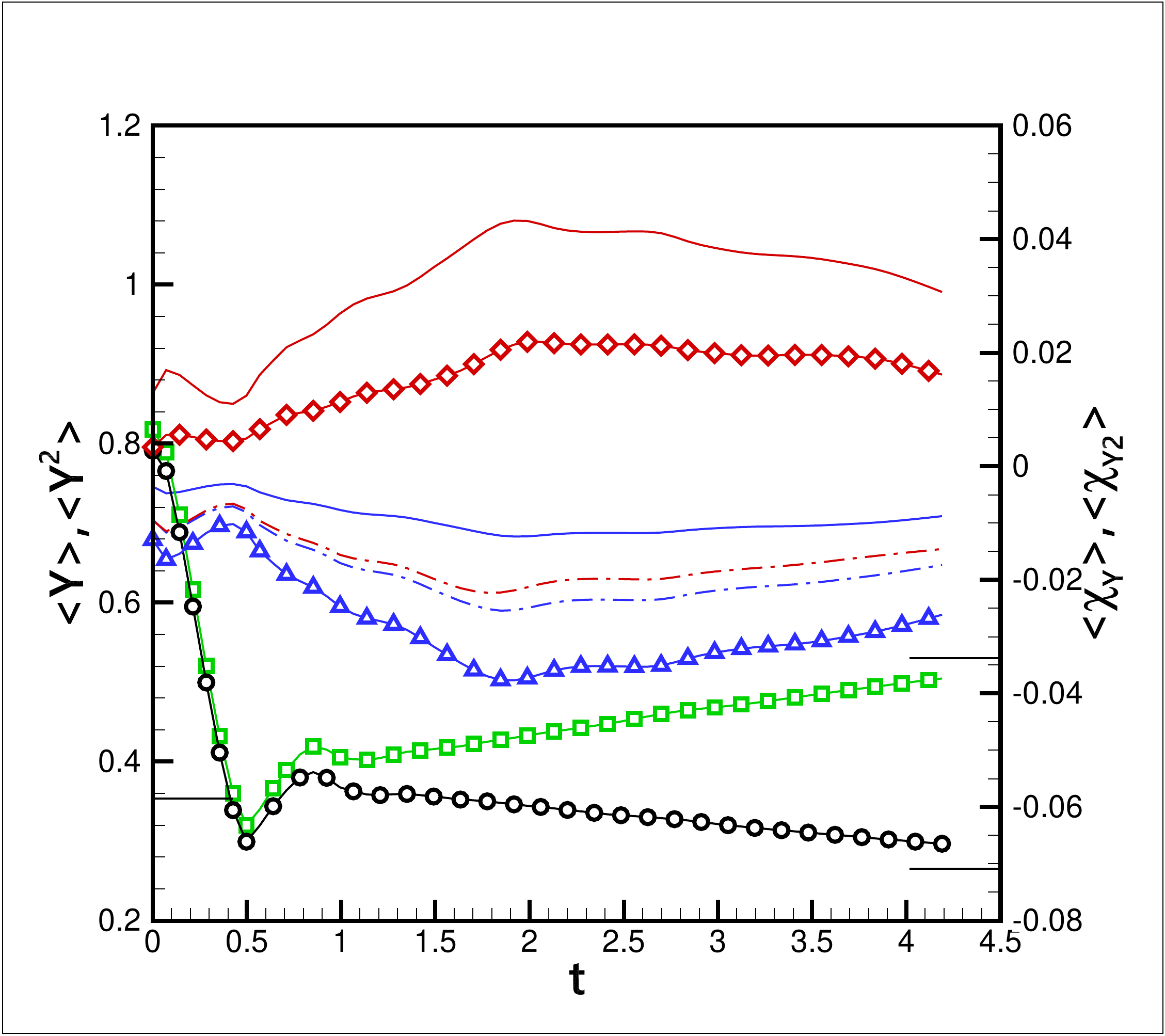}}
    \subfigure[]{
    \label{figres6-3} 
    \includegraphics[clip=true,trim=20 20 23 60,width=.32\textwidth]{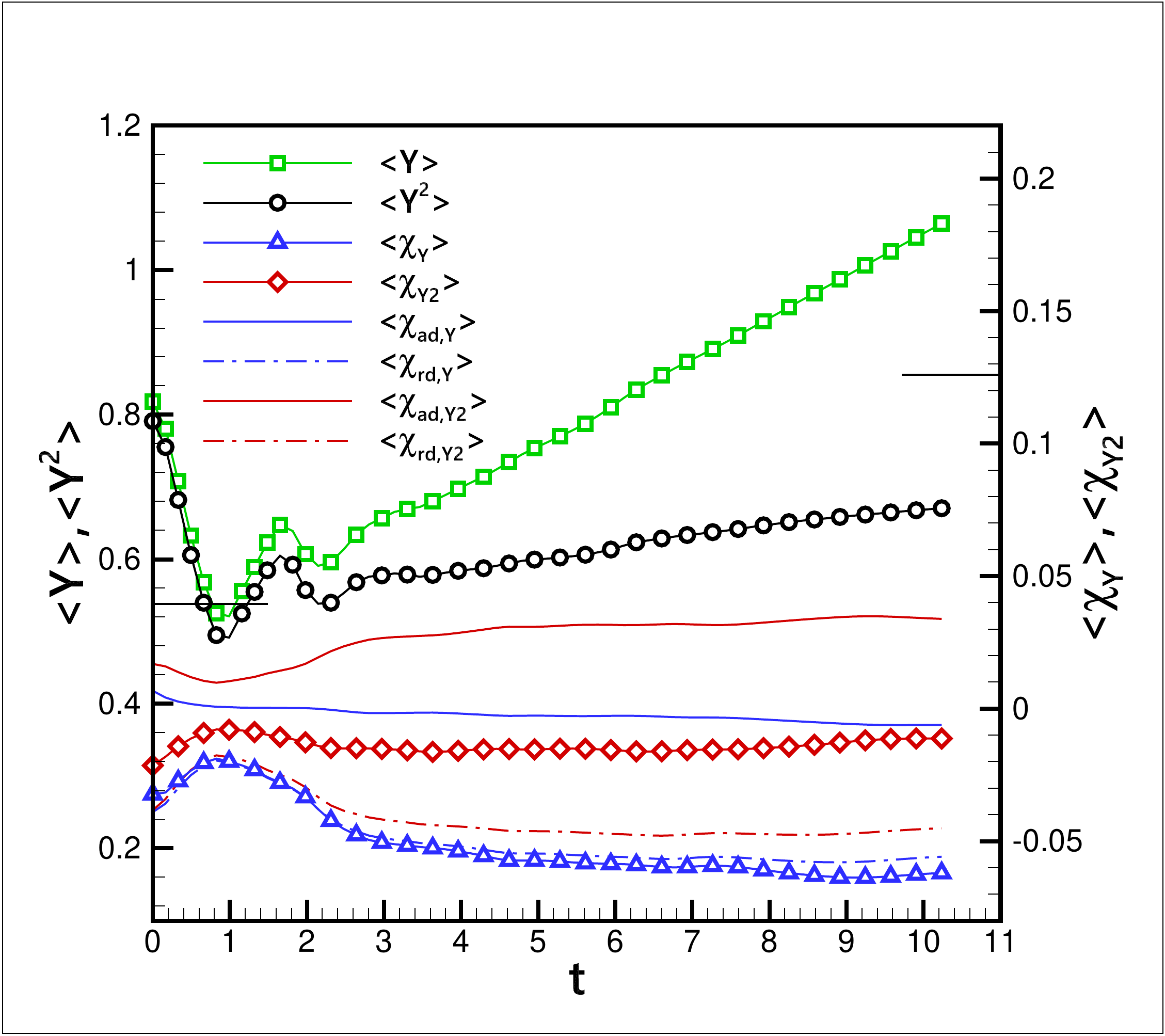}} \\
    \caption{Time evolution of mean mass fraction $\left<Y\right>$ and its mixing rate composed by DGAD and DGRD term, mean mass fraction energy $\left<Y^2\right>$ and its mixing rate composed by DGAD and DGRD term with (a) At = $-$0.21 (Ma = 2.4, CH$_4$), (b) At = 0.33 (Ma = 2.4, Kr) and (c) At = 0.69 (Ma = 1.22, SF$_6$). \label{Figres6} }
\end{figure}
As for At = $-0.21$, the mean mass fraction is nearly conservative, which is near to constant density mixing behavior, as depicted in Fig.~\ref{figres6-1}.
With the increase of At number, an interesting phenomenon, the opposite sign of mixing rate for mass fraction $\left<\chi_Y\right>$ and mass fraction energy $\left<\chi_{Y^2}\right>$, occurs.
Due to the variable-density effect on the coefficient of DGAD and DGRD as analyzed in Sec.~\ref{subsec:mixingrate}, the mean mass fraction increases after shock compression in heavy bubble cases (At $>0$) as shown in Figs.~\ref{figres6-2} and \ref{figres6-3}, contrary to the light bubble cases (At $<0$). However, the mean mass fraction energy still decreases with time at small At number (At = 0.33) while increases at a large At number (At = 0.69).
Hence, based on Eq.~(\ref{eq: dfdt}), we can infer that when the opposite sign of slope of mean mass fraction and its energy appears (i.e., $\mathrm{D}\left<Y\right>/\mathrm{D}t>0$ and $\mathrm{D}\left<Y^2\right>/\mathrm{D}t<0$), the mixing rate will be relatively higher.
The contribution of DGAD and DGRD to the mixing rate of mass fraction and its energy are also plotted in Fig.~\ref{Figres6}, showing that the DGRD term plays a dominant role in causing the increase of mass fraction energy when At number is higher.
Unlike the strictly positive characteristic of dissipation, it is hard to predicted the sign of non-zero diffusion term. Based on the generalized Green's theorem, we expand the DGRD term integration in the form of local dissipation in Appendix~\ref{sec:app10}. Therefore, the range of At number that leads to the increase of mass fraction and the decrease of mass fraction energy is proven as At $\in(0,0.2)$, which coincides with the best quadratic fitting's range with a high mixing rate in Fig.~\ref{Figres5-1}.

\begin{figure}
    \centering
    \subfigure[]{
    \label{figres7-1} 
    \includegraphics[clip=true,trim=20 28 70 60,width=.48\textwidth]{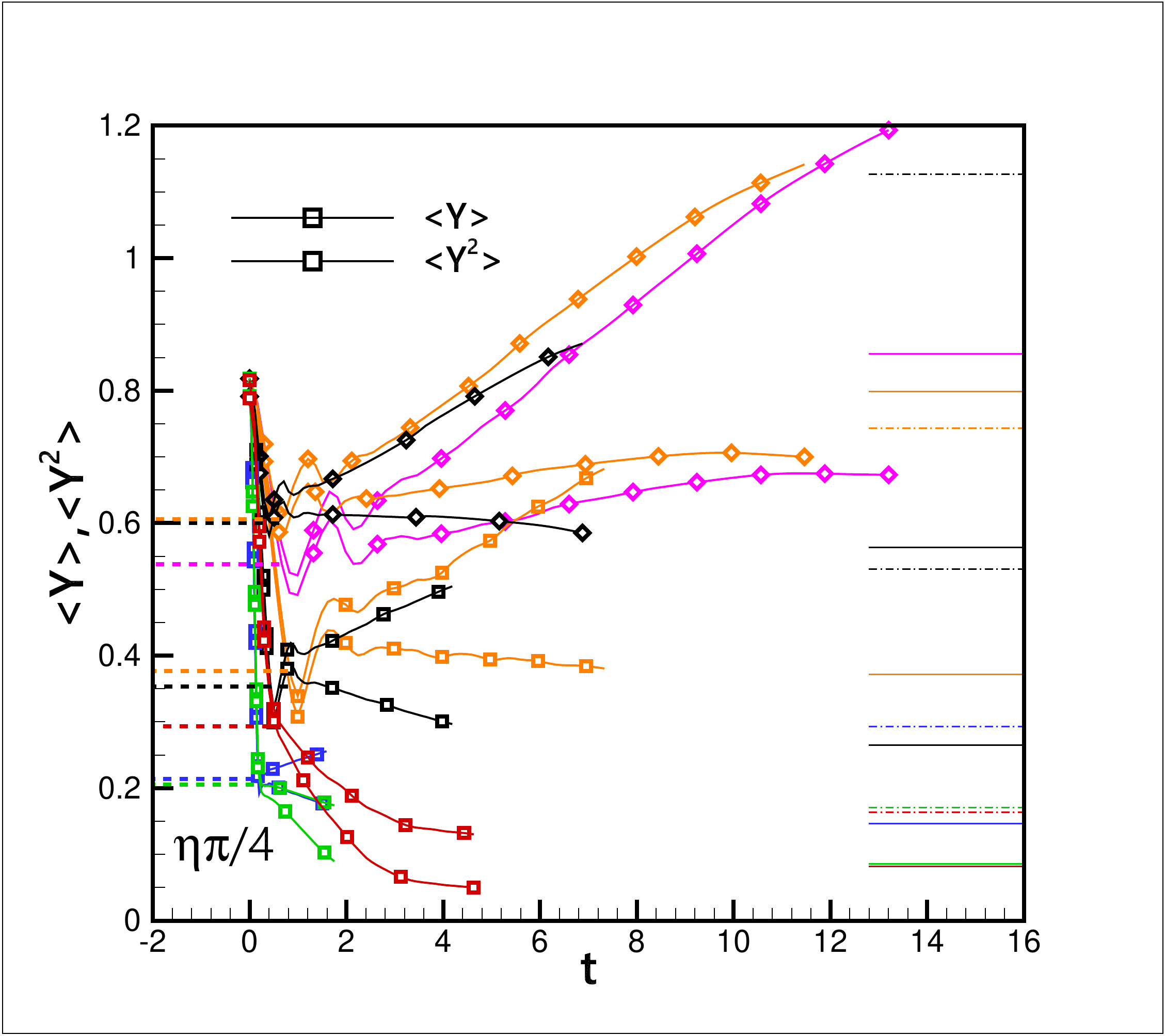}}
    \subfigure[]{
    \label{figres7-2} 
    \includegraphics[clip=true,trim=20 28 70 60,width=.48\textwidth]{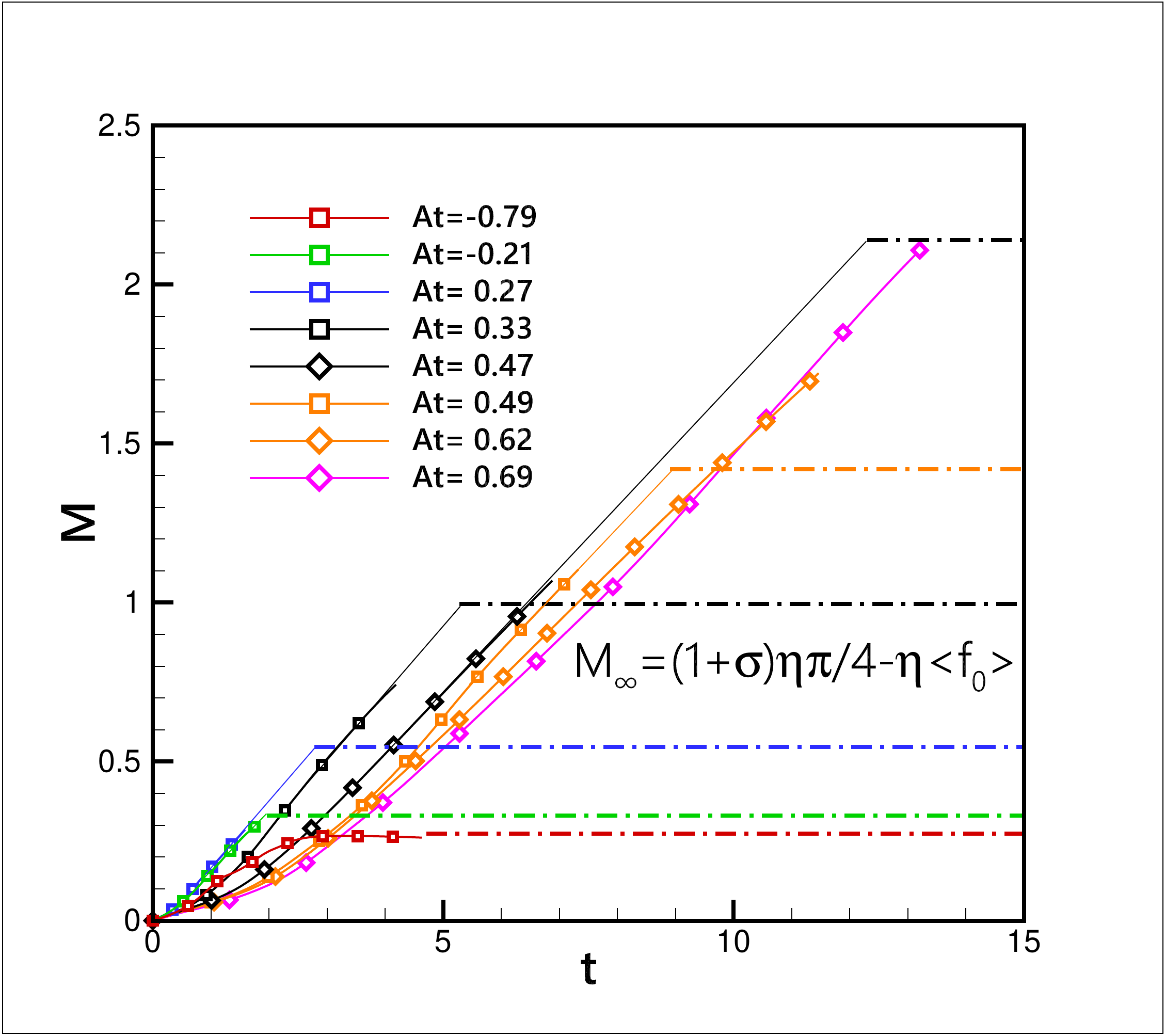}} \\
    \caption{Time evolution of mean mass fraction $\left<Y\right>$, mean mass fraction energy $\left<Y^2\right>$ (a) and mixing rate integral $\mathcal{M}$ (c) of different Atwood numbers. Theoretical values of $\left<Y_\infty\right>$, $\left<Y^2_\infty\right>$ in Eq.~(\ref{eq: meanmass}) and $\mathcal{M}_\infty$ in Eq.~(\ref{eq: finalmixed}) are indicated by the solid and dashed-dot lines. \label{Figres7} }
\end{figure}
Figure~\ref{figres7-1} further shows the time evolution of $\left<Y\right>$ and $\left<Y^2\right>$ in all variable-density cases. Those At number cases with an opposite sign between mean mass fraction and its energy growth overlap with the At number range where the high mixing rate happens, as depicted in Fig.~\ref{Figres5-heavyscale}(b).
Here, the At number range with the high-level mixing rate can be predicted heuristically in terms of the macroscopic well-mixed state model of $\left<Y_\infty\right>$ and $\left<Y^2_\infty\right>$ in Eq.~(\ref{eq: meanmass}). Due to the shock compression, both mass fraction and its energy will be compressed to $\eta\pi/4$ by compression rate at initial status. The prediction is acceptable from the comparison between modeled values depicted as dashed lines and the numerical results. Thus, when $\left<Y_\infty\right>>\eta\left<Y_0\right>$ and $\left<Y^2_\infty\right><\eta\left<Y^2_0\right>$, the opposite sign of $\left<Y\right>$ and $\left<Y^2\right>$ growth occurs:
\begin{equation}\label{eq: meanmassgrowth}
  \frac{(1+\sigma)\eta\pi}{8}>\frac{\eta\pi}{4},\quad \frac{(1+\sigma)\eta\pi}{16}<\frac{\eta\pi}{4} \Rightarrow 1<\sigma<3.
\end{equation}
This range in the form of At $\in (0,0.5)$ is plotted by a yellow region in Fig.~\ref{Figres5-heavyscale}(b), which coincides with the At number range with the high-level time-averaged mixing rate and overlaps the predicted range ($1<\sigma<3/2$) based on local DGRD behavior in Appendix~\ref{sec:app10}.

The temporal integrals of mixing rate $\mathcal{M}$ of all variable-density cases are shown in Fig.~\ref{figres7-2}.
Those mixedness profiles with high mixing rates grow linearly at a higher rate. Since the mixing still grows for heavy bubble cases, the final asymptotic limit $\mathcal{M}_\infty$ is presented as dashed-dot lines. Such a prediction can offer the characteristic mixing time when the well-mixed status is reached, if the mixing growth slope is prior known.

\section{Conclusions}
\label{sec:conclu}
We have investigated the behavior of convective mixing for RM-type shock bubble interaction through high-resolution simulation in this paper. From the start point, variable-density mixing characteristic of a shocked cylindrical bubble contained with helium is found by the hyperbolic conservation violation of mean mass fraction, which will not occur in the conservative passive scalar mixing problem.
The violation manifests the mean mass fraction decrease with time.

Further, by combining the compositional transport equation and the divergence relation for the miscible flows, we offer the exact mixing rate expression that suits mixing involving a wide range of density differences. The mixing rate shows two source terms from density gradient: density gradient accelerated dissipation (DGAD), and density gradient redistributed diffusion (DGRD). The first term dissipates the mass fraction at a rate that is higher than in passive scalar. The second term decreases the mixing content at late time steady mixing, and plays a vital role in dissipating mean mass fraction energy when heavy gas is concerned.
More precisely, we have examined the time evolution of the derived mixing rate $\left<\chi^*\right>$, the dependence of which on any system parameter can be extracted.
Two-stage mixing status can be identified, a quasi-linear growth stage of convective mixing due to the vortex roll-up and a steady mixing stage with a low mixing rate.

Then we pay attention to the dependence of the first-stage mixing growth rate on system parameters by analyzing several simulations for both cylindrical and spherical bubbles under a broad range of shock Mach numbers, Re numbers, and Pe numbers while keeping a constant density ratio as helium.
We have found a relatively weak dependence of time-averaged mixing growth rate $\overline{\left<\chi^*\right>}$ on Pe number by a scaling exponent $-0.185$ for cylindrical bubble and $-0.235$ for spherical bubble, and near independence on Re number and Ma number.
This leads to a robust scaling that time integral of mixing rate $\mathcal{M}(t)\sim t^1$ at the first stage and $\mathcal{M}(t)\sim t^{\alpha}$ at the second stage, where exponent $\alpha=-0.11$ for cylindrical cases and $\alpha=-0.2$ for spherical cases. Another interesting scaling shows that the asymptotic behavior of mixing is controlled jointly by shock compression rate and density ratio $\mathcal{M}_{\infty}\sim(1+\sigma)\eta$, which leads us to investigate the essential effect of density ratio on mixing rate.

The time-averaged mixing rate manifests a non-monotonic variation with the increase of the density ratio. The mixing rate will be relatively higher when the opposite sign of the growth rate of mean mass fraction and its energy occurs. The local mixing rate coefficient determines a range ($1<\sigma<3/2$) when the opposite sign emerges, which overlaps the predicted range from the macroscopic well-mixed model as $1<\sigma<3$, in the form of Atwood number as $\textrm{At}=0\sim0.5$. The theoretical prediction from the local mixing rate coefficient or the global well-mixed model coincides with the observed Atwood number range with high time-averaged mixing rate.

The accelerated dissipation of variable-density mixing found in this paper implies the new standpoint for auto-ignition and extinction in non-premixed combustion in the extensive variable-density problems.
Moreover, the scaling behavior of the density gradient accelerated mixing may pave the new way for further examining the mixing behavior in variable-density flows and offer a quick estimation of the amount of mixedness in combustion applications.

\begin{acknowledgments}
This work is supported by the National Natural Science Foundation of China (NSFC) under Grants No. 91441205, No. 91941301, and National Science Foundation for Young Scientists of China (Grant No.51606120).
Besides, the authors would acknowledge the Center for High-Performance Computing of SJTU to provide the supercomputer $\pi$.
Moreover, the authors are grateful for D. Li and C. Zhang's support for the parallel realization and axisymmetric simulation.
This work benefited from fruitful discussions with L. Li, B. Zhang, M. He and the help from S. Zheng.
Finally, the authors would like to thank the anonymous referees for their valuable comments.
\end{acknowledgments}

\appendix
\section{Grid independence study}
\label{sec:app1}
Here, we examine the grid independence study of all cases studied in this paper. The grid resolution has a pronounced effect on the second-order differential of scalar dissipation~\cite{schumacher2005very}. Thus the grid resolution should be chosen cautiously.

\begin{figure}
    \centering
    \subfigure[]{
    \label{app1-circu1} 
    \includegraphics[clip=true,trim=18 20 50 30,width=.32\textwidth]{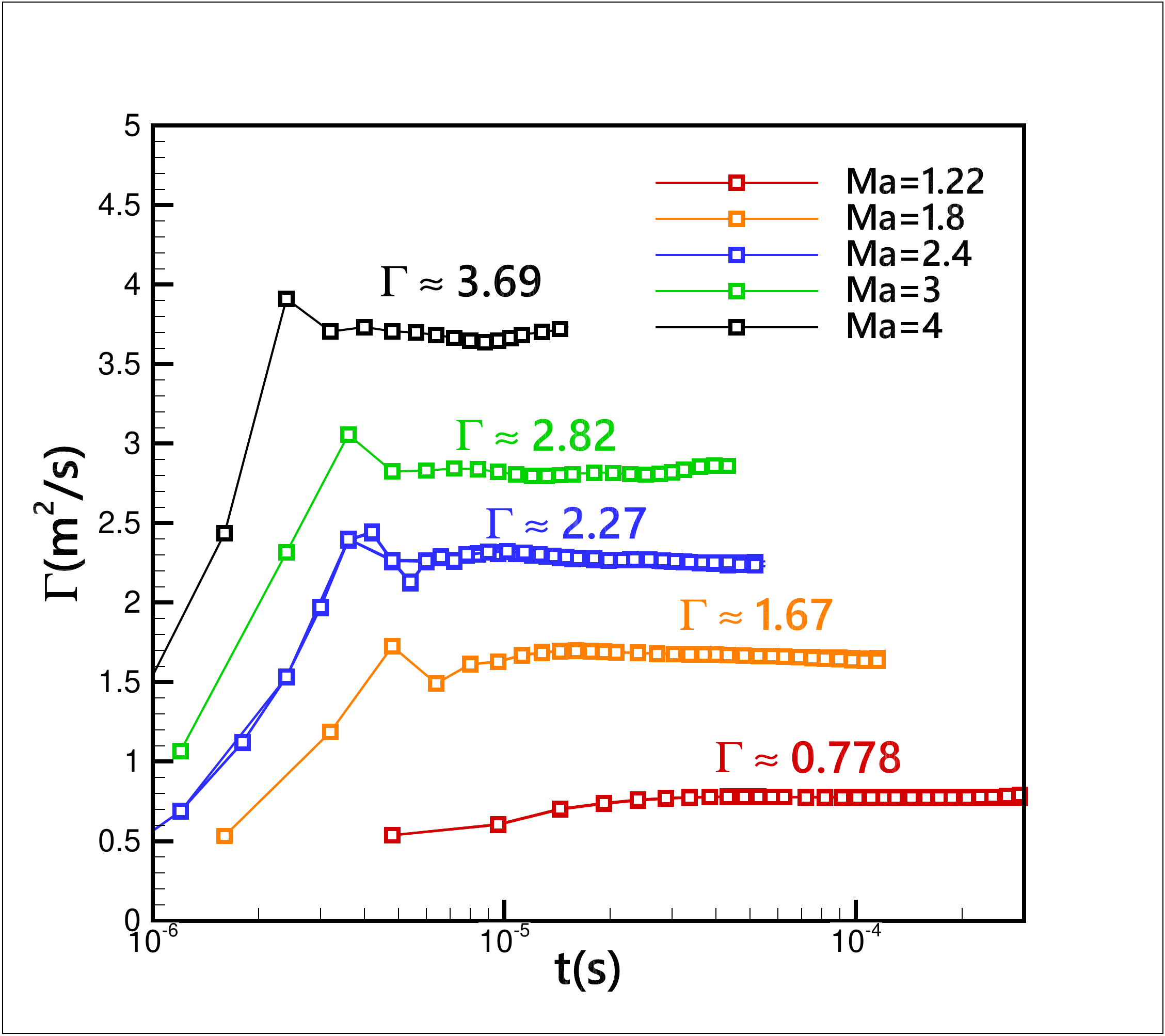}}
    \subfigure[]{
    \label{app1-circu2} 
    \includegraphics[clip=true,trim=18 20 50 30,width=.32\textwidth]{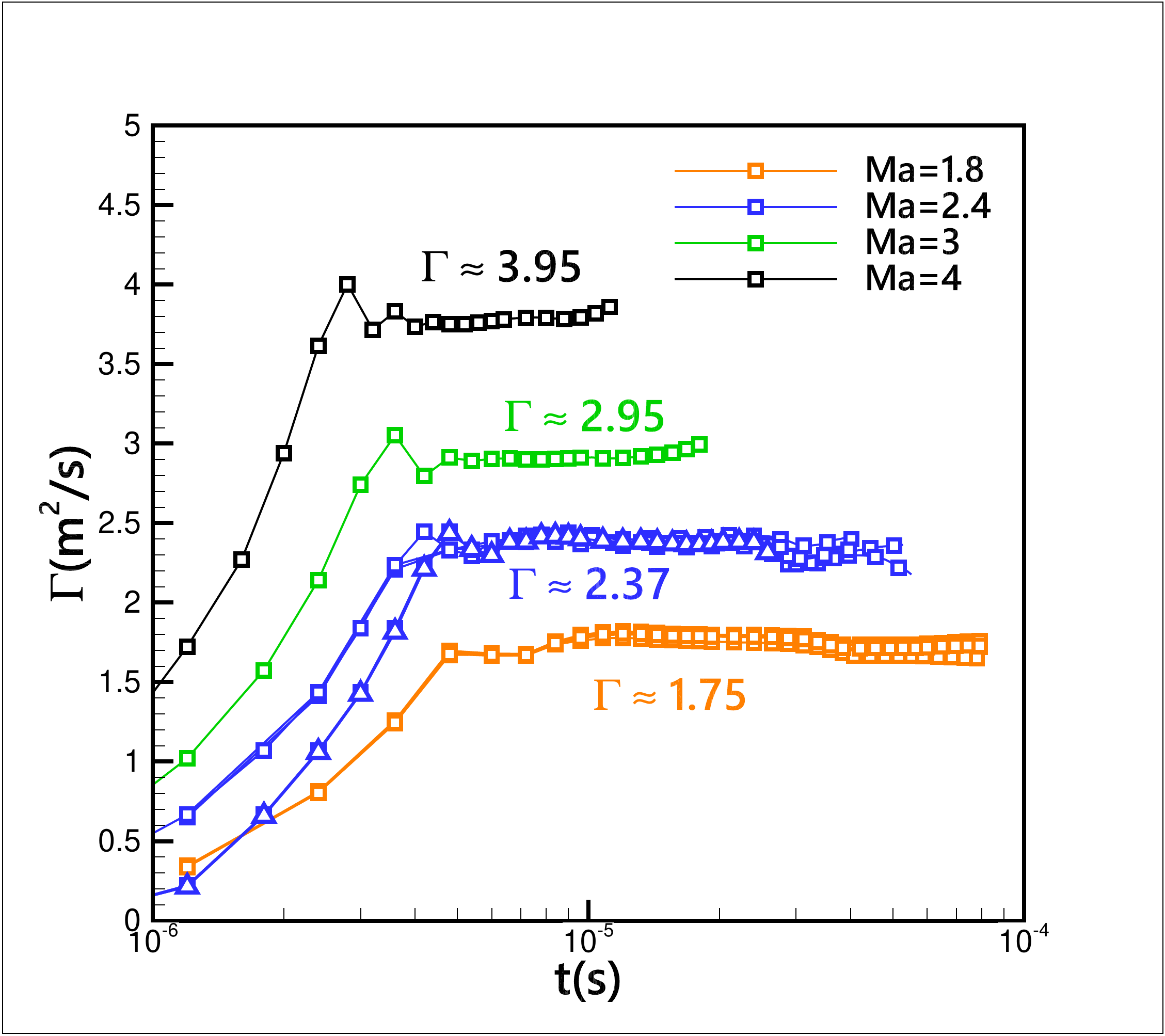}}
    \subfigure[]{
    \label{app1-circu3} 
    \includegraphics[clip=true,trim=18 20 50 30,width=.32\textwidth]{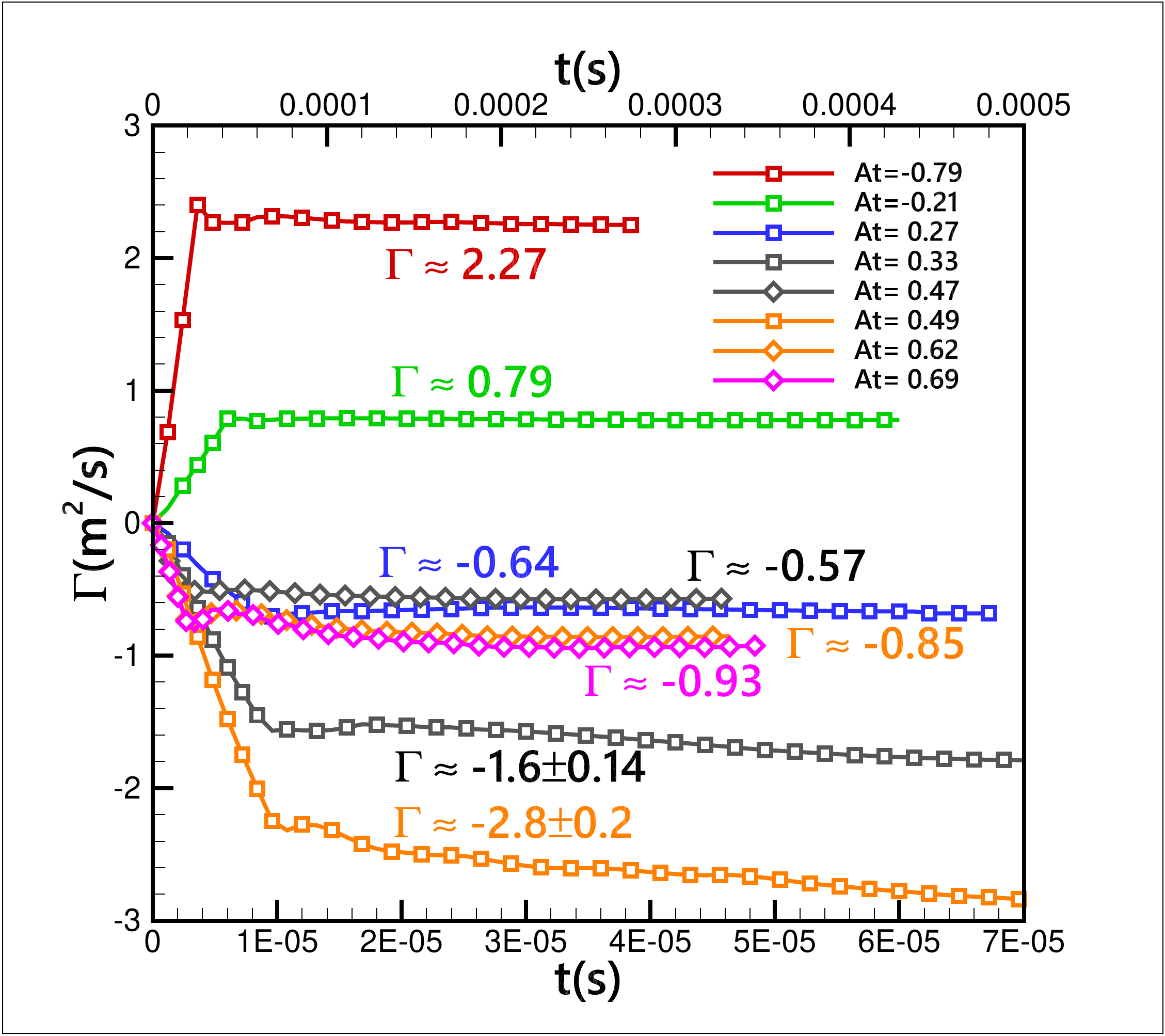}}
    \caption{Time history of circulation for all SBI cases. (a) Cylindrical bubble containing helium. (b) Spherical bubble containing helium. The 3D spherical bubble circulation of Ma = 2.4 is plotted as the blue line with triangle dots.
    (c) Cylindrical bubble containing variable density components. \label{app1-circu} }
\end{figure}

\begin{figure}
  \centering
  \includegraphics[clip=true,trim=0 140 0 0,width=.85\textwidth]{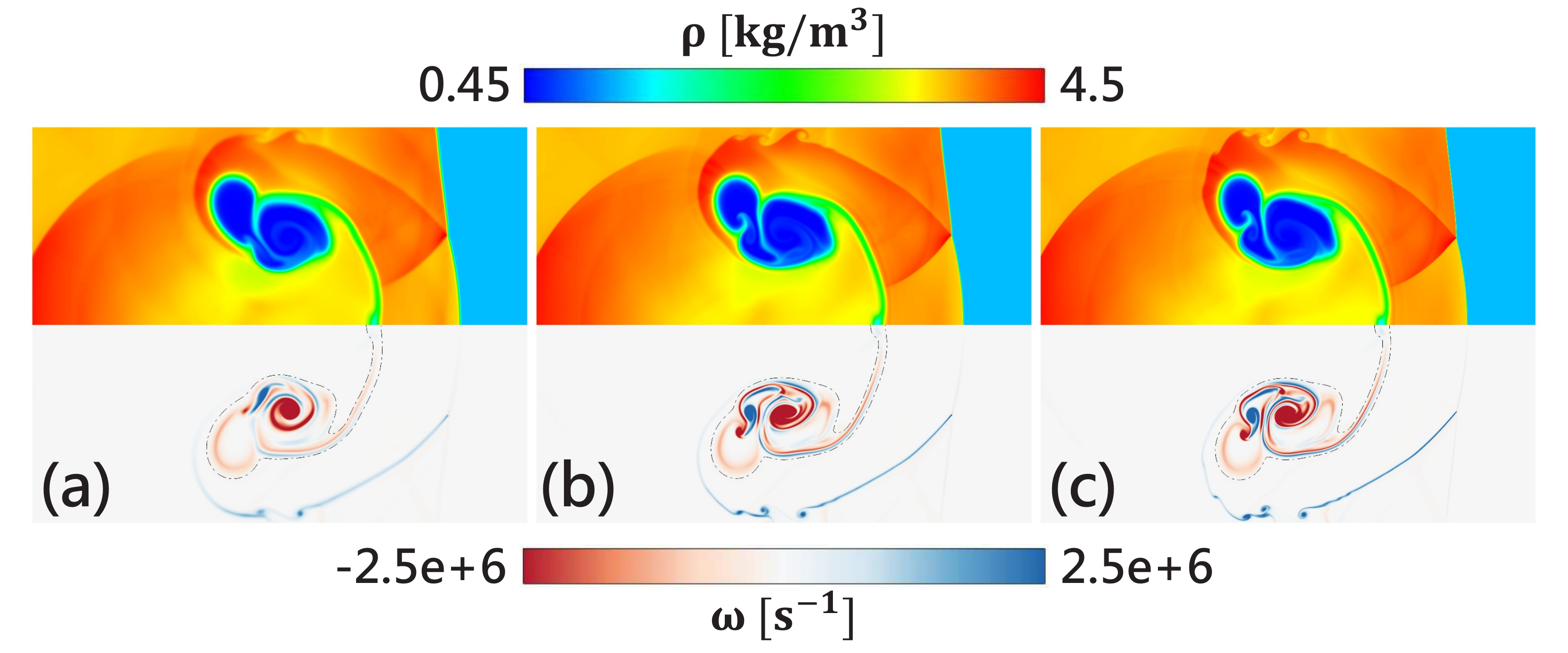}\\
  \caption{Density contour (up) and vorticity contour (bottom) of three different mesh resolutions. Ma = 2.4 cylindrical bubble case with Re = 38000 and Pe = 6400 at $t = 1.71$. (a) $\mathrm{Re}_\Delta=180$ and $\mathrm{Pe}_\Delta=30$; (b) $\mathrm{Re}_\Delta=90$ and $\mathrm{Pe}_\Delta=15$; (a) $\mathrm{Re}_\Delta=55$ and $\mathrm{Pe}_\Delta=9.2$. }\label{app1}
\end{figure}

Before testing the grid dependence, we need to choose the case with relative high Re number and Pe number to make the grid resolution, which sets the mesh resolution standard that should be reached by other concerned cases to guarantee the resolved grid number.
As for Reynolds number which is determined by circulation, Fig.~\ref{app1-circu} shows the circulation of all cases. The cases with higher shock Mach number and higher absolute Atwood number show higher circulation value. The conservative characteristic of circulation maintains well with time.
Thus, we choose Ma~=~2.4 cylindrical bubble case with Re~=~38000 and Pe~=~6400, to show the effect of grid resolution on concerned parameters. By defining mesh Reynolds number as $\mathrm{Re}_\Delta=u^*\Delta/\nu$ and mesh P\'eclet number as $\mathrm{Pe}_\Delta=u^*\Delta/\mathscr{D}$, where $\Delta$ is the mesh resolution, three kinds of mesh resolutions are studied qualitatively and quantitatively.
Figure~\ref{app1} shows the density and vorticity contour of three grid resolution. Small structures begin to appear in fine mesh with $\mathrm{Re}_\Delta=55$ and $\mathrm{Pe}_\Delta=9.2$, while gradient information is smeared by numerical viscosity in coarse mesh with $\mathrm{Re}_\Delta=180$ and $\mathrm{Pe}_\Delta=30$. General agreement from both density and vorticity is found between medium mesh and fine mesh.

\begin{figure}
  \centering
  \includegraphics[clip=true,trim=25 20 15 60,width=.6\textwidth]{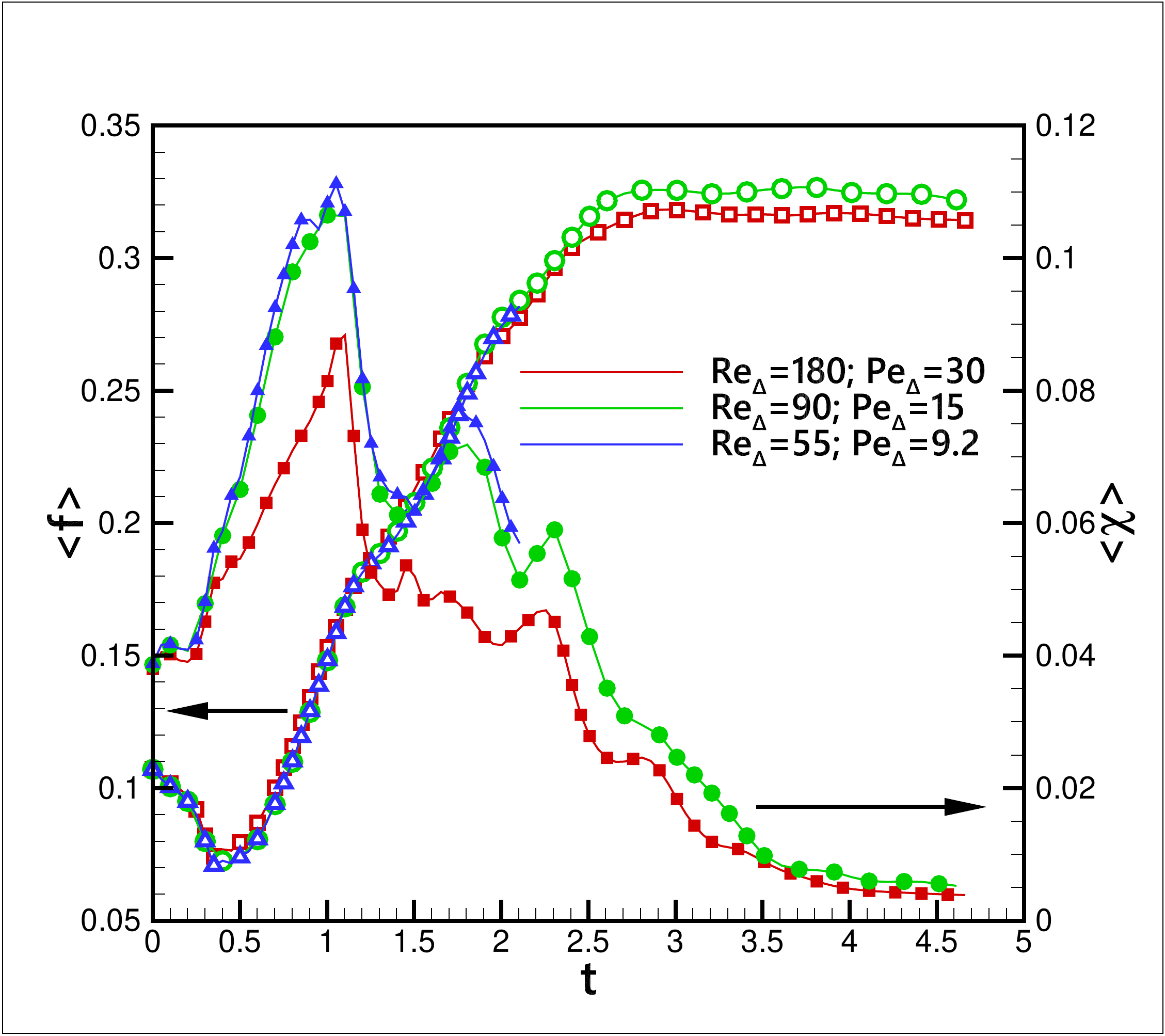}\\
  \caption{Mixedness and dissipation rate for three different mesh resolutions.}\label{app1-grid}
\end{figure}

Further checking influence of grid resolution on mixedness and dissipation, Fig.~\ref{app1-grid} illustrates the grid dependence on these two quantitative parameters. Coarse mesh fails to meet the requirement of capturing the correct value of dissipation, while the curve of medium mesh with $\mathrm{Re}_\Delta=90$ and $\mathrm{Pe}_\Delta=15$ shows the similarity with the one of fine mesh. Considering the computational burden and accuracy requirement of the simulation, we choose the medium-mesh resolution to convey the study, which is sufficient for capturing the mixing process correctly in a quantitative way.

\section{Some details for derivation of $\mathrm{D}Y/\mathrm{D}t$ and $\mathrm{D}(Y^2/2)/\mathrm{D}t$}
\label{sec:app2}
Here, more details for the derivation of $\mathrm{D}Y/\mathrm{D}t$ and $\mathrm{D}(Y^2/2)/\mathrm{D}t$ are offered.
Based on the NS equations for diffusion transport for scalar:
\begin{equation}\label{eq: diff of NS}
  \frac{\partial(\rho Y_m)}{\partial t}+\nabla(\rho Y_m \textbf{V})=\nabla\cdot\left(\mathscr{D}\rho\nabla Y_m\right), \quad m=1,2,\cdots ,s.
\end{equation}
Here we choose $m=2$ as concerned gas (helium) and breviate $Y_2$ as $Y$ in the following equations.
By using the mass conservation equation:
\begin{equation}\label{eq: mass conser}
  \frac{\partial\rho}{\partial t}+\nabla\cdot(\rho\textbf{V})=0 \Rightarrow
  \frac{\mathrm{D}\rho}{\mathrm{D}t}=-\rho\left(\nabla\cdot\textbf{V}\right),
\end{equation}
we can obtain the equation of $\mathrm{D}Y/\mathrm{D}t$ in the form of density $\rho$:
\begin{equation}\label{eq: dYdt}
  \frac{\mathrm{D}Y}{\mathrm{D}t}=\frac{1}{\rho}\nabla\cdot\left(\mathscr{D}\rho\nabla Y\right),
\end{equation}
where $\mathrm{D}\phi/\mathrm{D}t=\partial_t\phi+\textbf{V}\cdot\nabla\phi$.
It can be further derived that:
\begin{equation}\label{eq: dYdt-SDR-1}
  \left(\frac{\partial}{\partial t}+\textbf{V}\cdot\nabla-\mathscr{D}\nabla^2\right)Y=\frac{\mathscr{D}}{\rho}\nabla{\rho}\cdot\nabla {Y}.
\end{equation}
Again by using $1/\rho=Y/\rho'_2+(1-Y)/\rho'_1$:
\begin{equation}\label{eq: nabla rho}
  -\frac{\nabla\rho}{\rho^2}=\left(\frac{1}{\rho'_2}-\frac{1}{\rho'_1}\right)\cdot\nabla Y,
\end{equation}
then we can obtain the advection equation of mass fraction in variable-density flows as shown in Eq.~(\ref{eq: dYdt-SDR}) by dimensionless form:
\begin{equation}\label{eq: dYdt-SDR-KL}
  \left(\frac{\partial}{\partial t}+\textbf{V}\cdot\nabla-\mathscr{D}\nabla^2\right)Y=-\mathscr{D}\frac{1-\sigma}{(1-\sigma)Y+\sigma}\nabla Y\cdot\nabla Y.
\end{equation}
By multiplying the advection equation of mass fraction Eq.~(\ref{eq: dYdt-SDR-1}) by mass fraction $Y$, we obtain the advection equation of mass fraction energy $Y^2/2$ in the source of density:
\begin{equation}\label{eq: dY2dt-SDR-1}
  \left(\frac{\partial}{\partial t}+\textbf{V}\cdot\nabla-\mathscr{D}\nabla^2\right)\frac{1}{2}Y^2=\frac{\mathscr{D}Y}{\rho}\nabla{\rho}\cdot\nabla {Y}-\mathscr{D}\nabla Y\cdot\nabla Y,
\end{equation}
then the advection equation of mass fraction energy $Y^2/2$ as shown in Eq.~(\ref{eq: dY2dt-SDR}):
\begin{equation}\label{eq: dY2dt-SDR-KL}
  \left(\frac{\partial}{\partial t}+\textbf{V}\cdot\nabla-\mathscr{D}\nabla^2\right)\frac{1}{2}Y^2=-\mathscr{D}\left(2-\frac{\sigma}{(1-\sigma)Y+\sigma}\right)\nabla Y\cdot\nabla Y.
\end{equation}

\section{Some discussions on Eq.~(\ref{eq: Y-rho})}
\label{sec:app3}
\begin{figure}
    \centering
    \subfigure[]{
    \label{app3-com1} 
    \includegraphics[clip=true,trim=18 20 70 30,width=.32\textwidth]{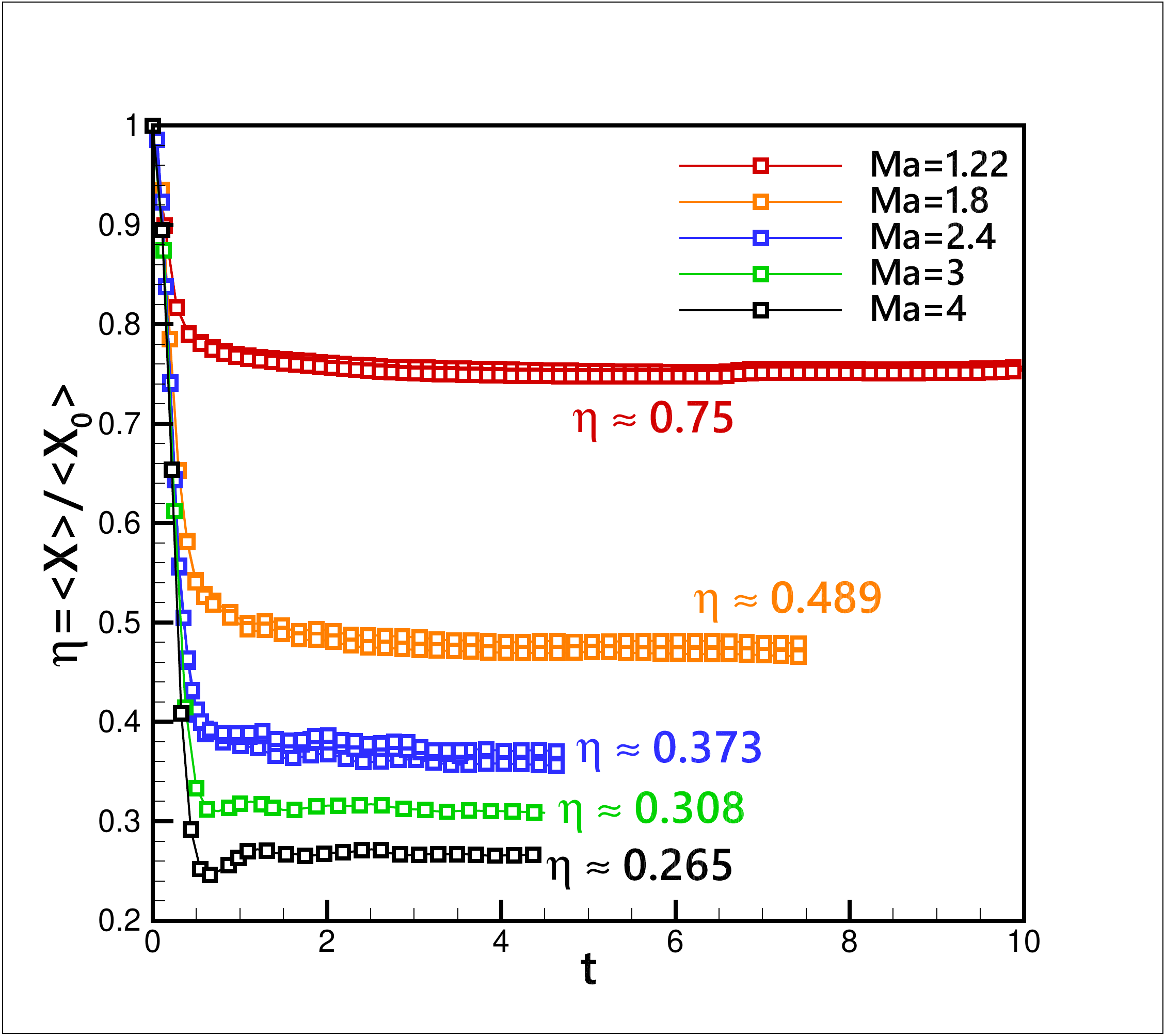}}
    \subfigure[]{
    \label{app3-com2} 
    \includegraphics[clip=true,trim=18 20 70 30,width=.32\textwidth]{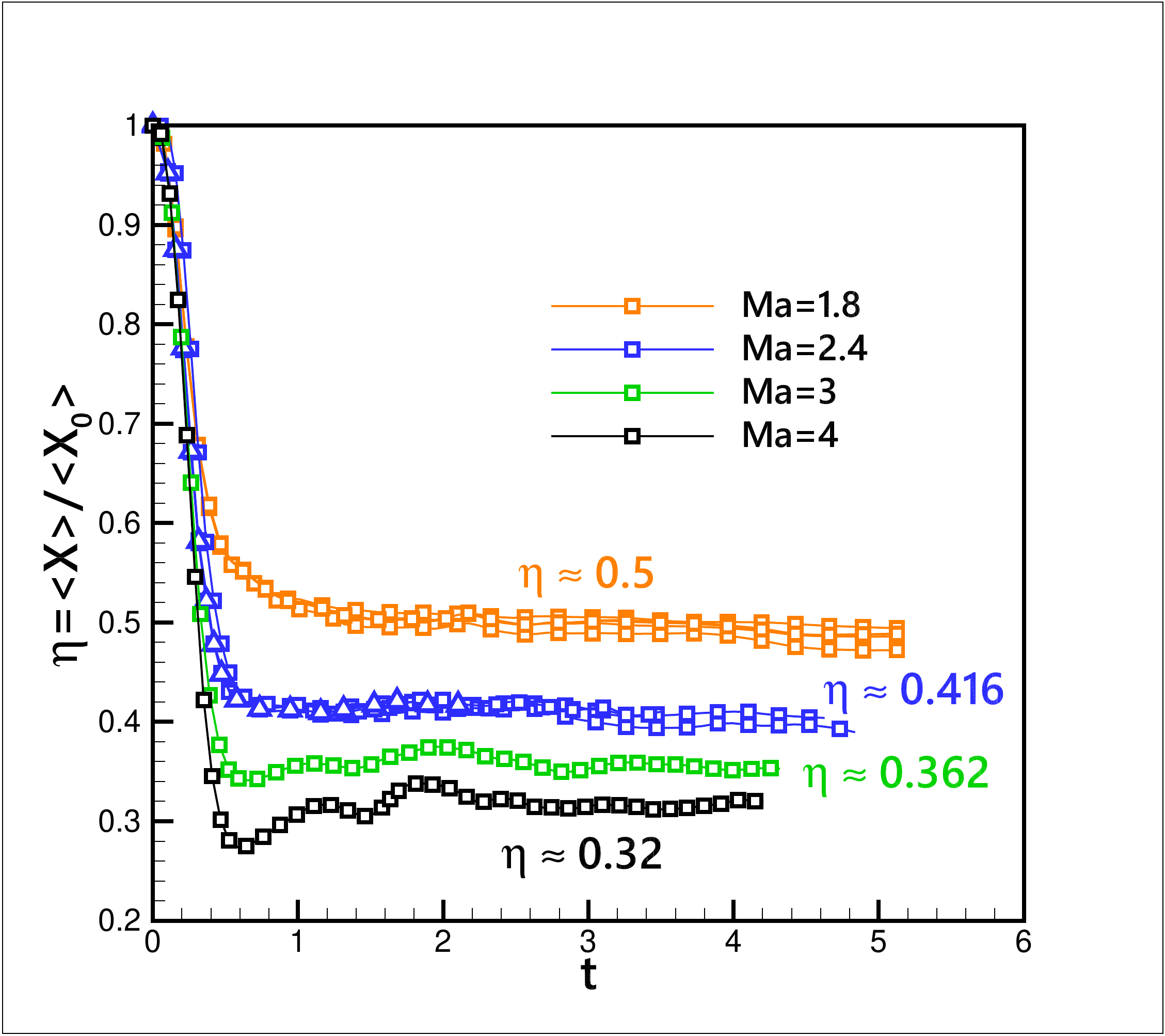}}
    \subfigure[]{
    \label{app3-com3} 
    \includegraphics[clip=true,trim=18 20 70 30,width=.32\textwidth]{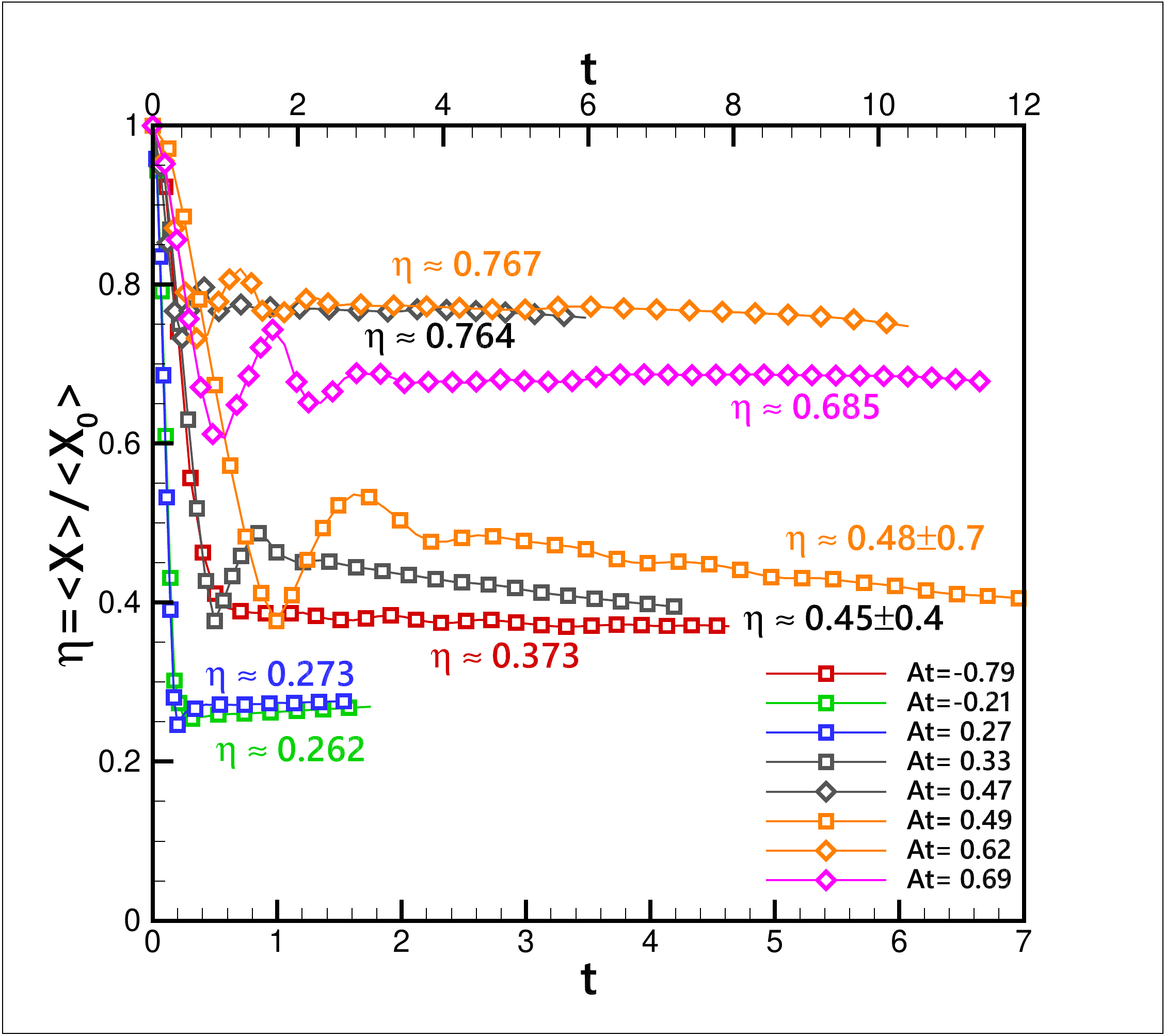}}
    \caption{Compression rate of all cases studied in the present paper. (a) Cylindrical bubble containing helium. (b) Spherical bubble containing helium. The 3D spherical bubble compression rate of Ma = 2.4 is plotted as the blue line with triangle dots.
    (c) Cylindrical bubble containing variable density components. \label{app3-com} }
\end{figure}
\begin{figure}
    \centering
    \subfigure[]{
    \label{app3-1} 
    \includegraphics[clip=true,trim=5 0 55 0,width=.31\textwidth]{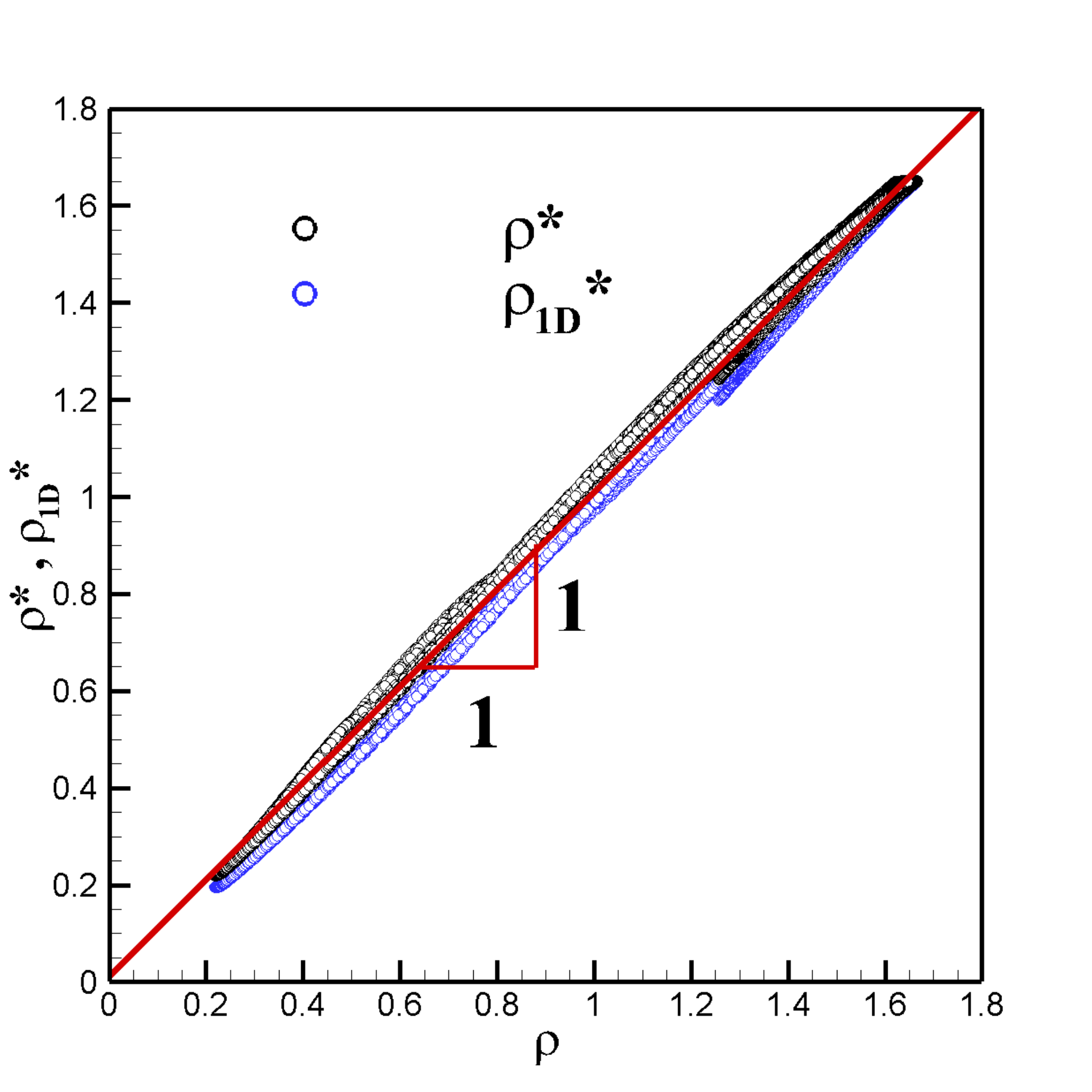}}
    \subfigure[]{
    \label{app3-2} 
    \includegraphics[clip=true,trim=5 0 55 0,width=.31\textwidth]{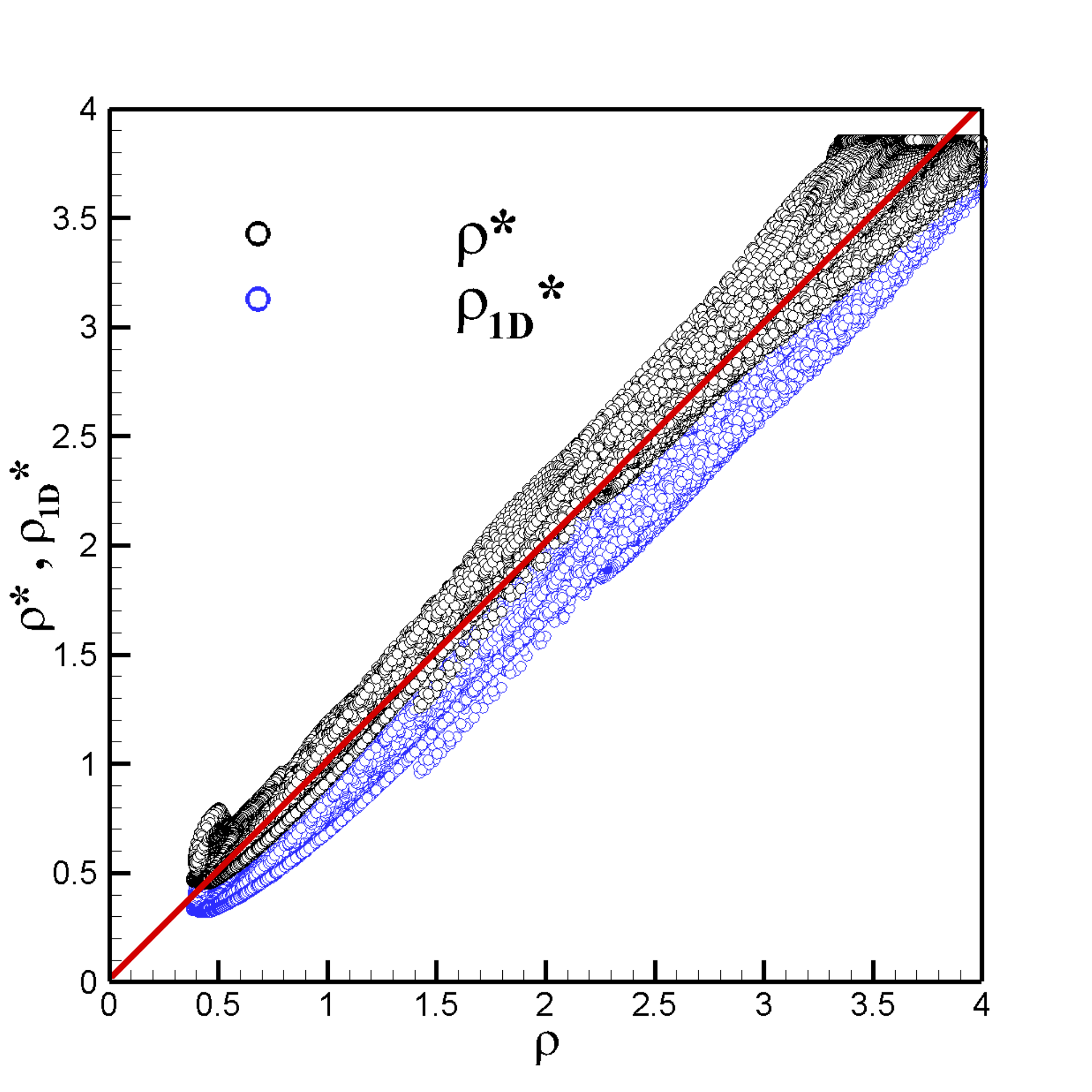}}
    \subfigure[]{
    \label{app3-3} 
    \includegraphics[clip=true,trim=5 0 55 0,width=.31\textwidth]{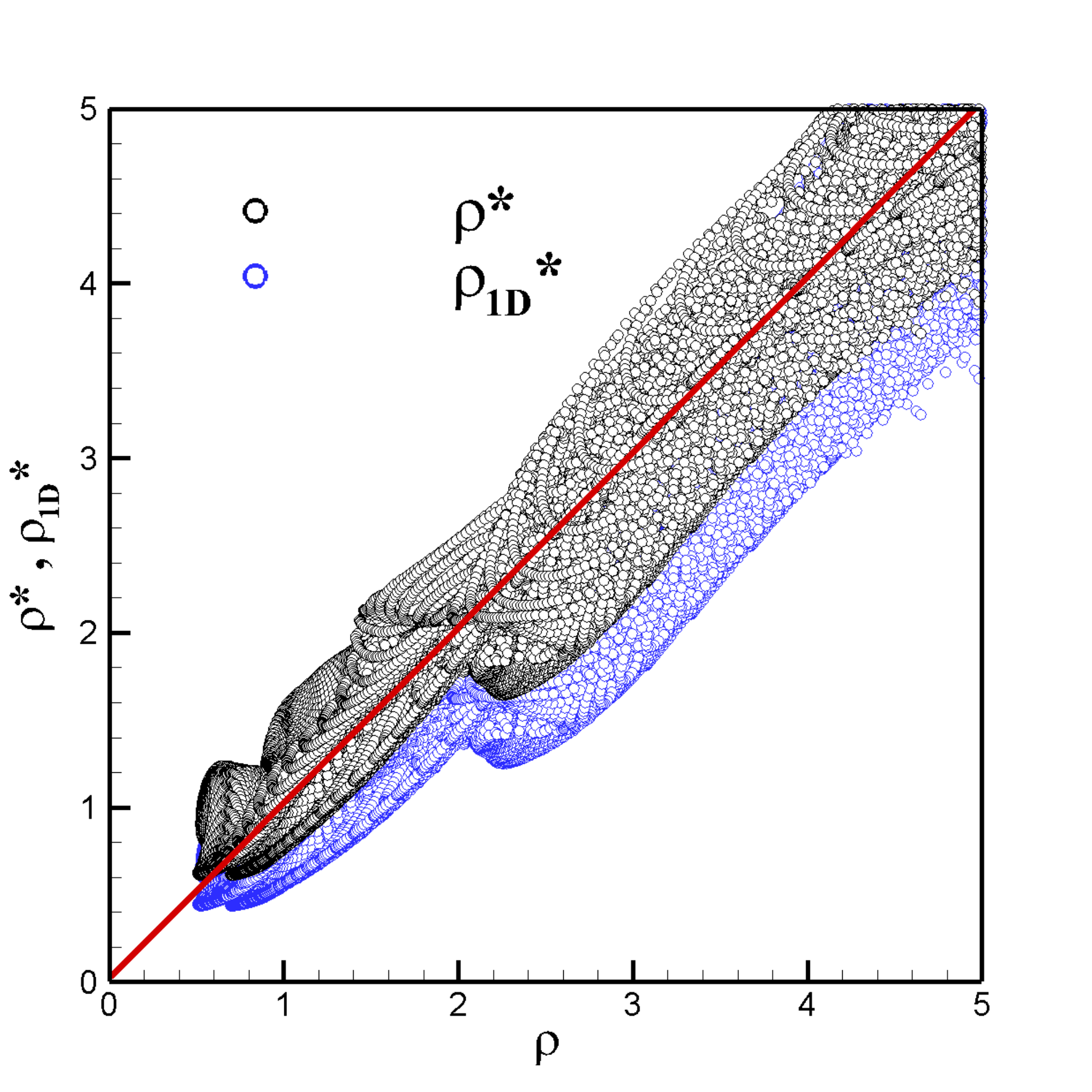}}
    \caption{Comparison between nominated density $\rho^*$ and exact real density $\rho$ at different shock Mach number in cylindrical cases. (a) Ma~=~1.22 (Re~=~5750, Pe~=~5500, $t=2.74$); (b) Ma~=~2.4 (Re~=~38000, Pe~=~6400, $t=1.71$); (c) Ma~=~4 ($t=1.42$). \label{app3} }
\end{figure}
Although Eq.~(\ref{eq: Y-rho}) is widely accepted in incompressible variable-density miscible flows, the compressible effect is needed to be carefully examined in RM-type flows, especially with high shock Mach number. The density of shocked air $\rho'_1$ can be directly calculated from one-dimensional shock dynamics. As for shocked gas, several reflect shock will occur immediately after the shock passage. Thus determining the macroscopic density of bubble $\rho'_2$ is essential. Here, we find the mass of helium bubble $\left<\rho Y\right>$ is essentially constant after shock, which can be proved as:
\begin{equation}\label{eq: mass int}
  \frac{\mathrm{D}\left<\rho Y\right>}{\mathrm{D}t}=\left<\frac{\mathrm{D}(\rho Y)}{\mathrm{D}t}\right>+\left<\rho Y(\nabla\cdot\textbf{V})\right>.
\end{equation}
From the diffusion equation of mass fraction in the form of NS equations Eq.~(\ref{eq: diff of NS}), we obtain:
\begin{equation}\label{eq: drhoY}
  \frac{\mathrm{D}(\rho Y)}{\mathrm{D}t}=-\rho Y\nabla\cdot\textbf{V}+\nabla\cdot\left(\mathscr{D}\rho\nabla Y\right).
\end{equation}
Then the time variation of mass of helium can be expressed as:
\begin{equation}\label{eq: drhoYdt}
  \frac{\mathrm{D}\left<\rho Y\right>}{\mathrm{D}t}=\left< \nabla\cdot\left(\mathscr{D}\rho\nabla Y\right) \right>=\oiint_\textrm{S}\mathscr{D}\rho\nabla Y\cdot \vec{\textbf{n}}d\textrm{S}=0,
\end{equation}
by using Gauss's flux theorem $\iiint_{\mathcal{V}}\nabla\cdot\vec{\phi}d\mathcal{V}=\oiint_\textrm{S}\vec{\phi}\cdot\vec{\textbf{n}}d\textrm{S}$.
Thus the density of post-shock helium in the bubble can be estimated as:
\begin{equation}\label{eq: rho}
  \rho'_2=\frac{\left<\rho Y\right>}{\mathrm{V}_\infty}\approx\frac{\rho_2\mathrm{V}_0}{\mathrm{V}_\infty}=\rho_2/\eta.
\end{equation}
Fortunately, compression rate $\eta$ collapses to a steady value for most cases, as shown in Fig.~\ref{app3-com}. The near-constant behavior of compression rate is also proven in Appendix~\ref{sec:app7}.
The compression rate is slightly higher in spherical bubble cases than in cylindrical cases under the same shock Ma number due to the weaker compression in axisymmetric shock than in symmetric shock~\cite{haas1987interaction}.
It is noteworthy that the compression rate declines at high At number with Ma=2.4, while conservative characteristic maintains well in other cases. The discrepancy is caused by strong shock focusing in the heavy bubble cases, leading to the continuous compression from the reflect shock inside the bubble~\cite{Zabusky1998shock}.

Here, we validate Eq.~(\ref{eq: Y-rho}) by defining an alternative density $\rho^*=1/\left(Y/\rho'_2+(1-Y)/\rho'_1\right)$ comparing with numerical results of $\rho$ as shown in Fig.~\ref{app3}. The alternative density from one-dimensional shock dynamics $\rho^*_{1D}=1/\left(Y/(\rho'_2)_{1D}+(1-Y)/\rho'_1\right)$ is also compared.
A linear relationship is obtained for even the Ma=4 case.
However, it still can be found that at low Mach number $\rho^*\approx\rho$ while a border width occurs at higher shock Mach number. This is due to the reflected shock that exists in higher shock Mach number. Comparing $\rho^*$ and $\rho^*_{1D}$, we find that $\rho^*_{1D}$ slightly underestimates density in high shock Mach number. Thus, it is better to use the compression rate to estimate the post-shock gas density. The theoretical model for the compression rate has already been built \cite{giordano2006richtmyer} and is recommended as the fundamental parameters that control mixing in RM-type flows.

\section{Some discussions and proof of $\left<\mathscr{D}\nabla^2Y\right>=0$}
\label{sec:app4}
\begin{figure}
    \centering
    \subfigure[]{
    \label{app4-1} 
    \includegraphics[clip=true,trim=0 0 0 0,width=.43\textwidth]{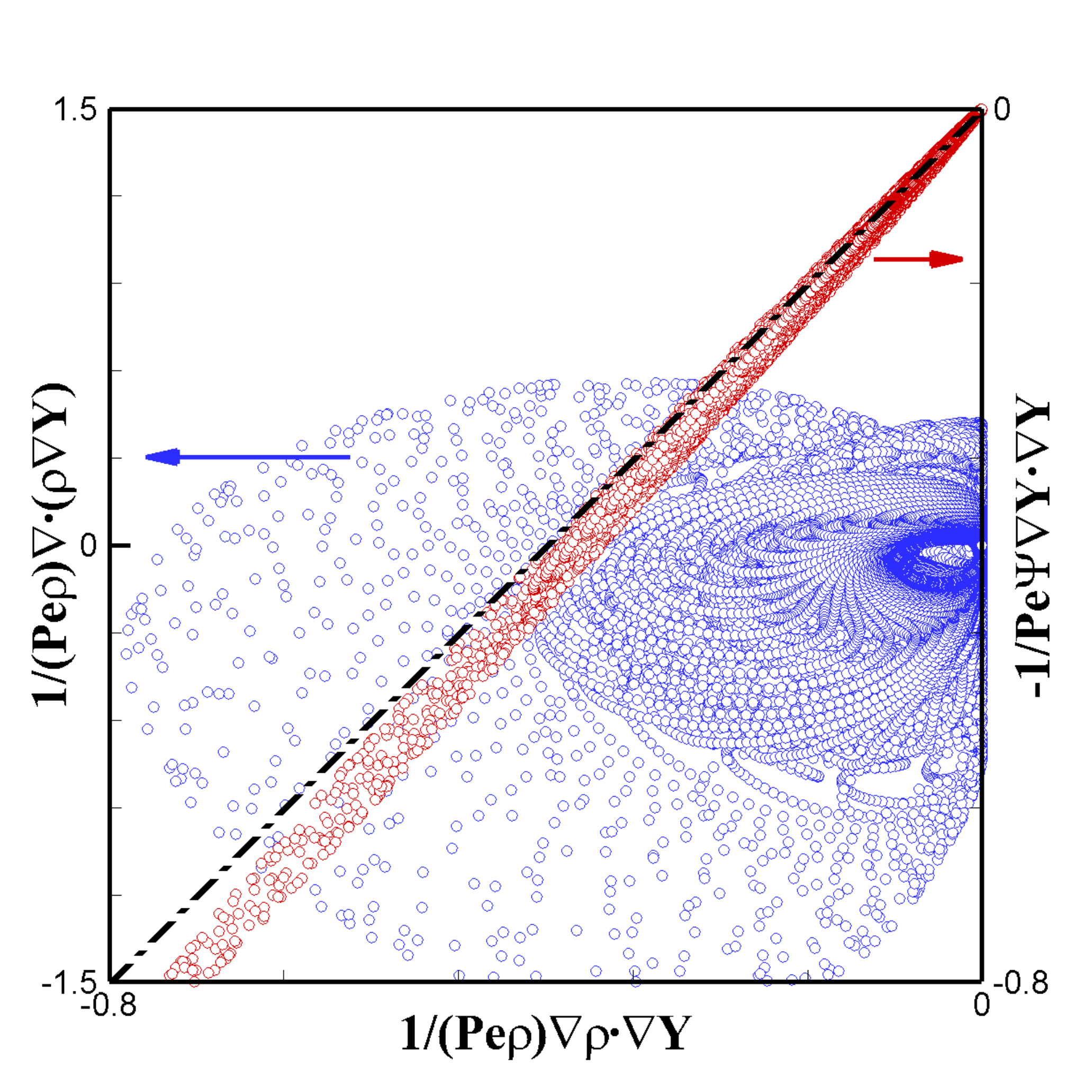}}
    \subfigure[]{
    \label{app4-2} 
    \includegraphics[clip=true,trim=20 20 15 25,width=.5\textwidth]{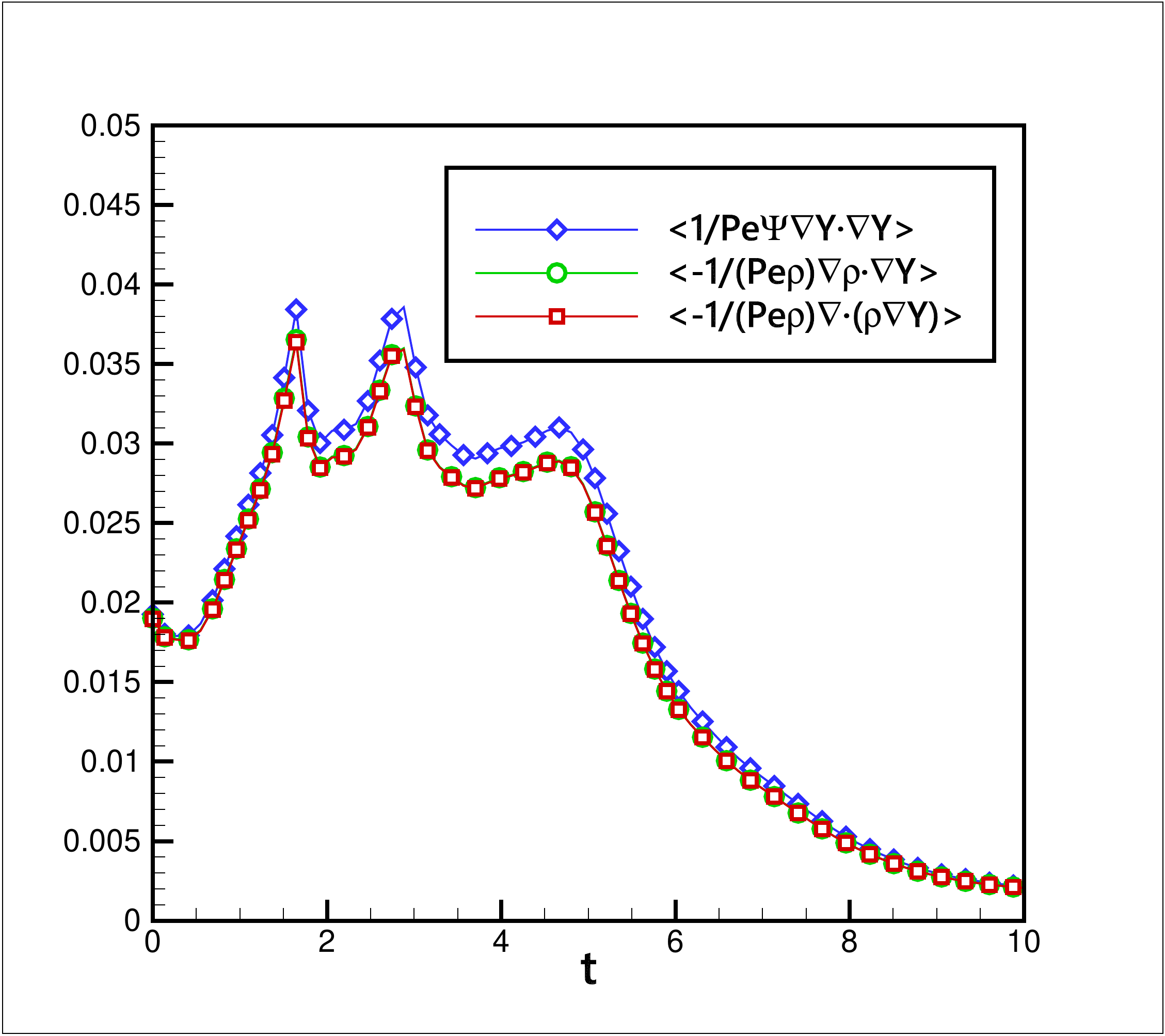}}
    \caption{Relation between the right hand term of Eqs.~(\ref{eq: dYdt}), (\ref{eq: dYdt-SDR-1}) and (\ref{eq: dYdt-SDR-KL}) where $\Psi=\frac{1-\sigma}{(1-\sigma)Y+\sigma}$ for Ma=1.22 case. (a) Scatter points of the three terms at $t=2.74$; (b) Time history of the volumetric integration of the three terms.   \label{app4} }
\end{figure}
Since the diffusion term is highly nonlinear and not strictly positive as dissipation, the characteristic of this term is briefly introduced in this appendix.
It is relatively simple to prove the zero value of the diffusion term as:
\begin{equation}\label{eq: diff_int}
  \left<\mathscr{D}\nabla^2Y\right>=  \mathscr{D}\left<\nabla\cdot(\nabla Y)\right>=\mathscr{D}\oiint_\textrm{S}\nabla Y\cdot\vec{\textbf{n}}d\textrm{S}=0.
\end{equation}
This characteristic is vital in deriving the integration of $\left<\mathrm{D}Y/\mathrm{D}t\right>$. Also, if we compare the right term of Eqs.~(\ref{eq: dYdt}), (\ref{eq: dYdt-SDR-1}), and (\ref{eq: dYdt-SDR-KL}), we can find the interesting phenomena that, as shown in Fig.~\ref{app4}:
\begin{equation}\label{eq: three-term relation}
\left\{ \begin{array}{l}
\frac{1}{\rho}\nabla\cdot\left(\mathscr{D}\rho\nabla Y\right)\ne\frac{\mathscr{D}}{\rho}\nabla{\rho}\cdot\nabla {Y}\approx-\mathscr{D}\Psi\nabla Y\cdot\nabla Y, \\
\left<\frac{1}{\rho}\nabla\cdot\left(\mathscr{D}\rho\nabla Y\right)\right>=\left<\frac{\mathscr{D}}{\rho}\nabla{\rho}\cdot\nabla {Y}\right>\approx-\left<\mathscr{D}\Psi\nabla Y\cdot\nabla Y\right>.
\end{array} \right.
\end{equation}
This local nonzero but global zero integration behavior of diffusion term may shows its effect in density gradient redistributed diffusion term as introduced in this paper.

\section{Some discussions on Eq.~(\ref{eq: div})}
\label{sec:app5}
\begin{figure}
    \centering
    \subfigure[]{
    \label{app5-1} 
    \includegraphics[clip=true,trim=50 70 110 110,width=.31\textwidth]{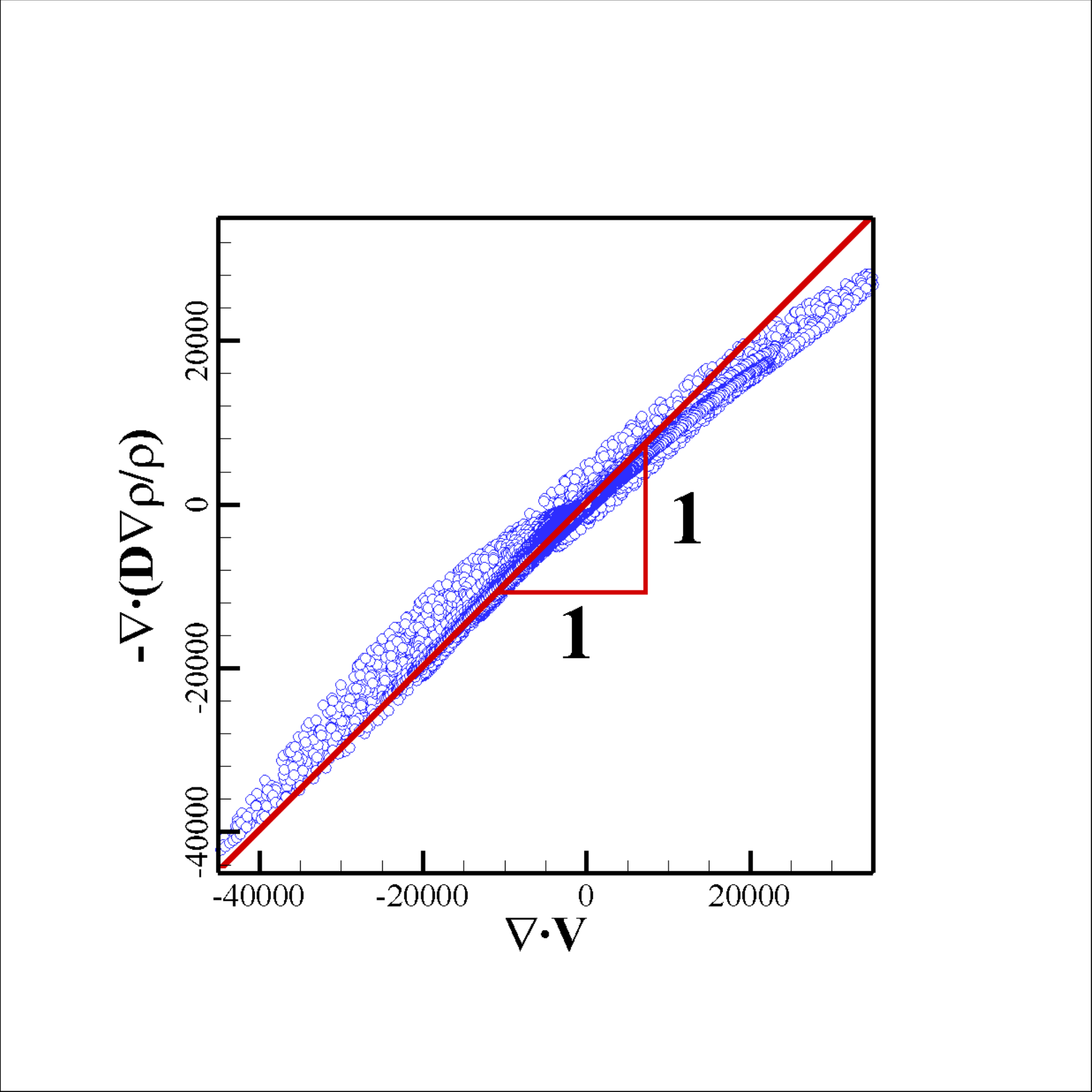}}
    \subfigure[]{
    \label{app5-2} 
    \includegraphics[clip=true,trim=50 70 110 110,width=.31\textwidth]{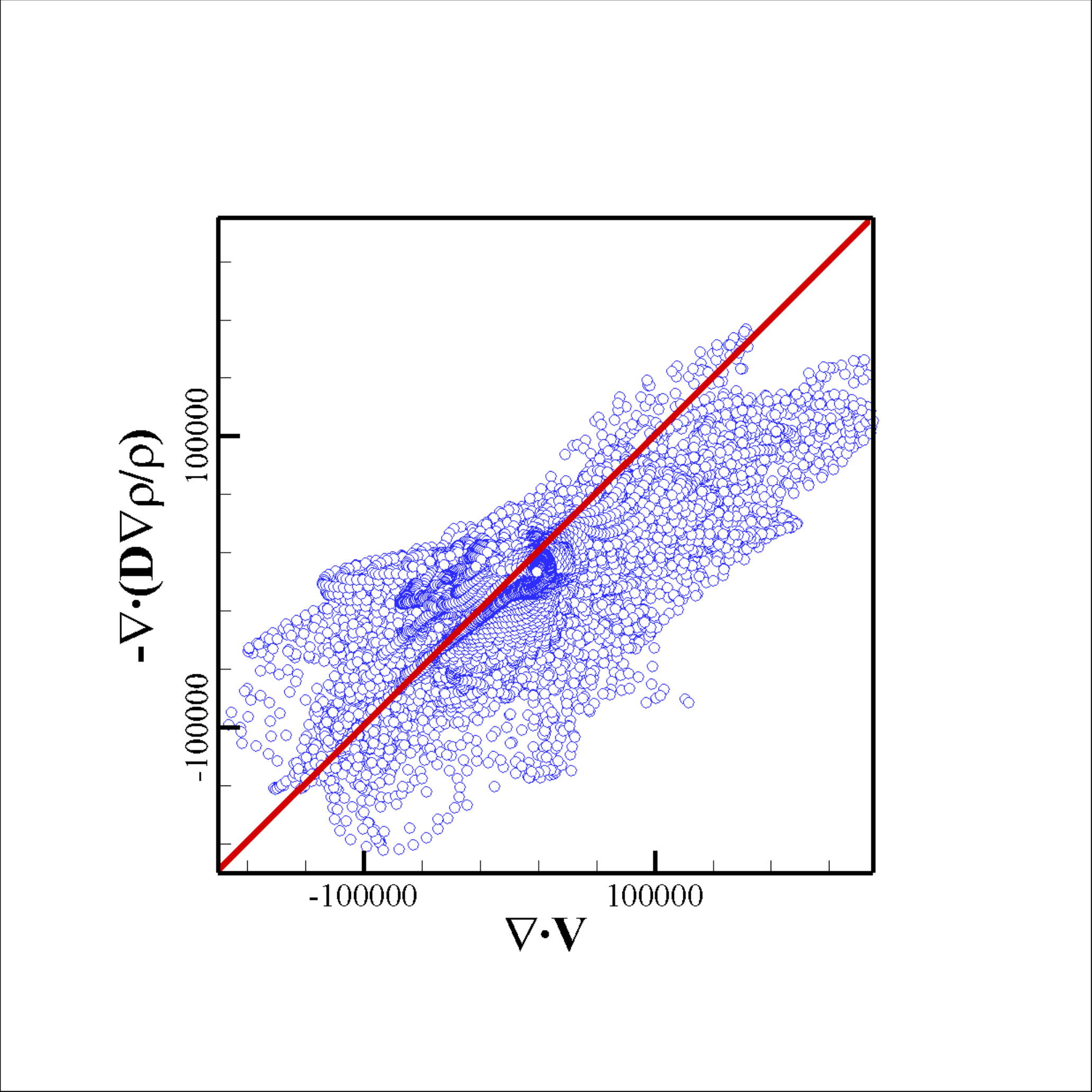}}
    \subfigure[]{
    \label{app5-3} 
    \includegraphics[clip=true,trim=50 70 110 110,width=.31\textwidth]{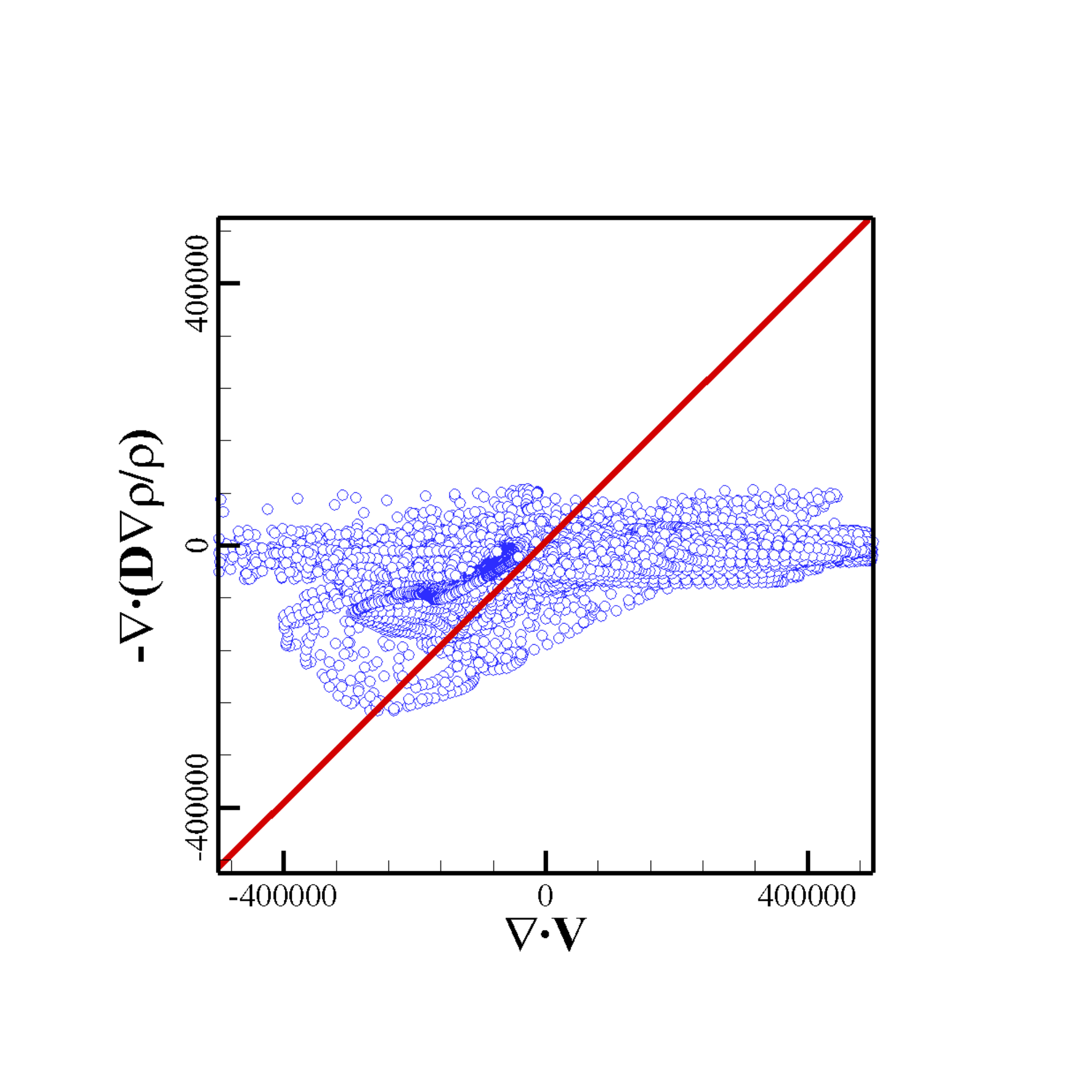}}
    \caption{Comparison between the velocity divergence and divergence of $-\mathscr{D}\nabla\rho/\rho$.
    (a) Ma~=~1.22 (Re~=~5750, Pe~=~5500, $t=2.74$); (b) Ma~=~2.4 (Re~=~38000, Pe~=~6400, $t=1.71$); (c) Ma~=~4 ($t=1.42$). \label{app5-dots} }
\end{figure}
\begin{figure}
  \centering
  \includegraphics[clip=true,trim=0 0 170 0,width=.8\textwidth]{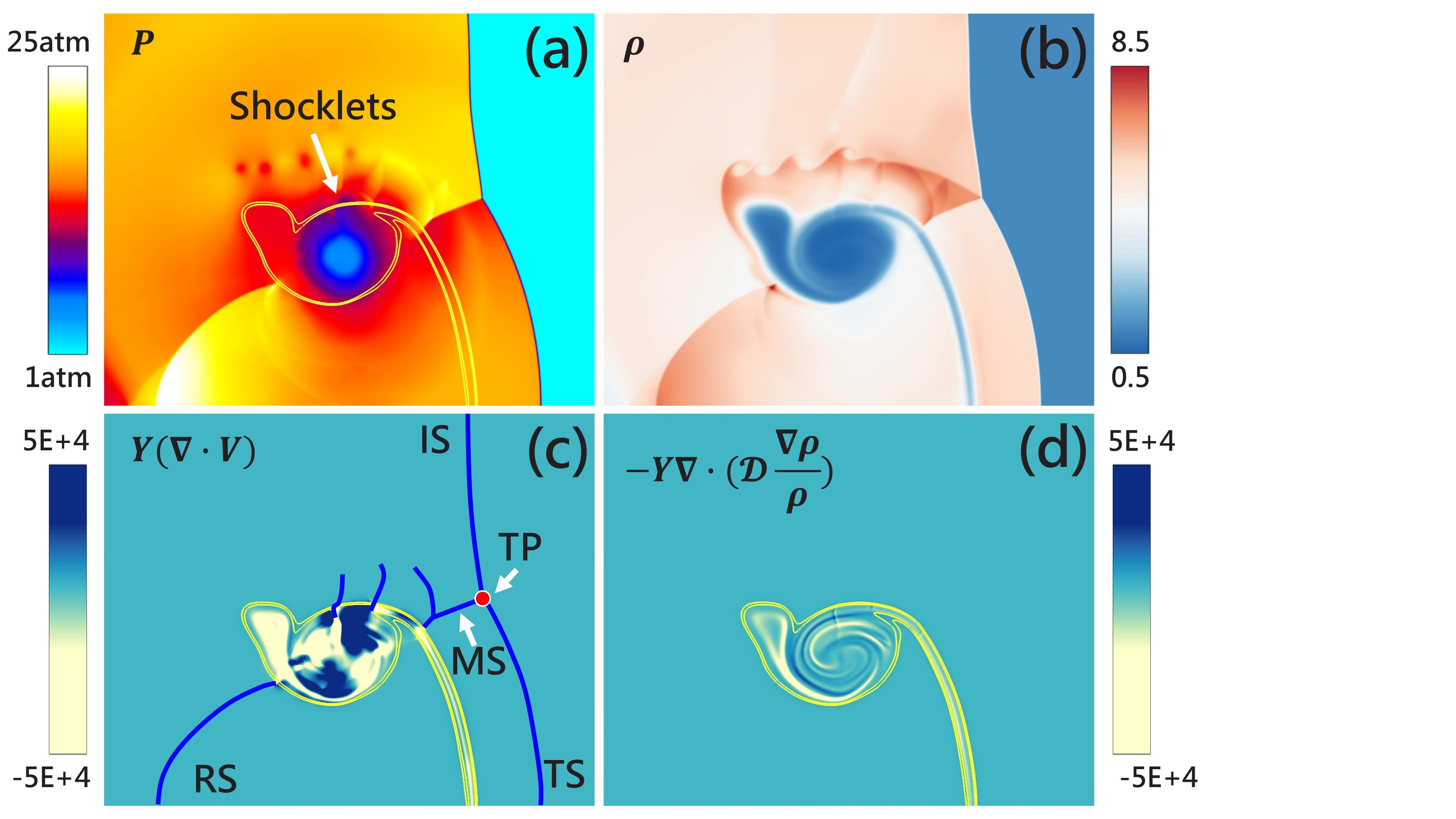}\\
  \caption{(a) Pressure contour; (b) density contour; (c) contour of $Y(\nabla\cdot\textbf{V})$ with the shock wave structures denoted by blue lines; (d) contour of $-Y\nabla\cdot\left(\mathscr{D}\nabla\rho/\rho\right)$. Ma~=~4 case at $t=1.42$. IS: incident shock; RS: reflected shock; MS: Mach stem; TS: transmitted shock.  \label{app5-contour}}
\end{figure}
Here, the divergence of velocity relates to density is discussed.
By using $1/\rho=Y/\rho'_2+(1-Y)/\rho'_1$, we can obtain:
\begin{equation}\label{eq: dYdt-drhodt}
  \left(\frac{1}{\rho'_2}-\frac{1}{\rho'_1}\right)\frac{\mathrm{D}Y}{\mathrm{D}t}=-\frac{1}{\rho^2}\frac{\mathrm{D}\rho}{\mathrm{D}t}.
\end{equation}
As for Eq.~(\ref{eq: dYdt}), by substituting $\mathrm{D}Y/\mathrm{D}t$ by Eq.~(\ref{eq: dYdt-drhodt}) and $\nabla Y$ by Eq.~(\ref{eq: nabla rho}) in the form of $\rho$, then we obtain:
\begin{equation}\label{eq: drhodt}
  \frac{\mathrm{D}\rho}{\mathrm{D}t}=\rho\nabla\cdot\left(\mathscr{D}\frac{\nabla\rho}{\rho}\right).
\end{equation}
By using the conservation law of mass, as shown in Eq.~(\ref{eq: mass conser}), we can obtain Eq.~(\ref{eq: div}):
\begin{equation}\label{eq: div-rho}
  \nabla\cdot \textbf{V}=-\nabla\cdot\left(\mathscr{D}\frac{\nabla\rho}{\rho}\right)=-\mathscr{D}
                       \left(\frac{\nabla^2\rho}{\rho}-\frac{1}{\rho^2}\nabla\rho\cdot\nabla\rho\right),
\end{equation}
in which the first term on the right, also the primary source of density gradient redistributed diffusion, can be expressed as:
\begin{equation}\label{eq: div-1}
  \frac{\nabla^2\rho}{\rho}=\left(\frac{\sqrt{2}(1-\sigma)}{(1-\sigma)Y+\sigma}\right)^2\nabla Y\cdot\nabla Y-
                    \left(\frac{1-\sigma}{(1-\sigma)Y+\sigma}\right)\nabla^2 Y,
\end{equation}
and the second term can be expressed as:
\begin{equation}\label{eq: div-2}
  \frac{1}{\rho^2}\nabla\rho\cdot\nabla\rho=\left(\frac{1-\sigma}{(1-\sigma)Y+\sigma}\right)^2\nabla Y\cdot\nabla Y.
\end{equation}

\begin{figure}
    \centering
    \subfigure[]{
    \label{app5-line2} 
    \includegraphics[clip=true,trim=25 20 25 60,width=.32\textwidth]{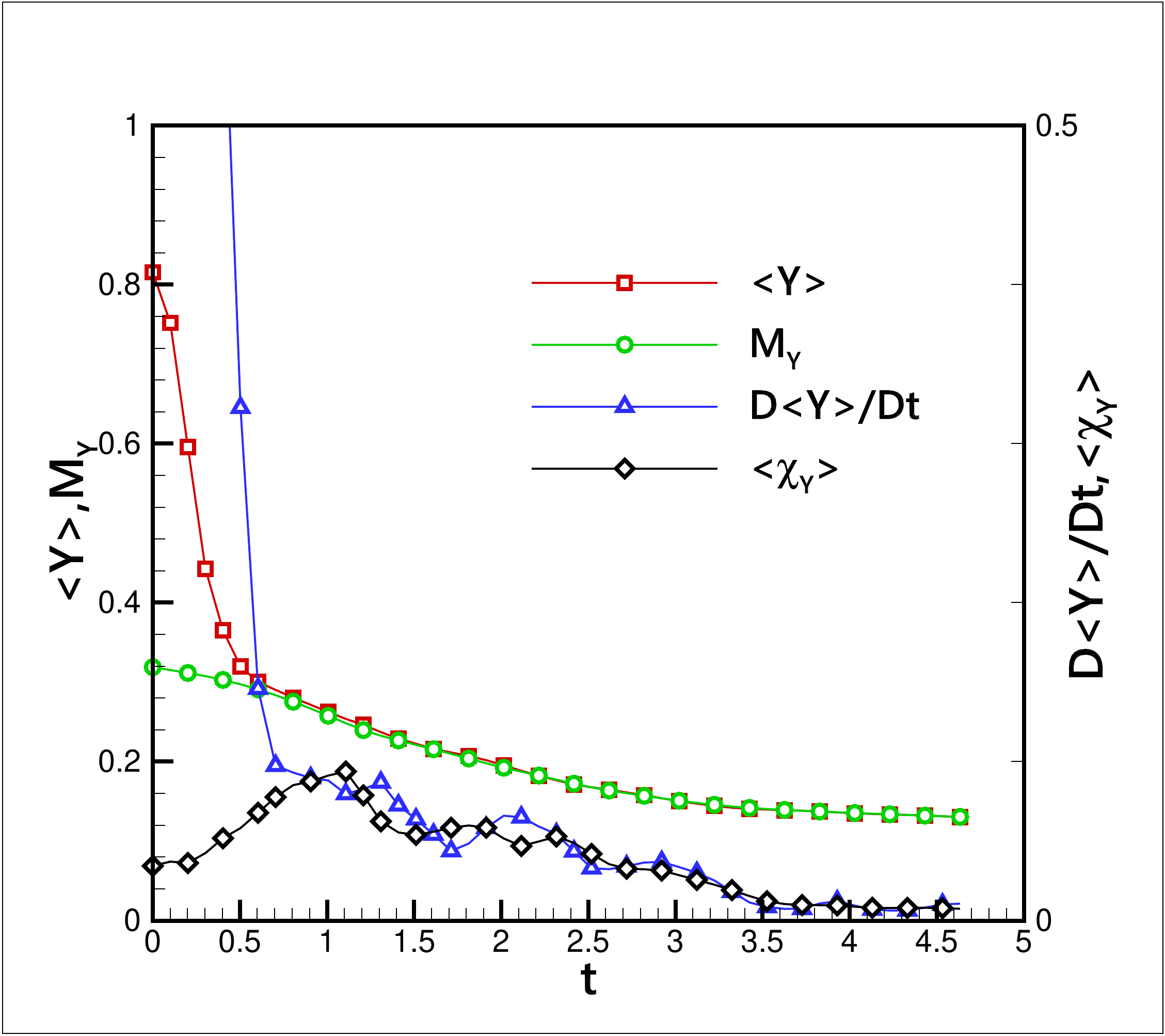}}
    \subfigure[]{
    \label{app5-line3} 
    \includegraphics[clip=true,trim=25 20 25 60,width=.32\textwidth]{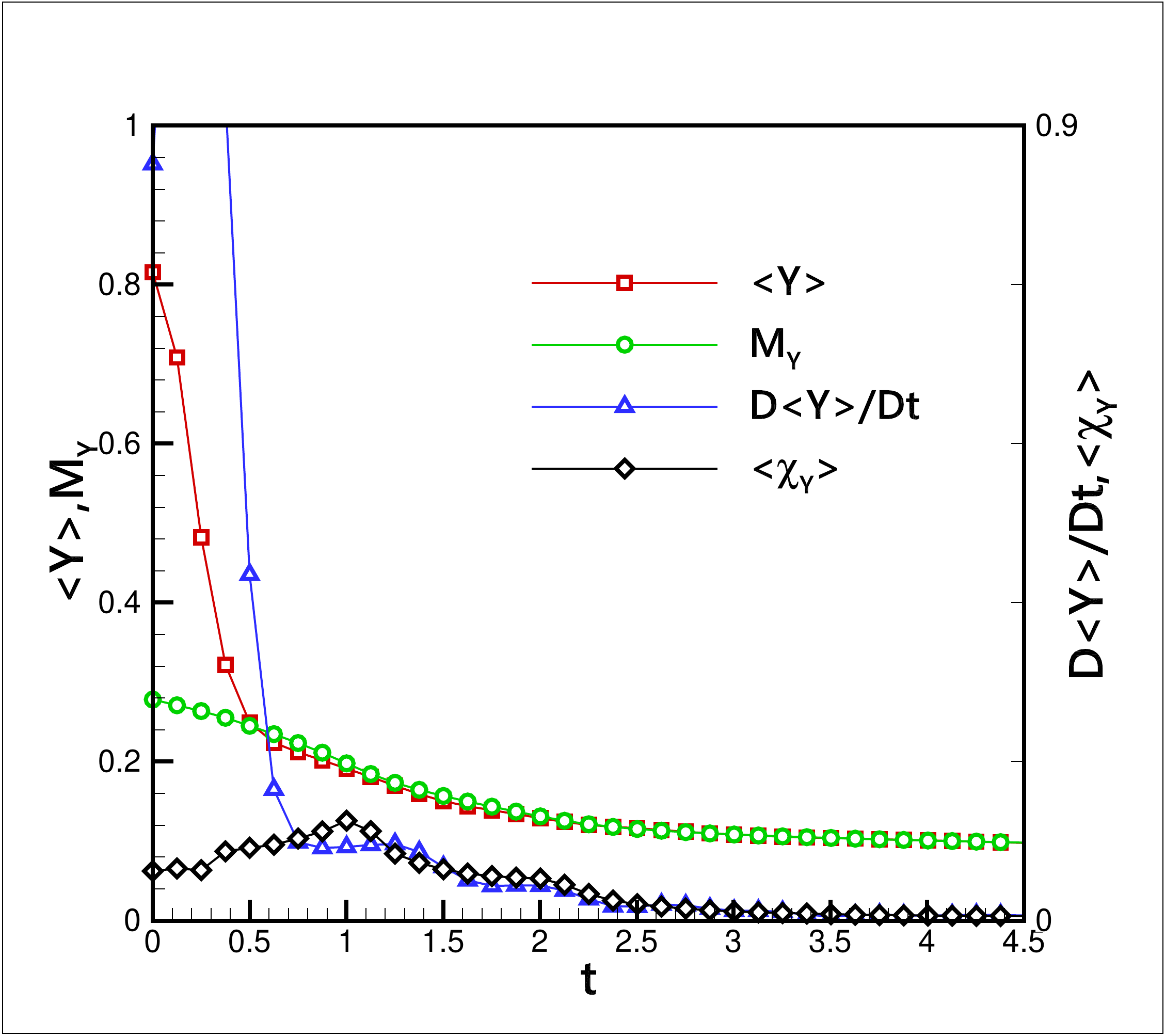}}
    \subfigure[]{
    \label{app5-line4} 
    \includegraphics[clip=true,trim=25 20 25 60,width=.32\textwidth]{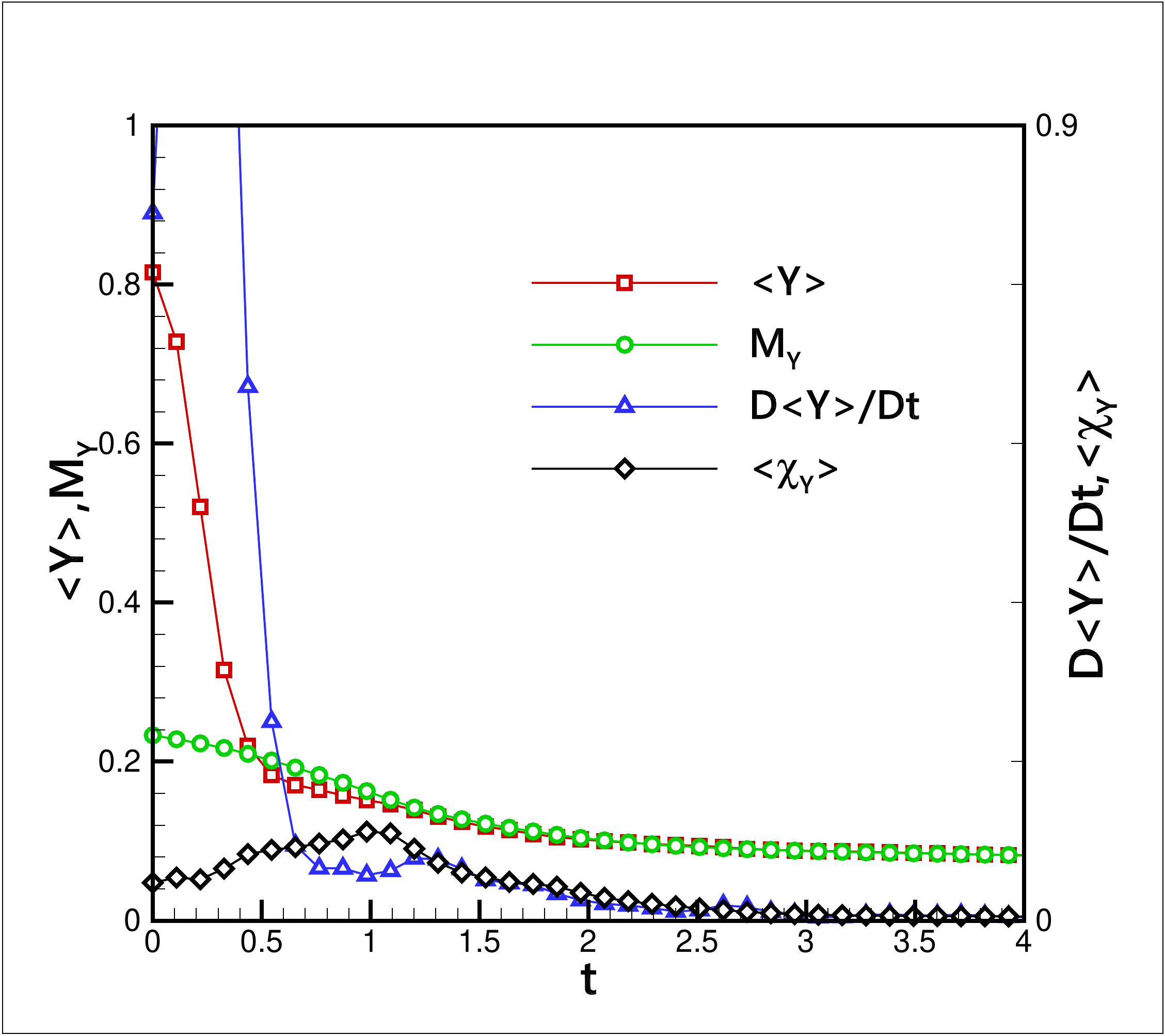}}
    \caption{Mixing rate of $\left<Y\right>$ and its decay validating Eq.~(\ref{eq: dYdt-SDR-com}) for different Ma number in cylindrical cases. (a) Ma~=~2.4 (Re~=~38000, Pe~=~6400); (b) Ma~=~3; (c) Ma~=~4. \label{app5-line} }
\end{figure}

Here, we compare the divergence of velocity and $\mathscr{D}\nabla\rho/\rho$ of different shock Mach numbers, as shown in Fig.~\ref{app5-dots}. It can be found that there exists a weak linear dependence between these two terms in Ma=4. To gain the reason for this dissimilarity, we further sort the quantitative comparison between these two terms, as shown in Fig.~\ref{app5-contour}. Pressure contour and density contour illustrate lots of shock structures in the Ma=4 case, including the shocklets in the shear layer, reflected shock, Mach stem, etc. These wave structures change the density distribution and add the source term of divergence of velocity that makes the deviation of prediction shown in Eq.~(\ref{eq: div-rho}).
However, we further show that this deviation will not largely change the mixing rate's magnitude, as shown in Fig.~\ref{app5-line}. The mean mass fraction decay of four higher Ma number cases are shown. It can be found that the derivation of $\mathrm{D}\left<Y\right>/\mathrm{D}t$ is still robust even in high shock Mach number in which only a small deviation of mean mass fraction and time integral of mixing rate. This may be explained by the fact that although the distribution of divergence of velocity in high shock Mach number is changed, the integral value is off-set for local compression and expansion co-exists in the field, as shown in Fig.~\ref{app5-contour}.

\section{Validation of $\left<f\right>$ and $\mathcal{M}$ in typical cases}
\label{sec:app6}
\begin{figure}
    \centering
    \subfigure[]{
    \label{app6-line1} 
    \includegraphics[clip=true,trim=25 20 25 60,width=.32\textwidth]{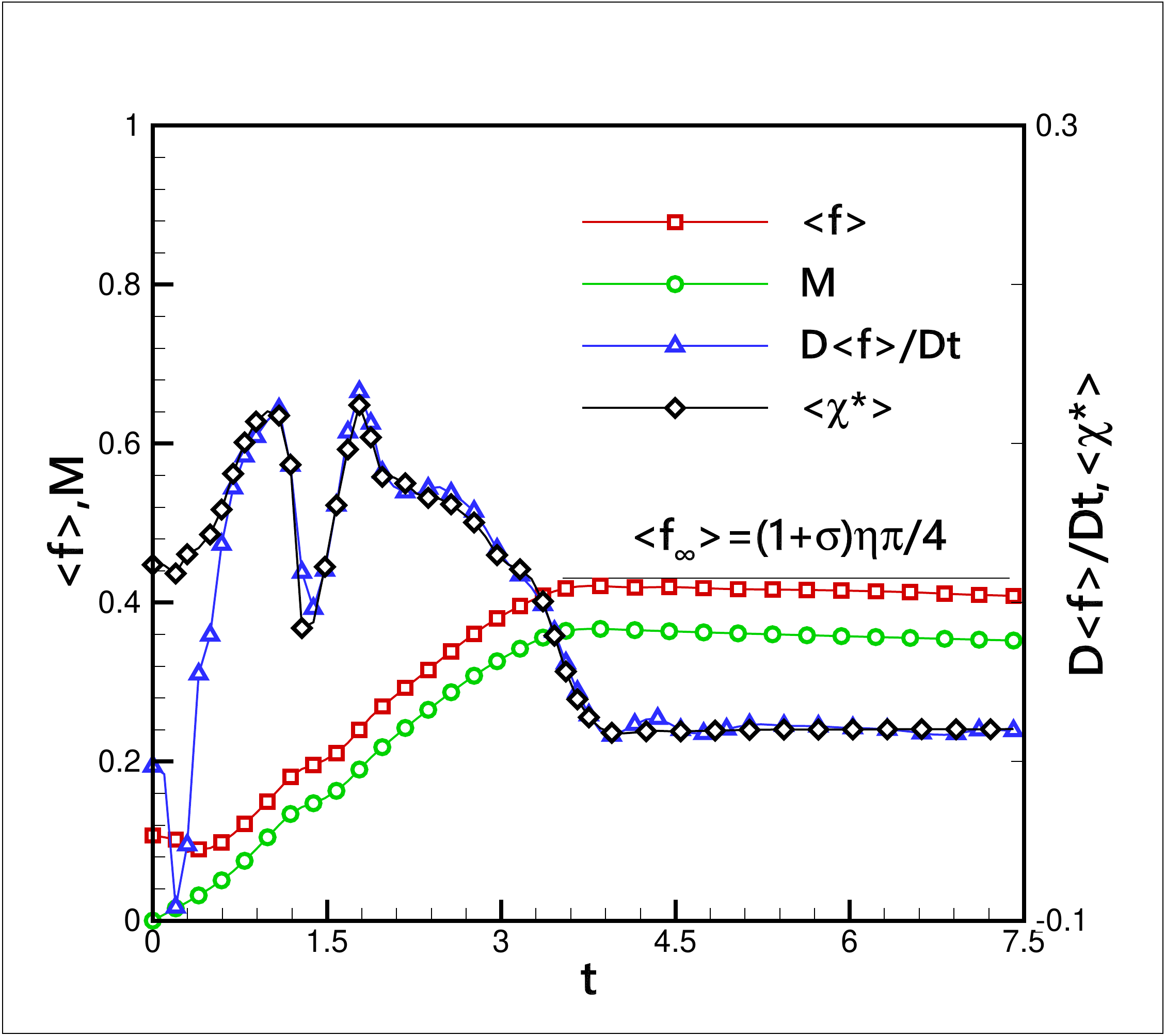}}
    \subfigure[]{
    \label{app6-line2} 
    \includegraphics[clip=true,trim=25 20 25 60,width=.32\textwidth]{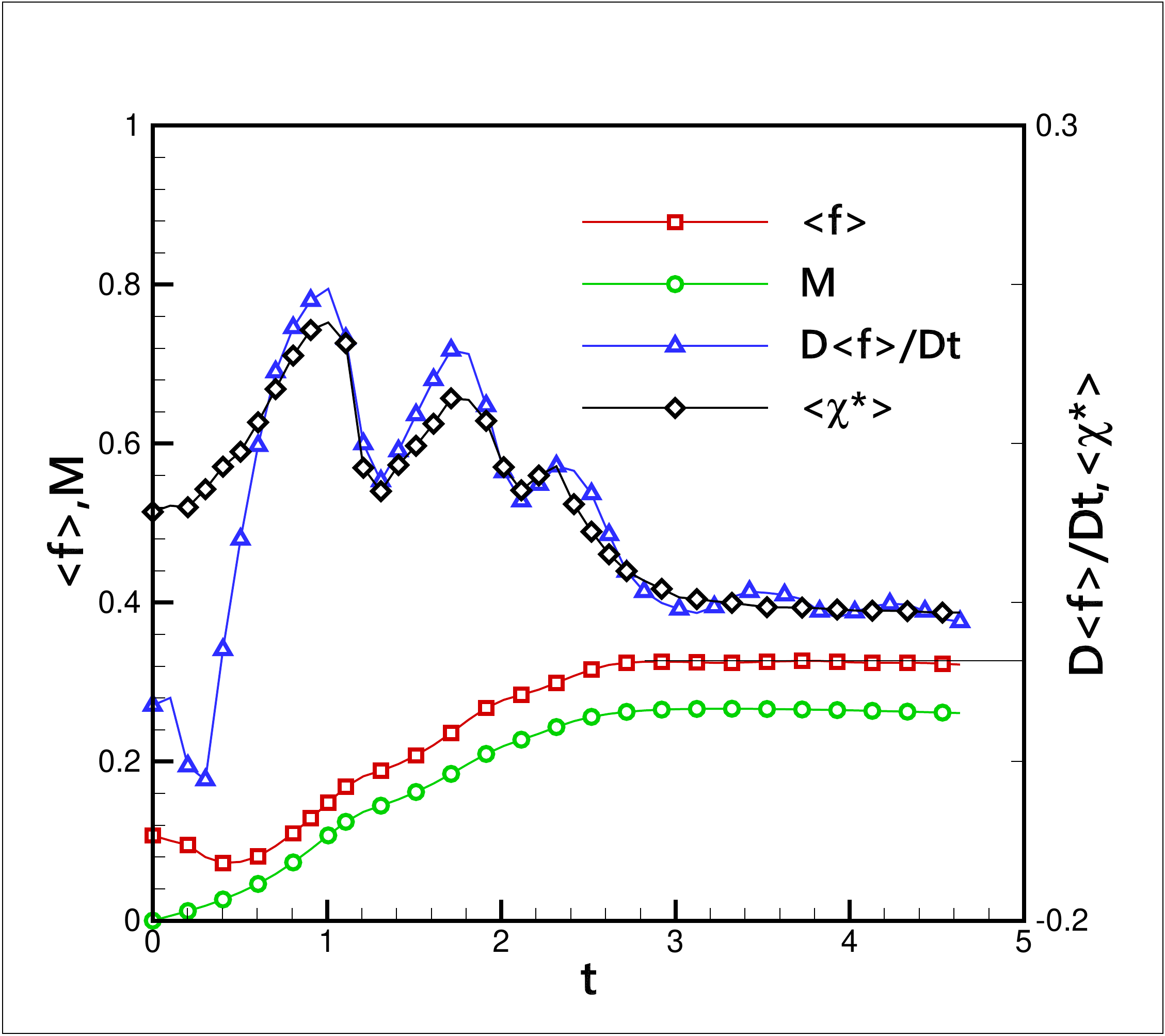}}
    \subfigure[]{
    \label{app6-line3} 
    \includegraphics[clip=true,trim=25 20 25 60,width=.32\textwidth]{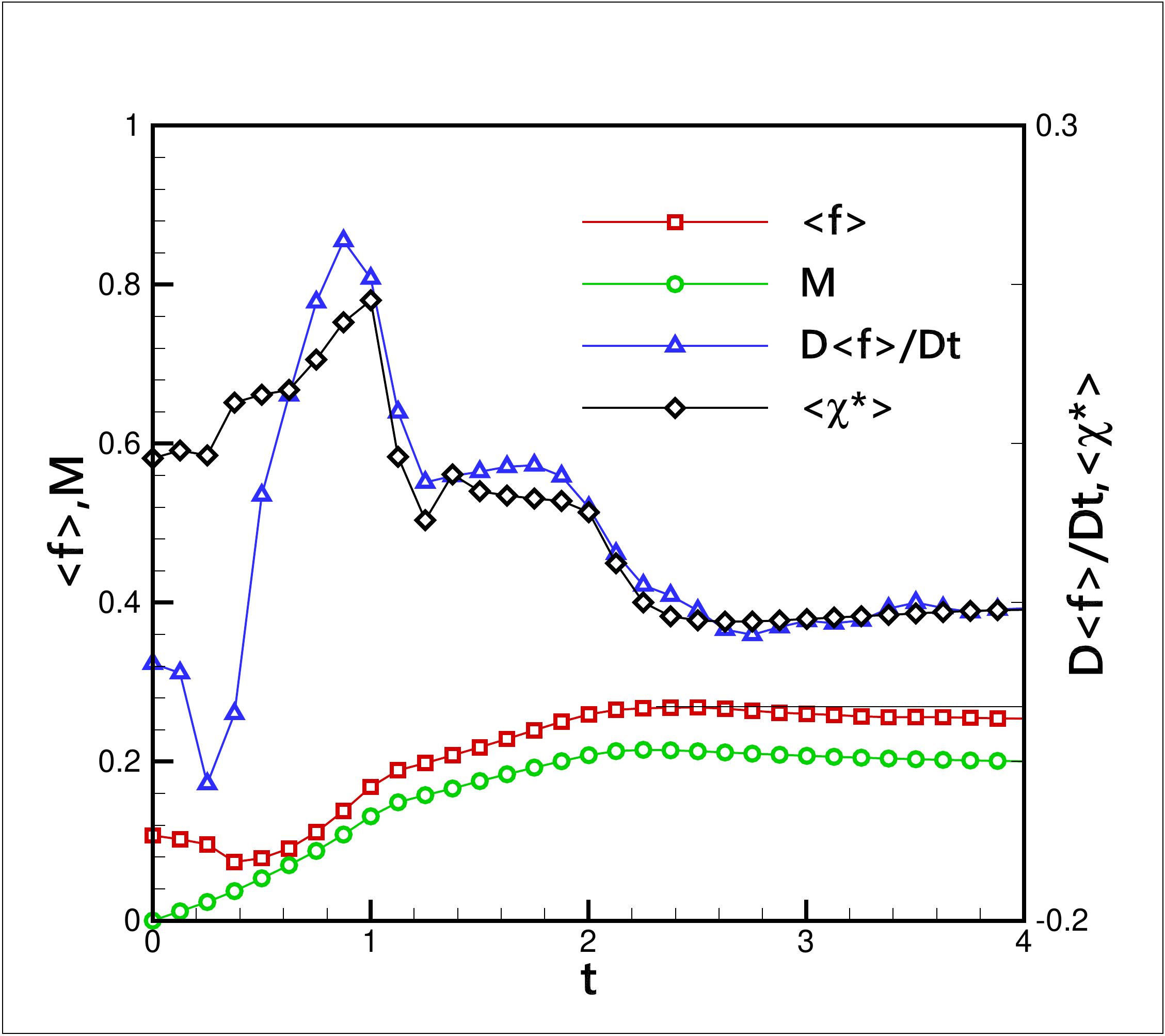}}
    \subfigure[]{
    \label{app6-line4} 
    \includegraphics[clip=true,trim=25 20 25 60,width=.32\textwidth]{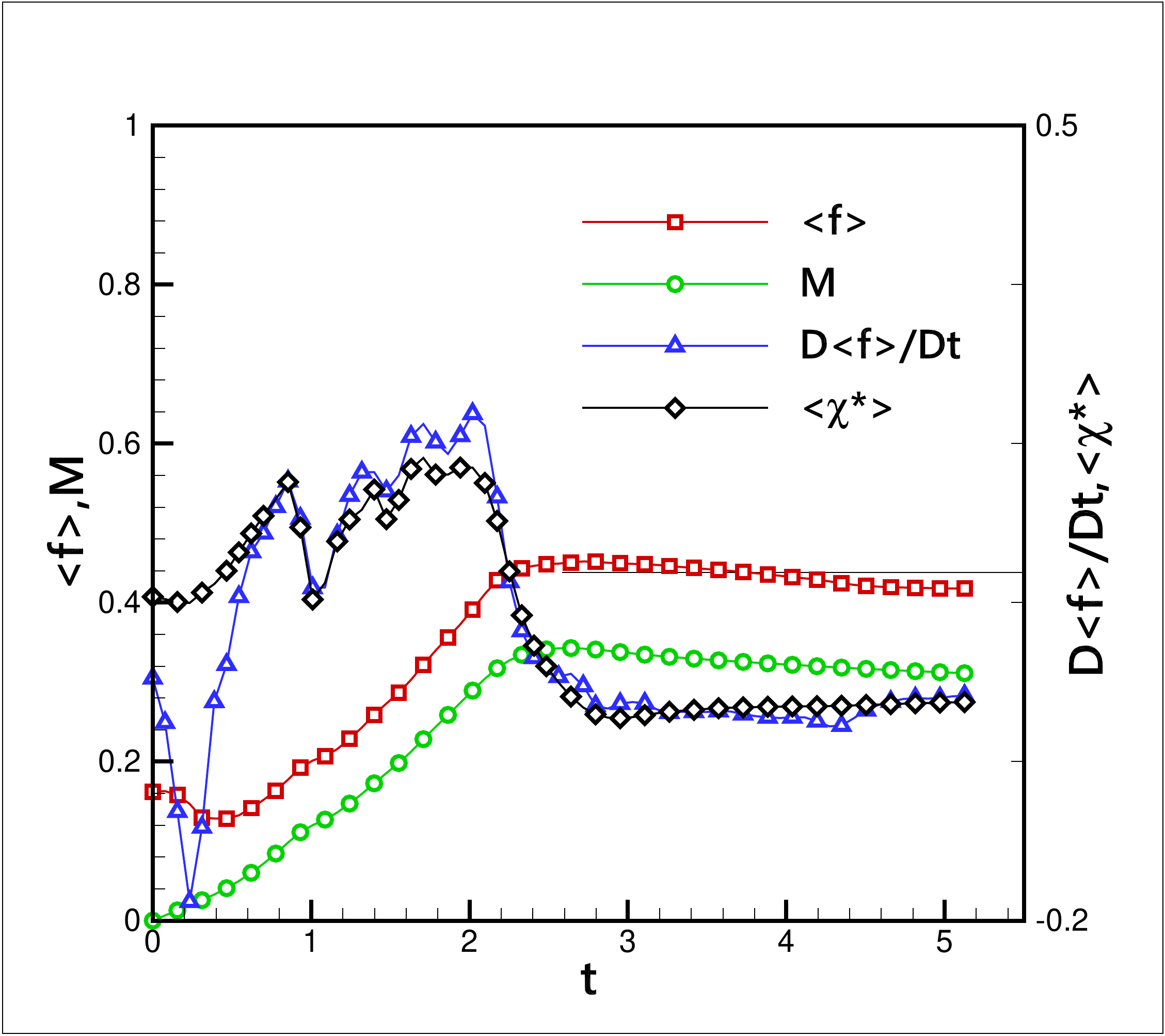}}
    \subfigure[]{
    \label{app6-line5} 
    \includegraphics[clip=true,trim=25 20 25 60,width=.32\textwidth]{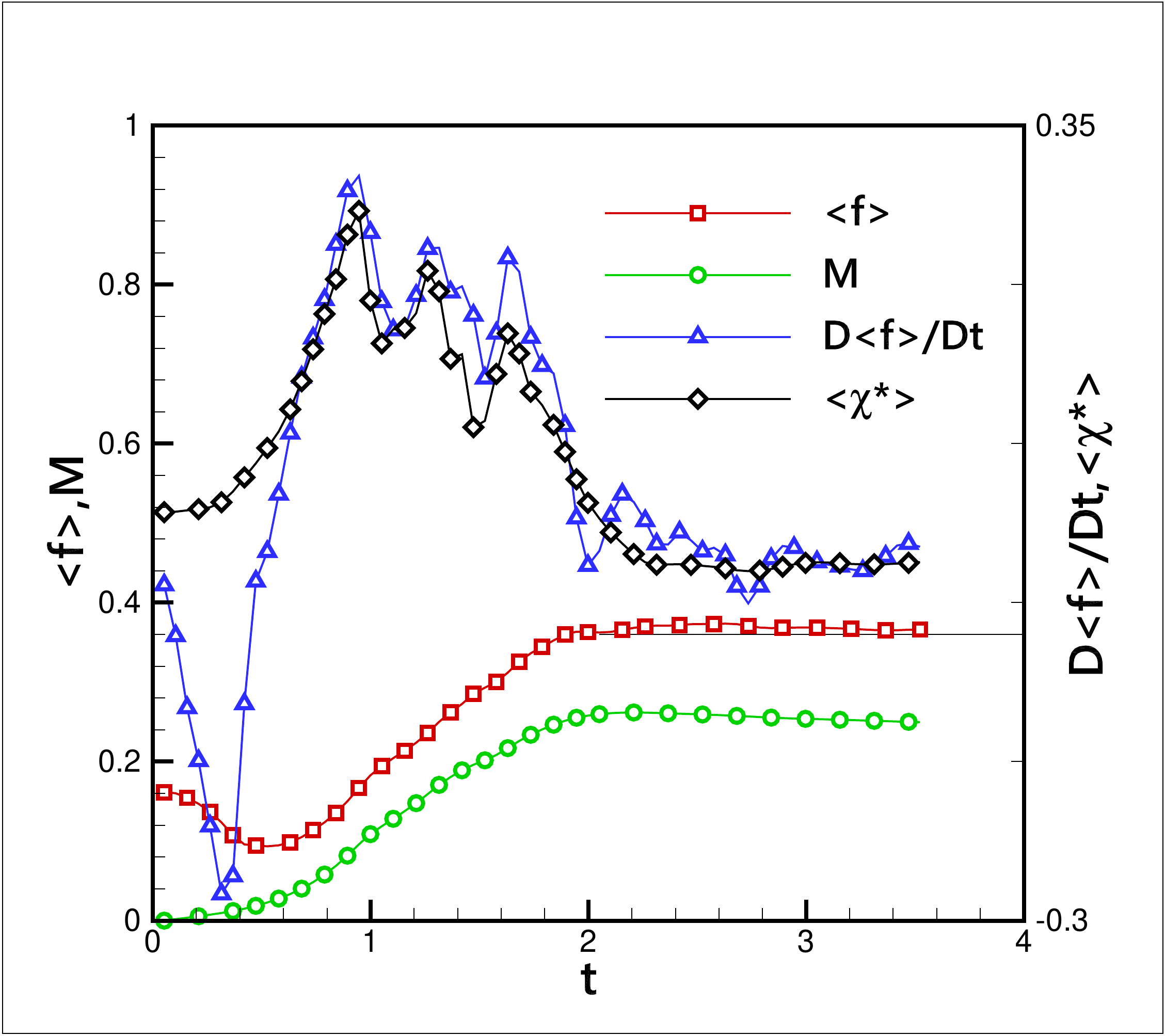}}
    \subfigure[]{
    \label{app6-line6} 
    \includegraphics[clip=true,trim=25 20 25 60,width=.32\textwidth]{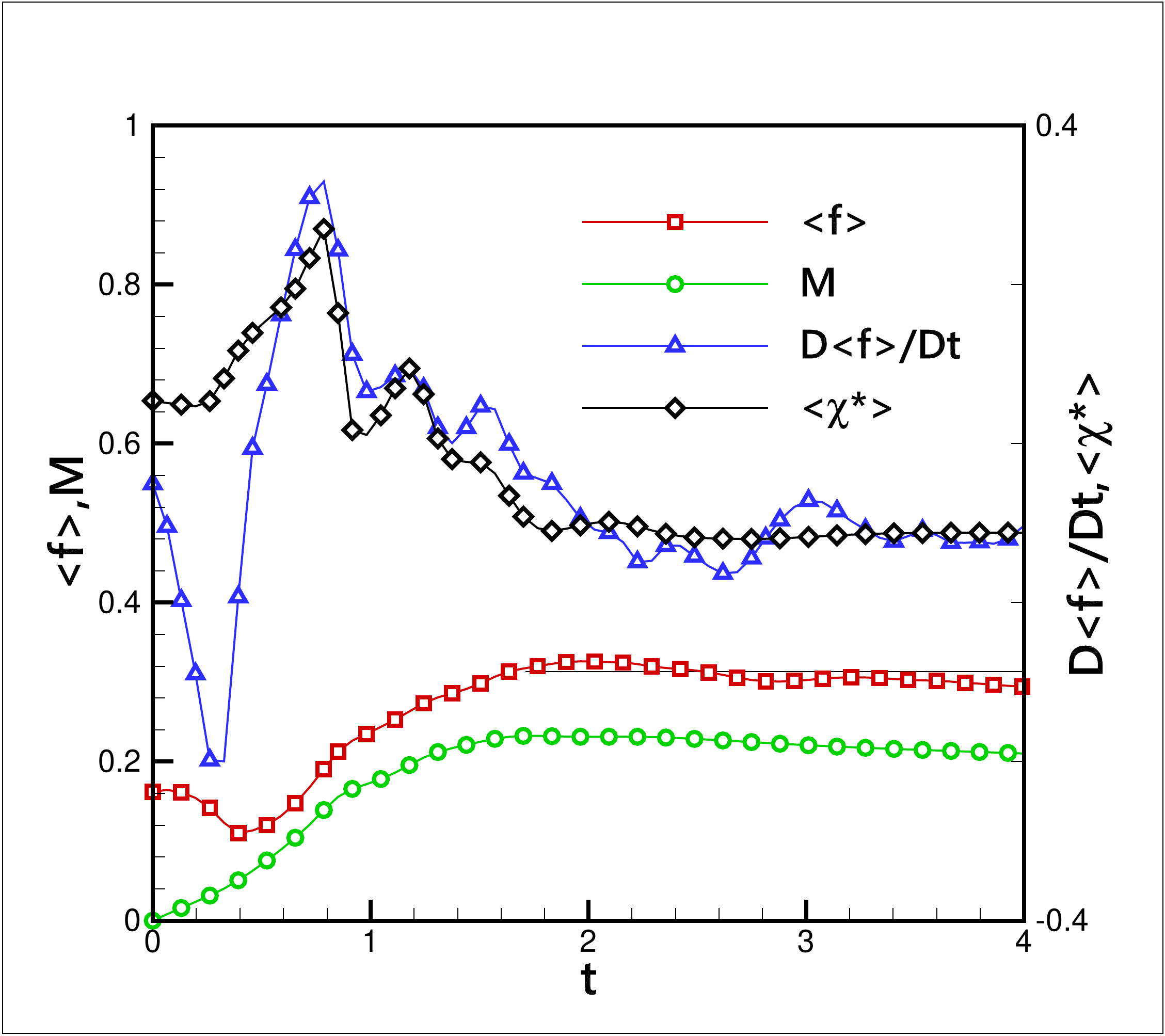}}
    \subfigure[]{
    \label{app6-line7} 
    \includegraphics[clip=true,trim=25 20 25 60,width=.32\textwidth]{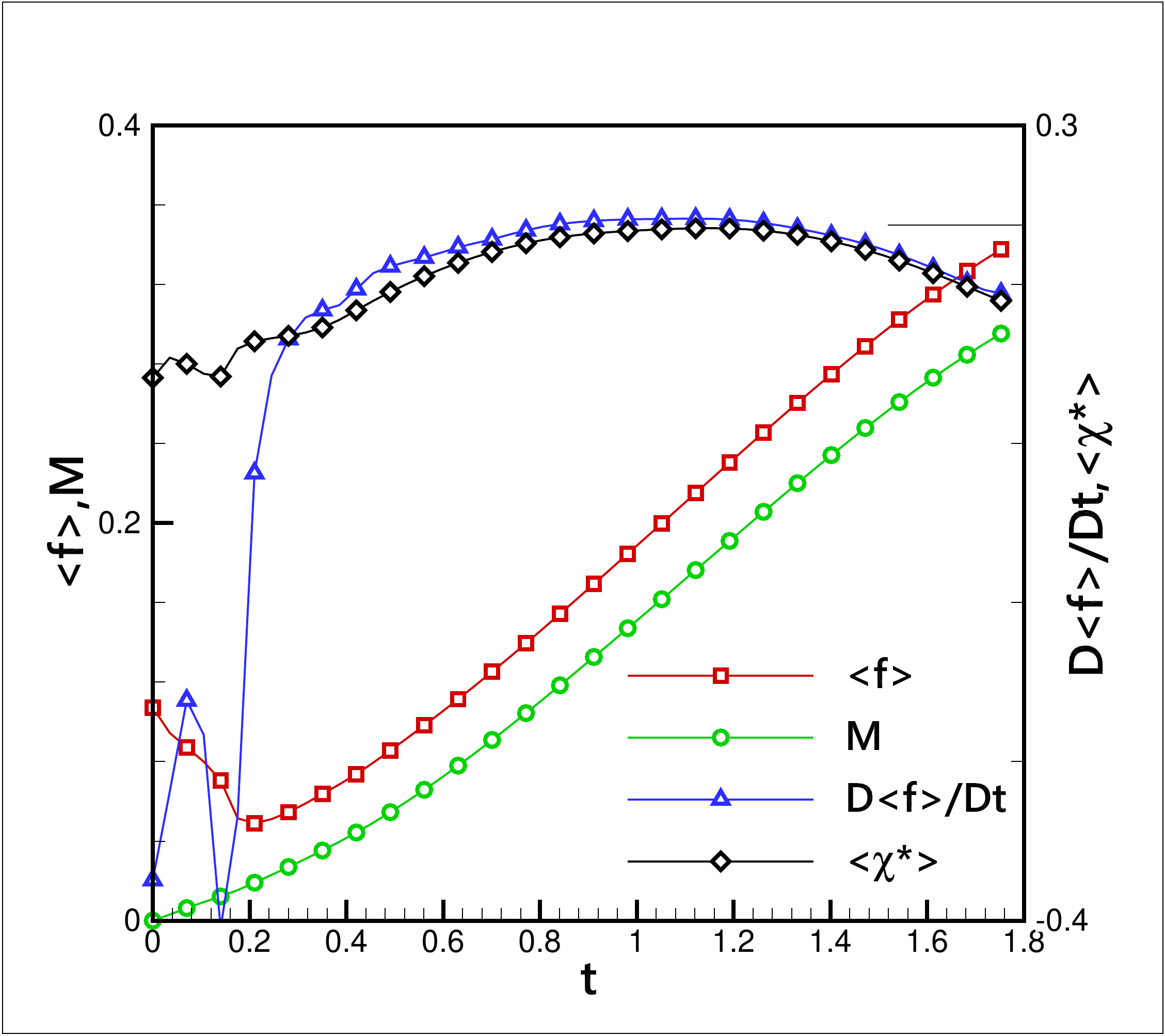}}
    \subfigure[]{
    \label{app6-line8} 
    \includegraphics[clip=true,trim=25 20 25 60,width=.32\textwidth]{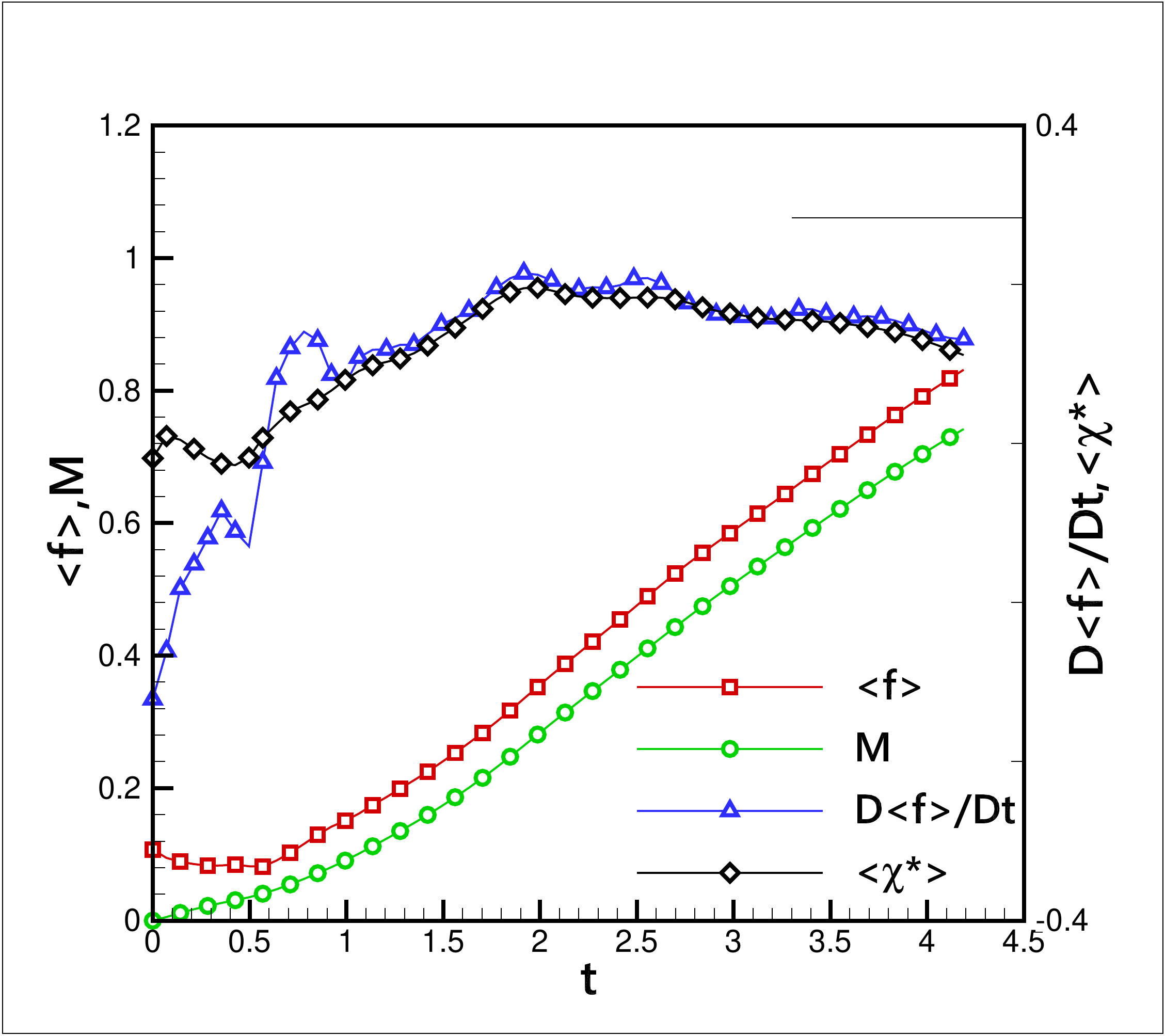}}
    \subfigure[]{
    \label{app6-line9} 
    \includegraphics[clip=true,trim=25 20 25 60,width=.32\textwidth]{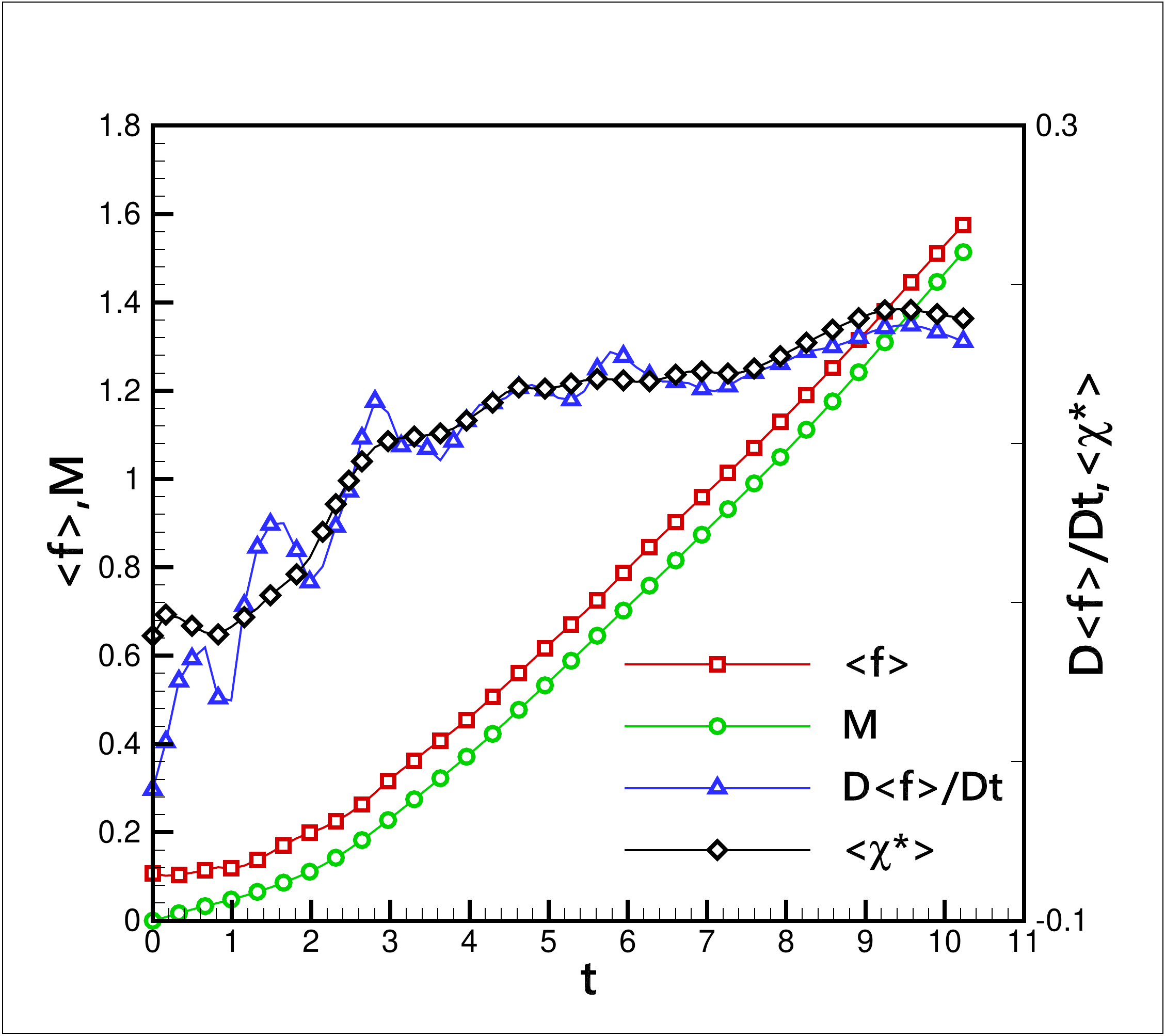}}
    \caption{Time evolution of mixedness and its time derivative comparing with $\left<\chi^*\right>$ in typical cases.
    (a-c): Cylindrical bubble containing helium with Ma=1.8 (Re=20700); Ma=2.4 (Re=38000, Pe=6400); Ma=3.
    (d-f): Spherical bubble containing helium with Ma=1.8 (Re=6900, Pe=6400); Ma=2.4 (Re=39600, Pe=15300); Ma=3.
    (g-i): Cylindrical bubble containing variable density components with At=$-$0.21 (Ma=2.4, CH$_4$); At=0.33 (Ma=2.4, Kr); At=0.69 (Ma=1.22, SF$_6$).
     \label{app6-line} }
\end{figure}
In this section, we validate Eq.~(\ref{eq: dfdt-SDR}) for typical cases concerned. As shown in Fig.~\ref{app6-line}, the time history of mixedness and its time derivative comparing with density gradient accelerated mixing rate $\left<\chi^*\right>$ are plotted. The general agreement can be found in all cases, even for high shock Mach numbers. Two-stage mixing is shown in the cylindrical and spherical bubbles containing helium.
From Eq.~(\ref{eq: meanmass}), the asymptotic limit of mixedness can be estimated as $\left<f_\infty\right>=4\left(\left<Y_\infty\right>-\left<Y^2_\infty\right>\right)=(1+\sigma)\eta\pi/4$, marked by a solid line. The model predicts well in all helium bubble cases.
As introduced in Sec.~\ref{subsec:scaling3}, the mixing in variable-density cases continues in accordance to the linear growth of mixedness, as shown in Figs.~\ref{app6-line}(g-i).
For time integral of mixing rate $\mathcal{M}(t)$, it shows a similar trend as mixedness and is only different in the start point of due the initial diffusion layer that makes the non-zero of the initial value of mixedness $\left<f_0\right>$, which has been considered in Sec.~\ref{subsec:scaling2}.

\section{An effective proof of $\mathrm{D}\left<\mathcal{X}\right>/\mathrm{D}t\approx0$ after shock impact}
\label{sec:app7}
\begin{figure}
  \centering
  \includegraphics[clip=true,trim=20 20 15 25,width=.65\textwidth]{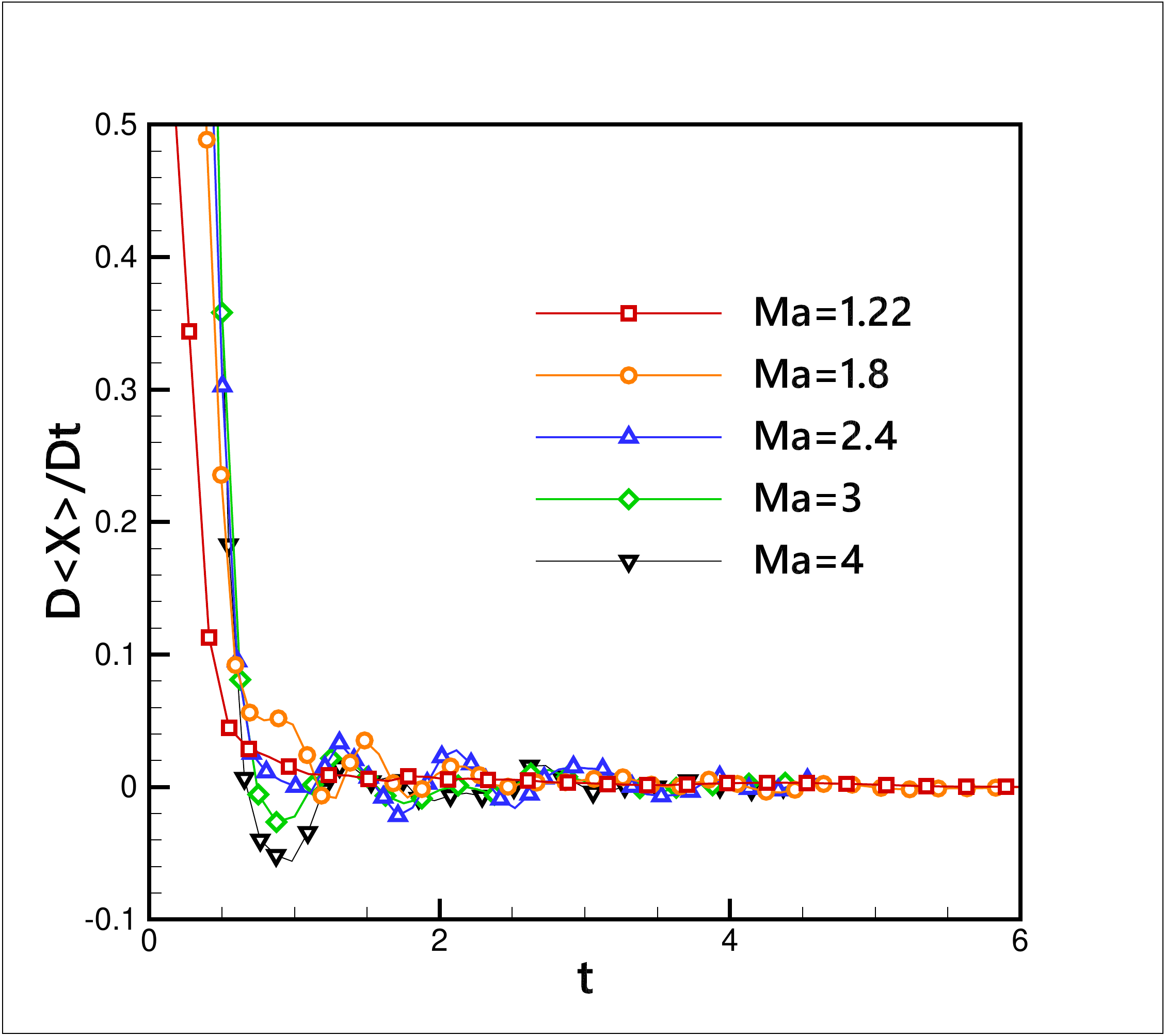}\\
  \caption{Time derivative of the mean volume fraction tends to be zero immediately after the shock passage for cylindrical bubble cases. }\label{App7-DXDt}
\end{figure}
In the compressible flows, it always uses a normalized mole fraction \cite{ruan2020density} to define the volume fraction, which is different from the definition in incompressible variable-density flows as $\mathcal{X}=(\rho-\rho_1)/(\rho_2-\rho_1)$ \cite{linden1994molecular}.
Moreover, in combustion flows, the mole fraction is also used to calculate the reaction rate \cite{diegelmann2016pressure}.
Therefore, it is crucial to understand the variation of mole/volume fraction, not only the mass fraction. From the profile of compression rate as shown in Fig.~\ref{app3-com}, the mean mole fraction $\left<\mathcal{X}\right>$ seems to be conservative contrary to the decay of mean mass fraction $\left<Y\right>$, which makes compression rate a robust controlling parameter for mixing. Figure~\ref{App7-DXDt} plots the time derivative of the mean mole fraction $\mathrm{D}\left<\mathcal{X}\right>/\mathrm{D}t$. While oscillation is found due to the reflected shock (especially in high shock Ma number cases), the values are near zero for all cases at the medium-late time.

\begin{figure}
    \centering
    \subfigure[]{
    \label{app7-1} 
    \includegraphics[clip=true,trim=0 0 0 0,width=.31\textwidth]{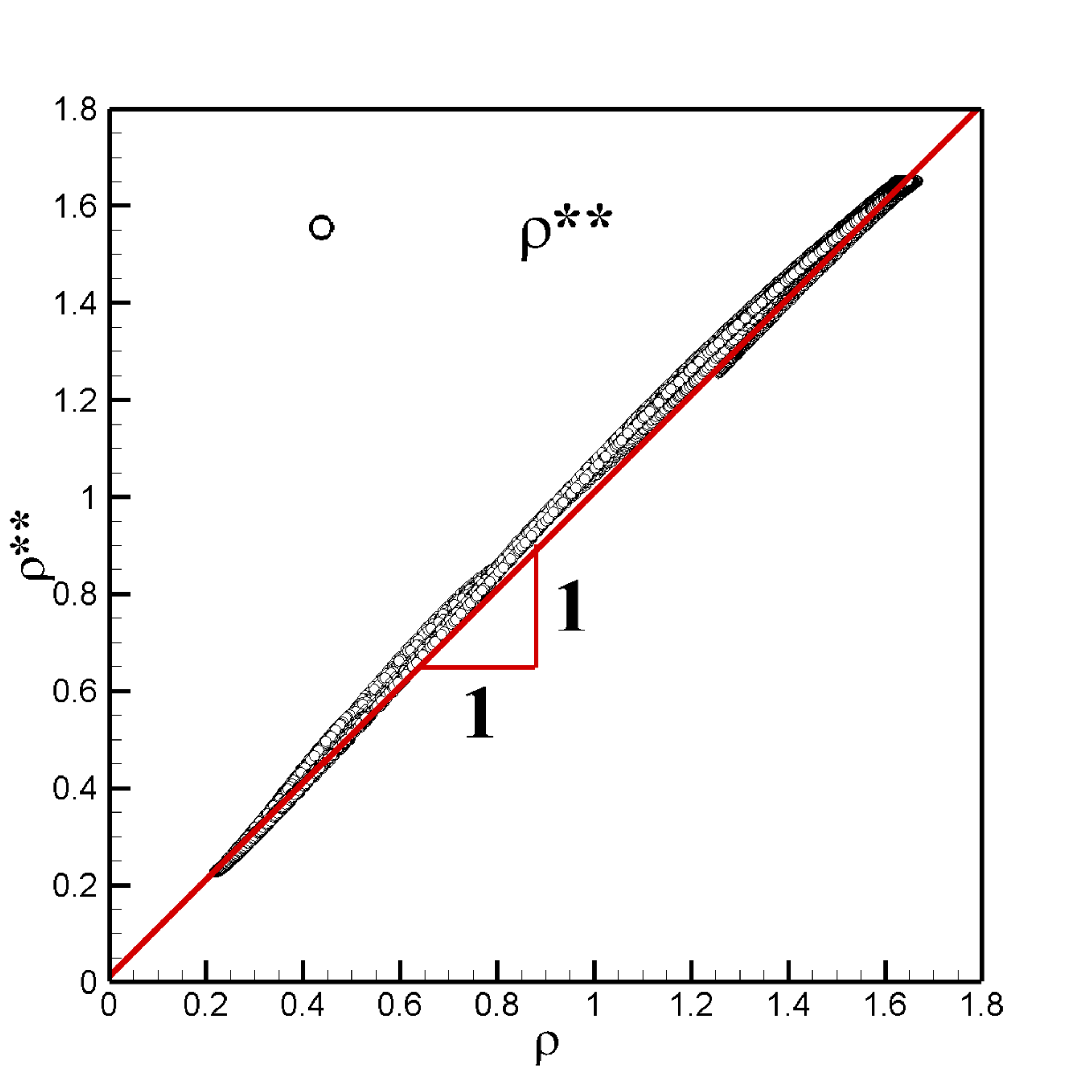}}
    \subfigure[]{
    \label{app7-2} 
    \includegraphics[clip=true,trim=0 0 0 0,width=.31\textwidth]{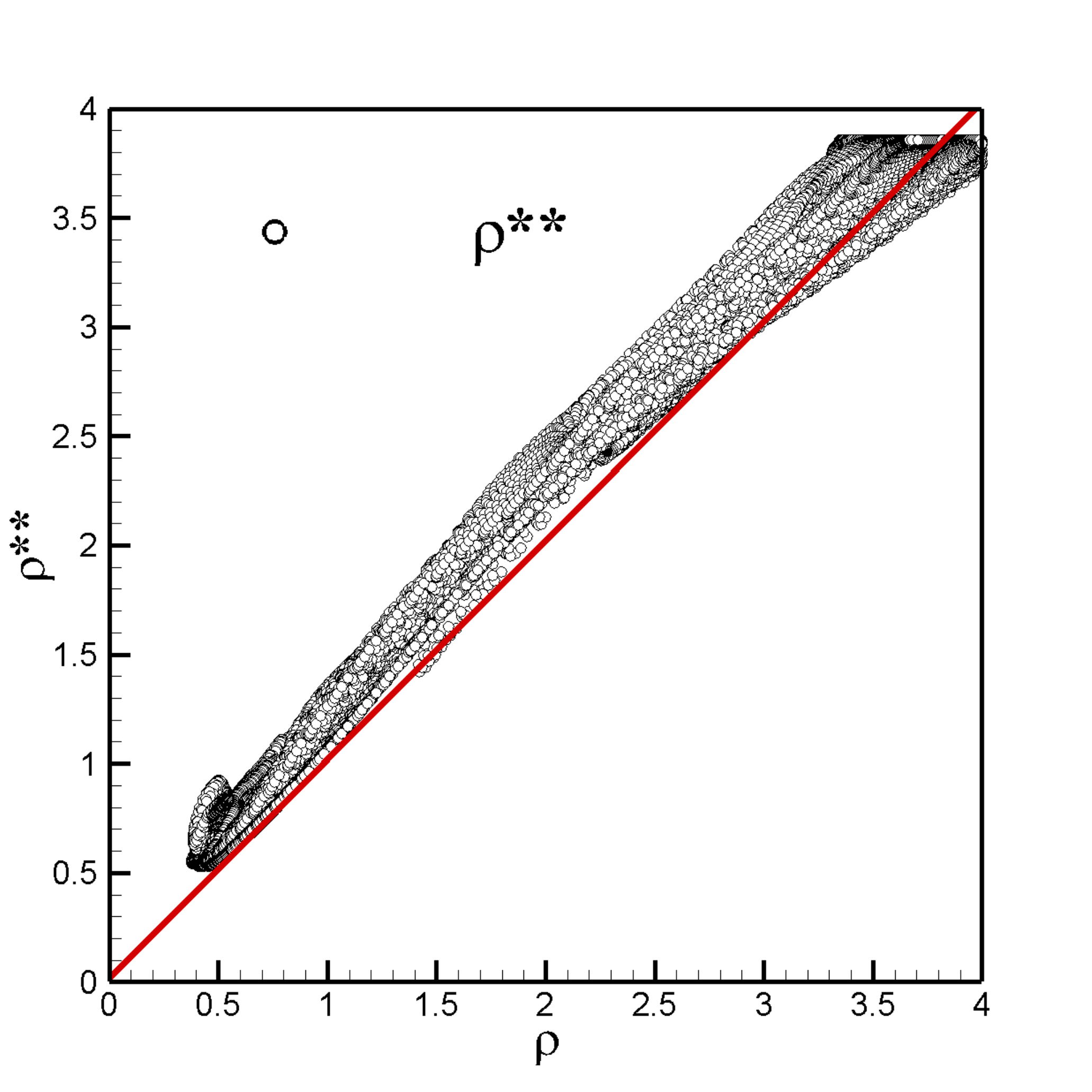}}
    \subfigure[]{
    \label{app7-3} 
    \includegraphics[clip=true,trim=0 0 0 0,width=.31\textwidth]{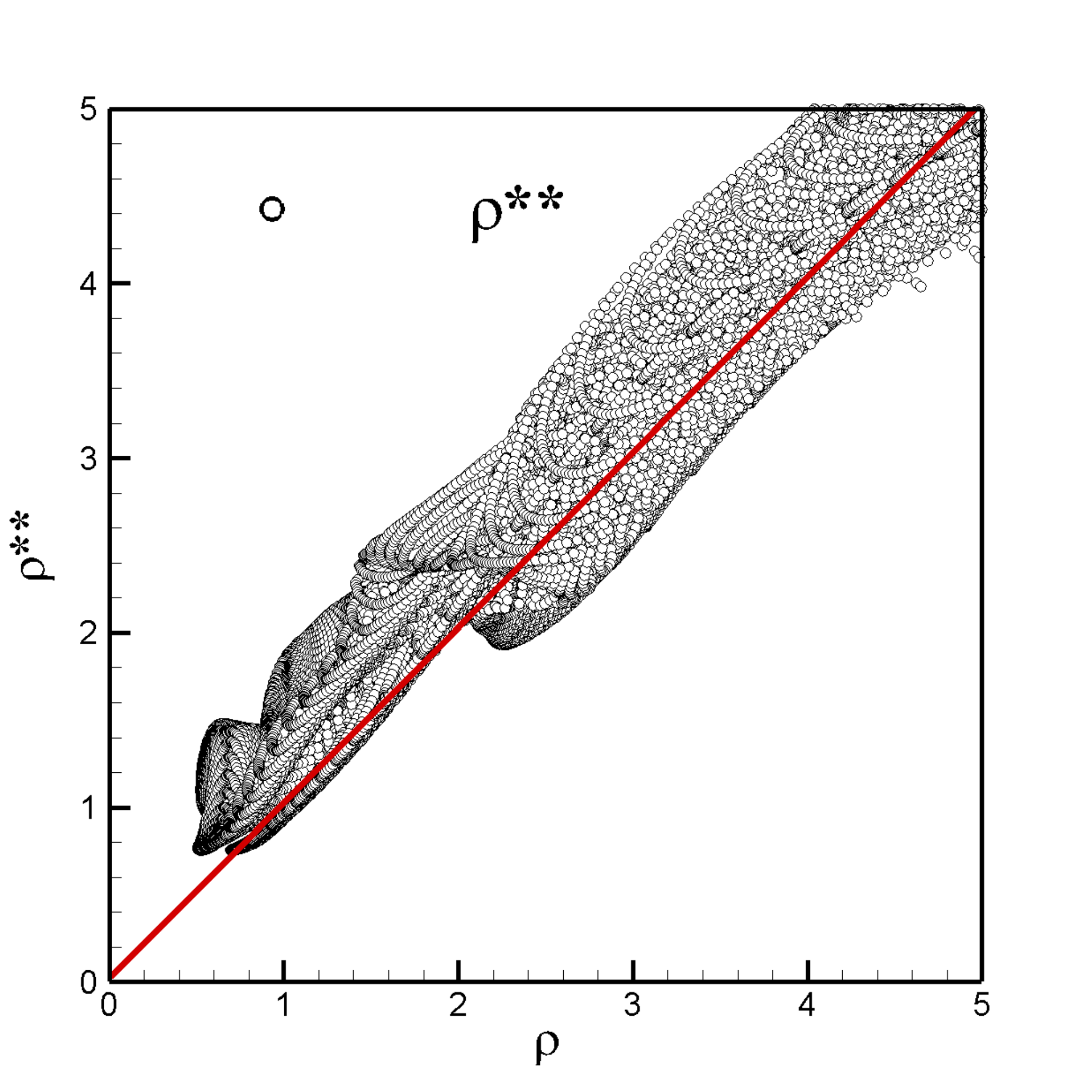}}
    \caption{Comparison of nominated density $\rho^{**}$ and exact real density $\rho$ at different Mach number for cylindrical helium bubble cases.
    (a) Ma = 1.22 (Re = 5750, Pe = 5500, $t = 2.74$); (b) Ma = 2.4 (Re = 38000, Pe = 6400, $t = 1.71$); (c) Ma = 4 ($t = 1.42$).  \label{app7-dots} }
\end{figure}

Although it can be assumed that the volume of the bubble is conservative after shock passage from a physical standpoint, it is rather complicated to prove the conservation of bubble volume from the mathematical point in a rigorous way. Here an effective proof is provided by neglecting the impact of first shock compression. Again by using Eq.~(\ref{eq: derivative}) for mole fraction, we get:
\begin{equation}\label{eq: dXdt}
  \frac{\mathrm{D}\left<\mathcal{X}\right>}{\mathrm{D}t}=\left<\frac{\mathrm{D}\mathcal{X}}{\mathrm{D}t}\right>
               +\left<\mathcal{X}\left(\nabla\cdot\textbf{V}\right)\right>.
\end{equation}
From the definition of the mole fraction in the form of mass fraction, we get:
\begin{equation}\label{eq: Dxdt-1}
  \mathcal{X}=\frac{Y}{(1-\alpha)Y+\alpha}\Rightarrow
  \frac{\mathrm{D}\mathcal{X}}{\mathrm{D}t}=\frac{\alpha}{\left((1-\alpha)Y+\alpha\right)^2}\frac{\mathrm{D}Y}{\mathrm{D}t},
\end{equation}
where $\alpha=M_2/M_1=\rho_2/\rho_1$ of preshock conditions.
Then Eq.~(\ref{eq: dXdt}) can be rewritten as:
\begin{equation}\label{eq: dXdt-2}
  \frac{\mathrm{D}\left<\mathcal{X}\right>}{\mathrm{D}t}=\left<\frac{1}{(1-\alpha)Y+\alpha}\underbrace{\left(
           \frac{\alpha}{(1-\alpha)Y+\alpha}\frac{\mathrm{D}Y}{\mathrm{D}t}+Y\left(\nabla\cdot\textbf{V}\right) \right)}_{\mathscr{A}}   \right>.
\end{equation}
Note that:
\begin{equation}\label{eq: volfrac}
  \frac{\alpha}{(1-\alpha)Y+\alpha}=1-(1-\alpha)\mathcal{X}.
\end{equation}
By using Eq.~(\ref{eq: dYdt-drhodt}) and conservation equation of mass Eq.~(\ref{eq: mass conser}), then $\mathscr{A}$ can be expressed as in the form of density $\rho$ and mole fraction $\mathcal{X}$:
\begin{equation}\label{eq: math-A}
  \mathscr{A}=-\frac{1}{(1/\sigma-1)\rho^2}\frac{\mathrm{D}\rho}{\mathrm{D}t}\left(
     \rho'_1(1-\mathcal{X})+\alpha\rho'_1\mathcal{X}-\rho                \right)\approx 0.
\end{equation}
Here a new alternative density $\rho^{**}=\rho'_1(1-\mathcal{X})+\alpha\rho'_1\mathcal{X}$ can be defined. It can easily deduce that when $\alpha\approx\sigma$, $\rho^{**}\approx\rho$ that makes the conservative of mole fraction. The validation of $\rho^{**}\sim\rho$ relationship is shown by the scatter points in Fig.~\ref{app7-dots}. Similar to the relationship of $\rho^*\sim\rho$, a higher Mach number makes the broader width of the scatter points, while the linear relation is also evident. That explains the near conservative behavior of mole fraction $\left<\mathcal{X}\right>$, which is also believed to exist in spherical cases and variable-density cases.
If the conservative behavior is solid, it means the mole fraction shows a similar characteristic as passive scalar mixing, which is another story not covered by this paper while it is worthy in the future study.

\section{Some discussions on spherical bubble cases}
\label{sec:app8}
\begin{figure}
  \centering
  \includegraphics[clip=true,trim=0 250 0 0,width=.99\textwidth]{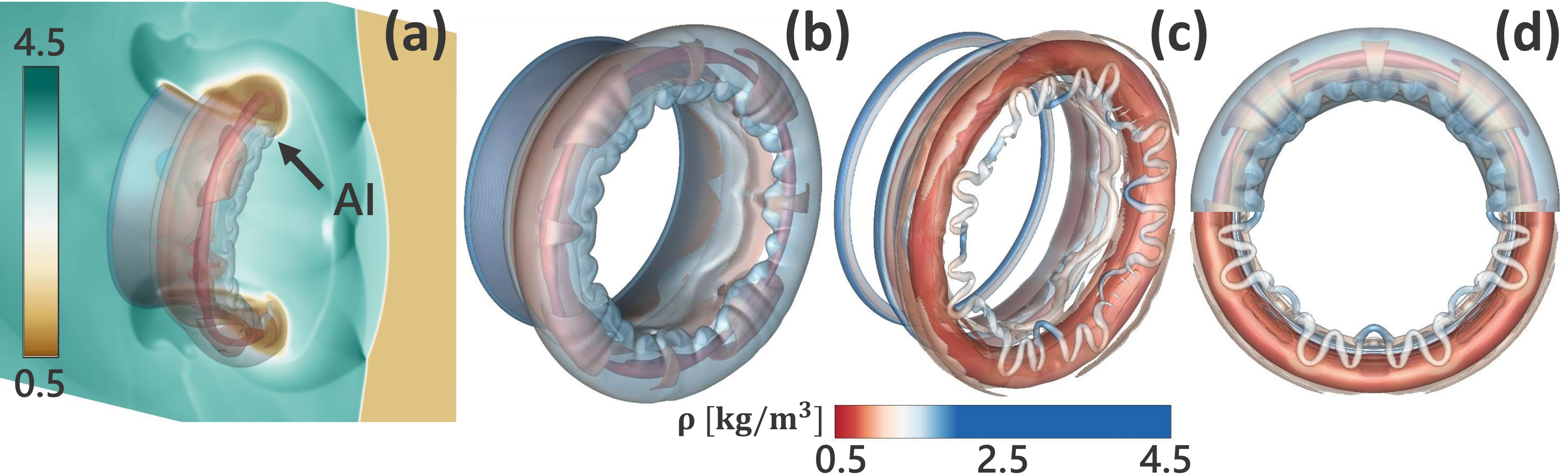}\\
  \caption{Flow structures of 3D spherical bubble cases at Ma $=2.4$ with Re = 39600 and Pe = 6700 at $t = 1.79$. (a) Iso-contours of the mass fraction at $Y$ = 0.2 and 0.6 with its center slice of density contour; (b) amplification of mass fraction iso-contours colored by density magnitude; (c) iso-contour of $Q$ criterion colored by density magnitude; (d) front view of iso-contour of density (upper half) and $Q$ criterion (bottom half). AI: azimuthal instability.}\label{App8-bub-contour}
\end{figure}
The two-dimensional axisymmetric bubble is compared and validated by a full three-dimensional bubble simulation to show the 3D effect on mixing.
Characteristic instantaneous flow structures of 3D results are depicted in Fig.~\ref{App8-bub-contour}. From the iso-contours of mass fraction in Fig.~\ref{App8-bub-contour}(b), azimuthal instability occurs ahead of the main supersonic vortex ring. Further examining the iso-contour of $Q$ criterion \cite{jeong1995identification}, the vortex ring and vessel-like coherent structures can be extracted in Fig.~\ref{App8-bub-contour}(c). This vessel-like vortex structure is formed from secondary baroclinic vorticity, as analyzed later. The comparison of iso-contours of mass fraction and $Q$ criterion from the front view as in Fig.~\ref{App8-bub-contour}(d) shows clearly that the azimuthal instability comes from the secondary vortex structures. The axisymmetric characteristic maintains well in full 3D results in general.

\begin{figure}
  \centering
  \includegraphics[clip=true,trim=0 0 20 0,width=.8\textwidth]{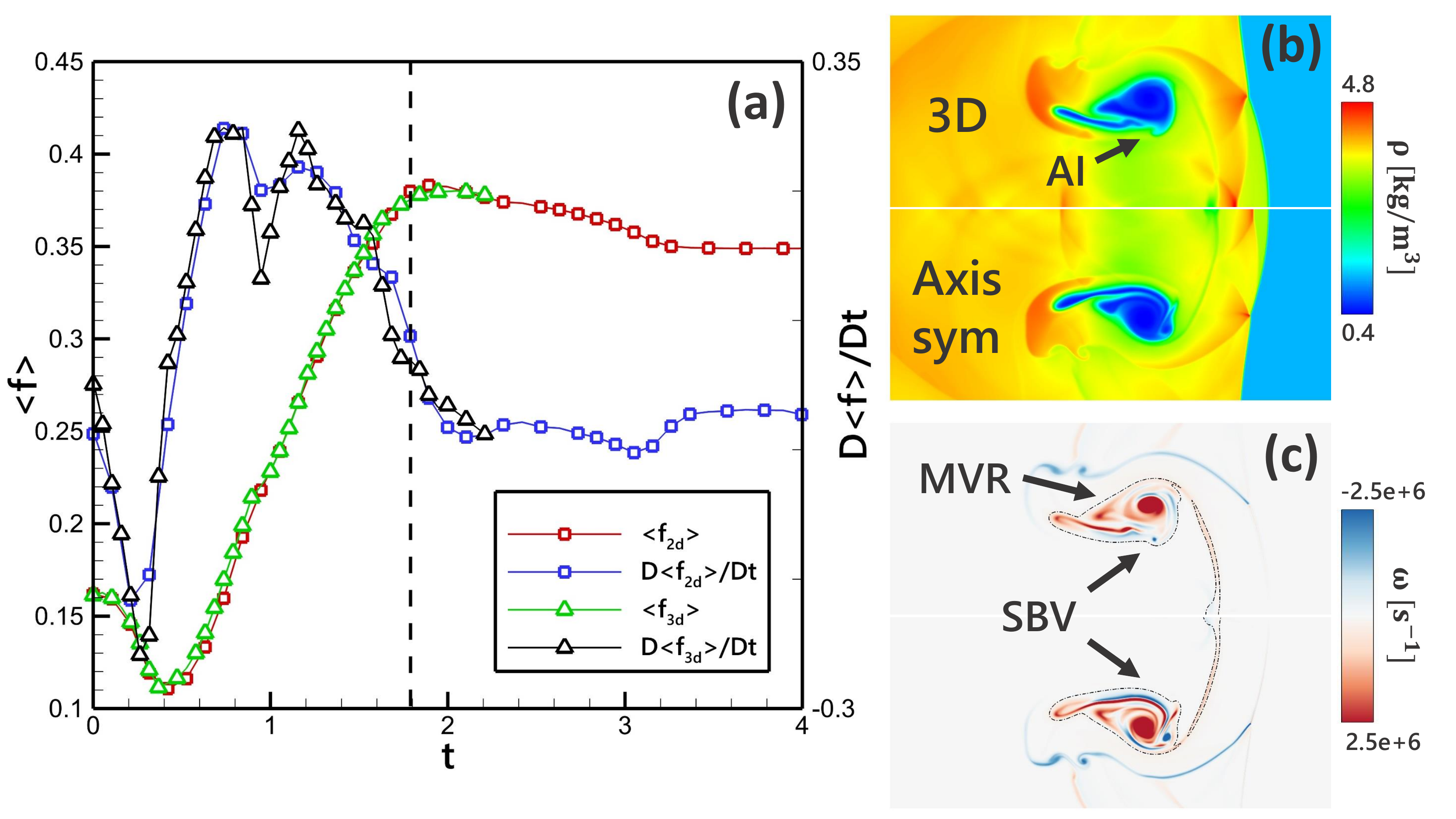}\\
  \caption{(a) Comparison of time evolution of mixedness $\left<f\right>$ and its time derivative $\mathrm{D}\left<f\right>/\mathrm{D}t$ between 3D spherical case and 2D axisymmetric case. (b) Comparison of density contour between 3D and 2D spherical case at $t = 1.79$, as indicated by the dashed line in (a). (c) Comparison of vorticity contour with isoline of $Y=0.01$ (dashed-dot line) between 3D and 2D spherical cases. AI: azimuthal instability; MVR: main vortex ring; SBV: secondary baroclinic vorticity.  }\label{App8-bub-line}
\end{figure}
A quantitative comparison of mixedness and its time derivative between the 3D spherical case and 2D axisymmetric case at the same conditions is shown in Fig.~\ref{App8-bub-line}(a). Although the azimuthal instability exists in the 3D case, the integral results conclude that the axisymmetric characteristic dominates the flow structures. The qualitative comparisons of density and vorticity contour between 3D and 2D axisymmetric results are illustrated in Figs.~\ref{App8-bub-line}(b) and (c). General consistency is obtained. The vorticity contour shows that the secondary baroclinic vorticity \cite{peng2021mechanism} is the cause of azimuthal instability ahead of the main vortex ring, which also appears in oblique shock-jet interaction \cite{yu2020two}.

In short, the scaling law revealed from the axisymmetric simulations in the present paper may support the mixing pattern in full 3D SBI. Detailed analysis on 3D effect in even higher Reynolds number deserves future study.
Besides, the axisymmetric simulation can also be validated through this comparison.


\section{Turbulence effect on mixedness and mixing rate}
\label{sec:app9}
\begin{figure}
  \centering
  \includegraphics[clip=true,trim=0 200 0 0,width=.9\textwidth]{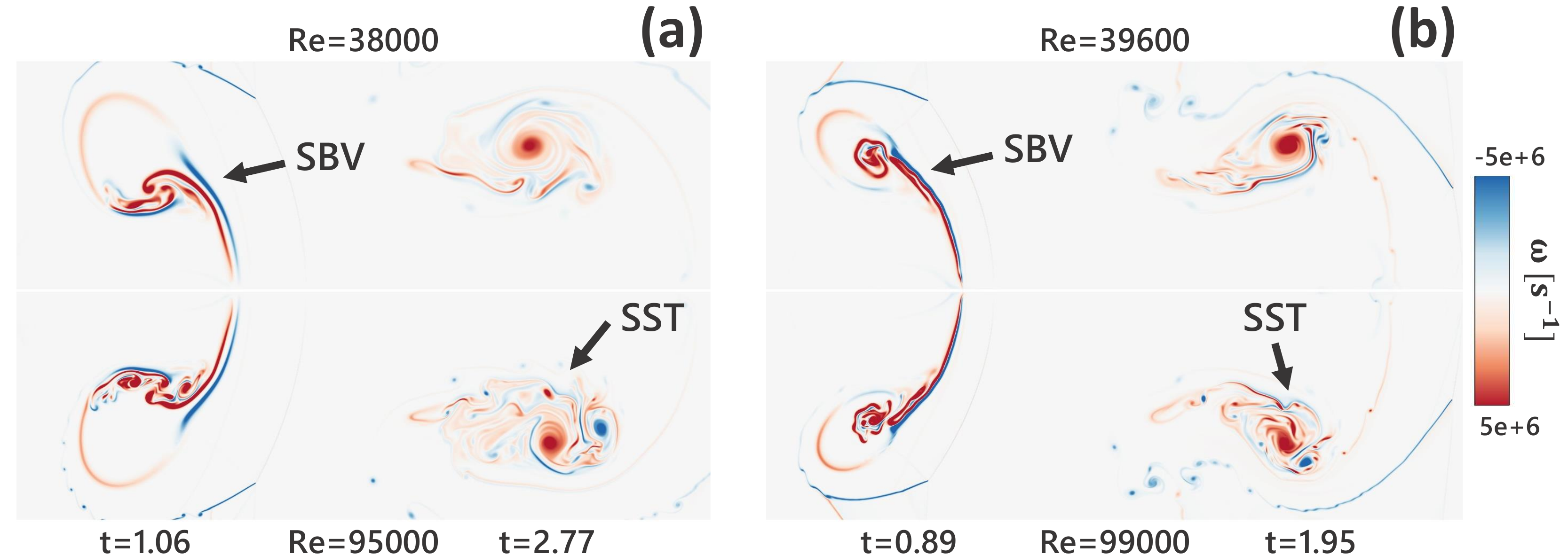}\\
  \caption{The comparison of vorticity contour between medium Re number (upper half) and high Re number (bottom half) for cylindrical cases (b) and spherical cases (b). SBV: secondary baroclinic vorticity; SST: small-scale turbulence. }\label{App9-Tur-cont-vor}
\end{figure}
Figures~\ref{App9-Tur-cont-vor}(a) and (b) compare both cylindrical and spherical bubble cases under the same Pe number (Pe~=~15000 for cylinder and Pe~=~15300 for sphere) but two different Re numbers.
Two instantaneous vorticity contours are compared. As for Re = 38000 cylindrical case, a large vortical structure dominates the flow, although secondary baroclinic structures form early and dissipate at a late time. However, small-scale turbulence occurs and rips the main vortex into disturbance status in Re~=~95000 cylindrical case, which also appears in the high Re number spherical bubble case.

\begin{figure}
    \centering
    \subfigure[]{
    \label{App9-line1} 
    \includegraphics[clip=true,trim=18 20 30 60,width=.48\textwidth]{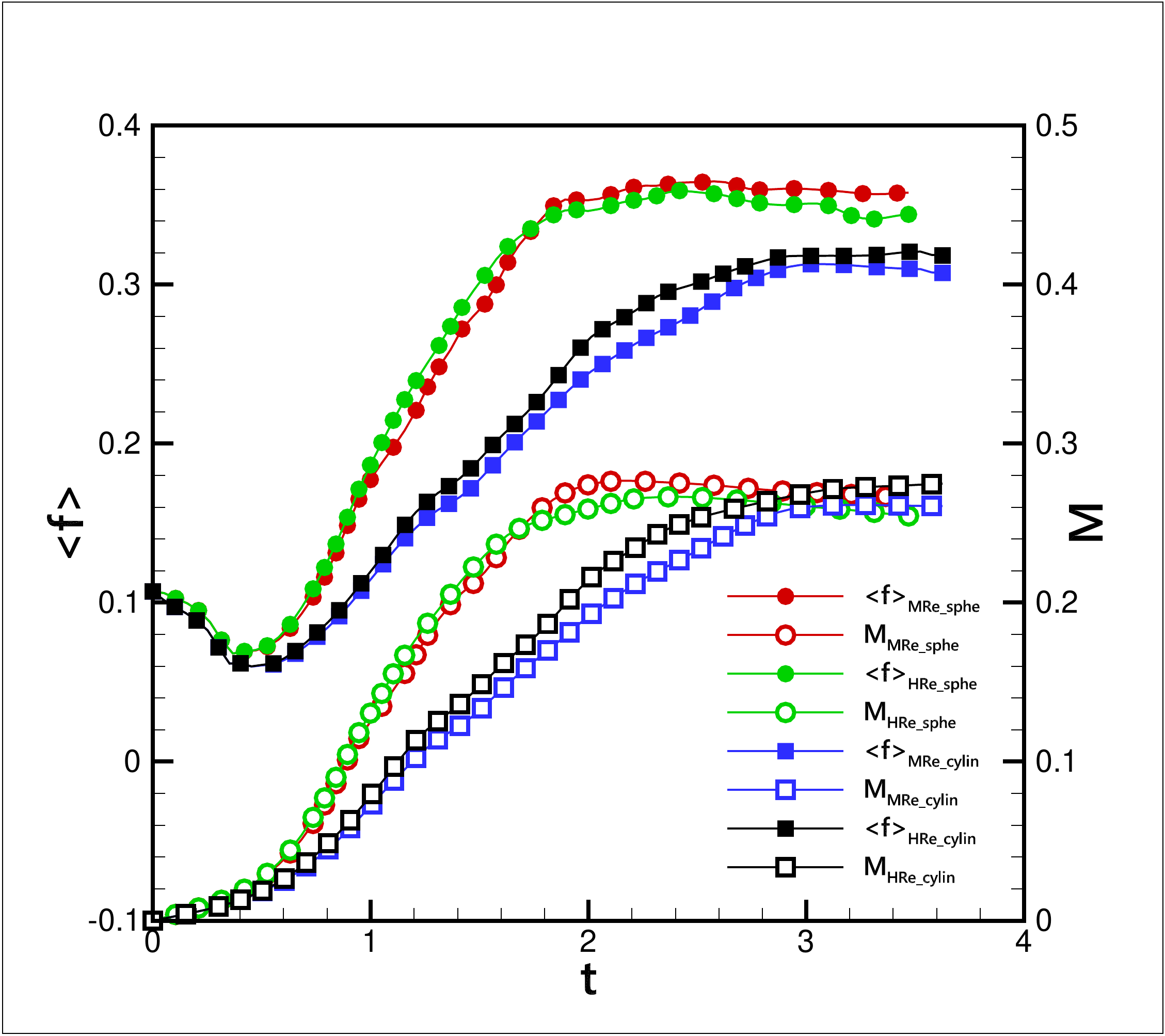}}
    \subfigure[]{
    \label{App9-line2} 
    \includegraphics[clip=true,trim=18 20 30 60,width=.48\textwidth]{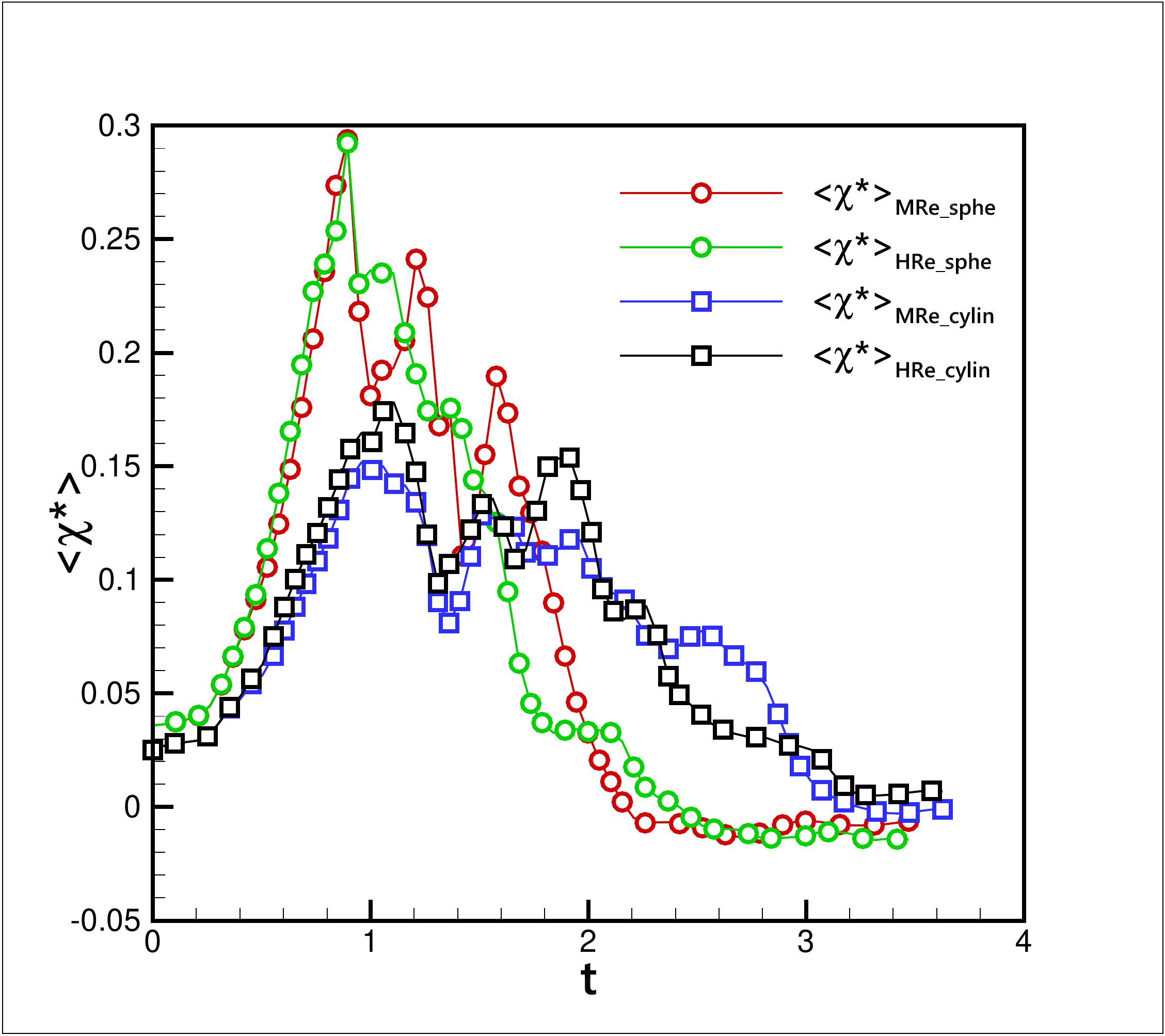}} \\
    \caption{(a) Comparison of time history of mixedness between medium Re number and high Re number with same Pe number. Subscripts \emph{sphe} and \emph{cylin} represent spherical and cylindrical bubbles, respectively. (b) Comparison of time history of mixing rate $\left<\chi^*\right>$ between medium Re number and high Re number. \label{app9-line} }
\end{figure}
Figure~\ref{app9-line} quantitatively compares the effect of small-scale structures on mixedness and mixing rate. The existence of turbulence in both cylindrical and spherical bubbles slightly increases the mixedness growth rate, which can also be discovered from the mixing rate evolution profile. However, the turbulence has a limited effect on mixing in accordance to the independent scaling of mixing rate on Re number, as revealed in Sec.~\ref{subsec:scaling1}.

Besides, the evolutionary difference of mixing behavior between the cylindrical and spherical case can be observed in Fig.~\ref{app9-line}. The spherical bubble experiences a faster mixing rate than the cylindrical bubble. The difference in mixing rate between the two configurations can be attributed to two reasons. First, from Fig.~\ref{app1-circu}, the circulation of spherical bubble is 5\% larger than cylindrical cases, leading to a faster stretching rate. Second, self-induced velocity in vortex ring \cite{shariff1992vortex} tends to a faster vortex evolution in the spherical bubble than in the cylindrical bubble. The velocity model for shock spherical bubble and cylindrical bubble interaction \cite{rudinger1960behaviour,yu2020two} also reflects the faster motion of a vortex ring than a vortex pair.
The mechanisms causing the difference in mixing behavior between spherical and cylindrical geometry are worthy of investigation in future work.

\section{Some discussions on DGRD term}
\label{sec:app10}
\begin{figure}
    \centering
    \subfigure[]{
    \label{app10-1} 
    \includegraphics[clip=true,trim=20 15 25 20,width=.48\textwidth]{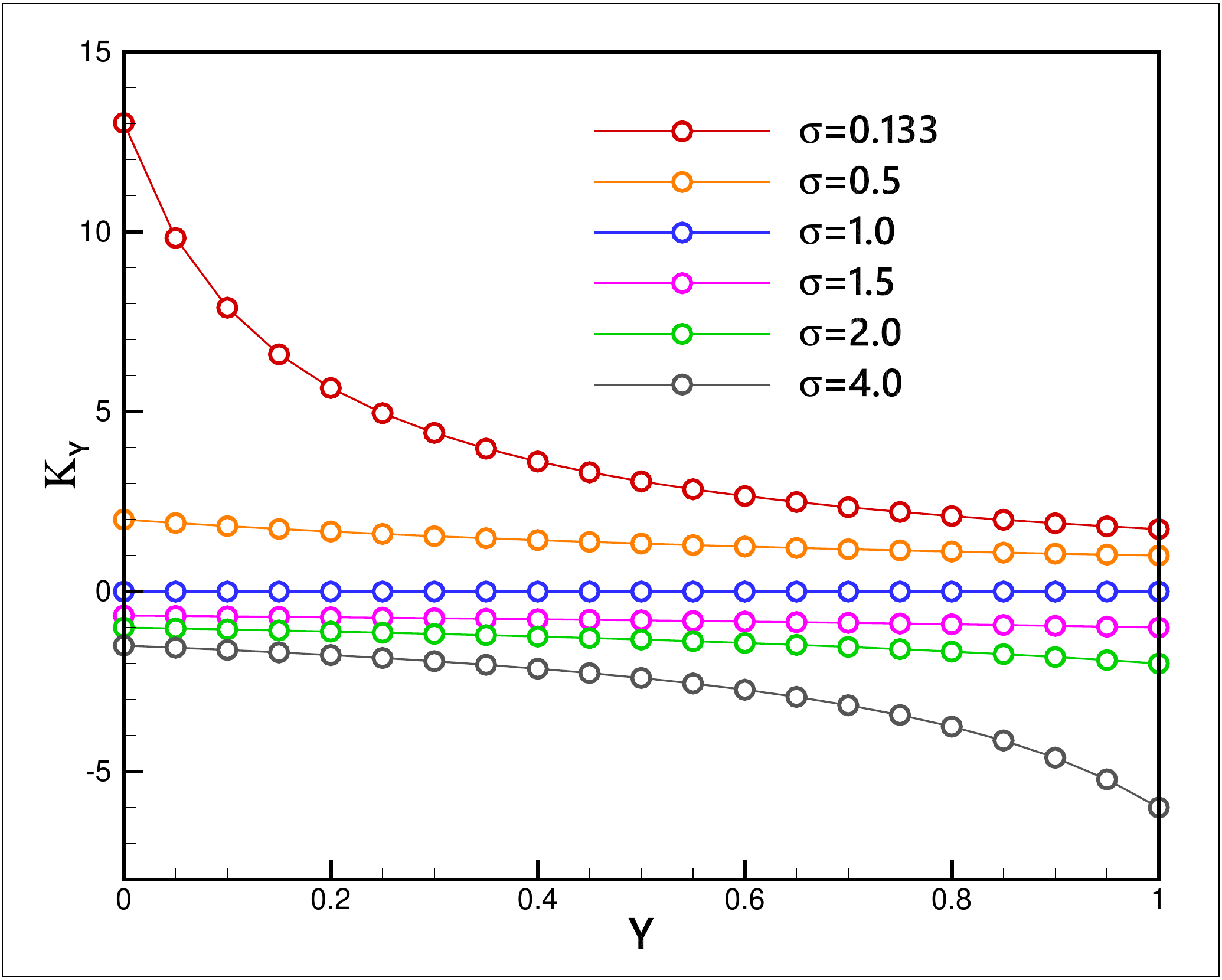}}
    \subfigure[]{
    \label{app10-2} 
    \includegraphics[clip=true,trim=20 15 25 20,width=.48\textwidth]{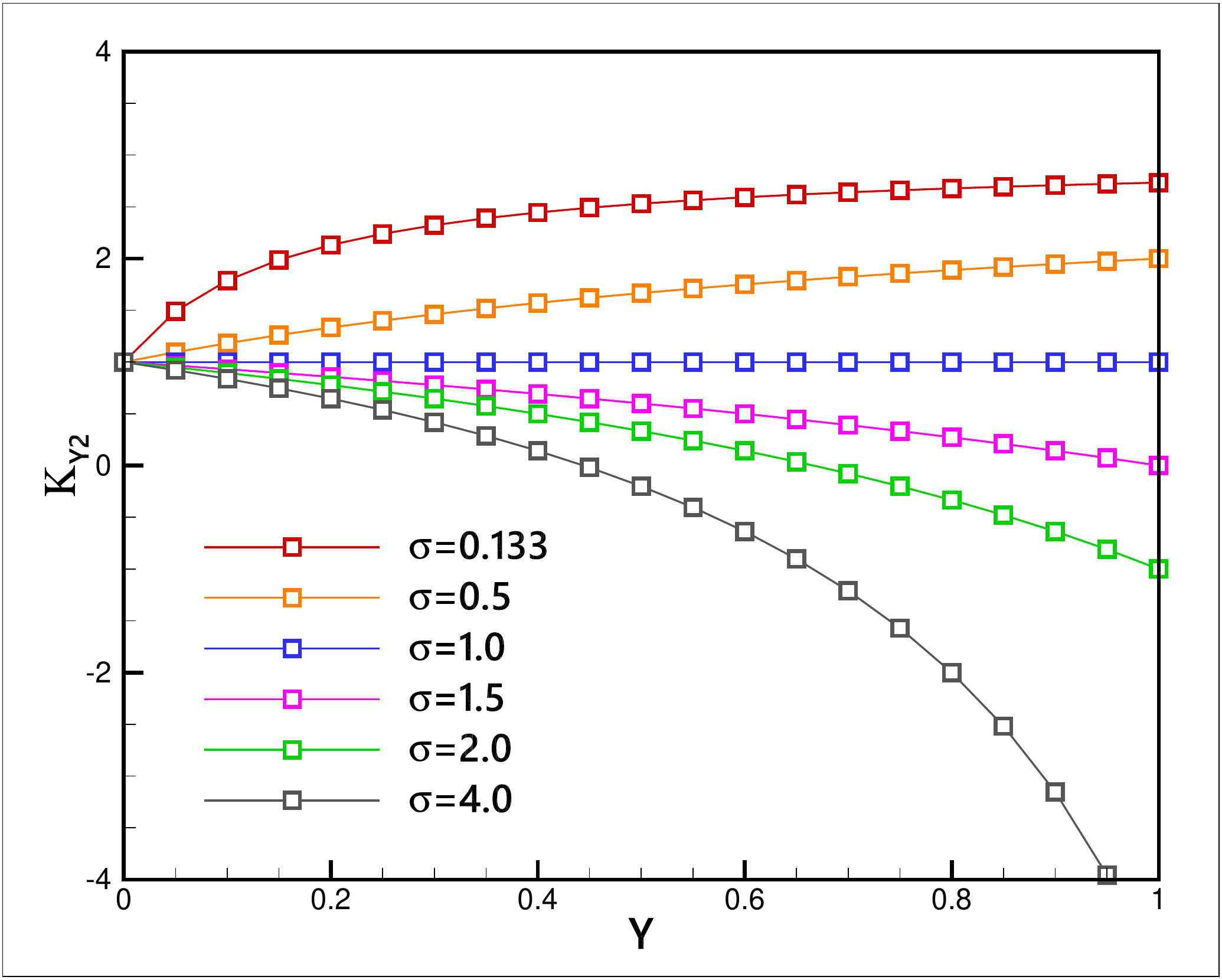}}
    \caption{
    Coefficients of accelerated dissipation $\mathscr{K}_{Y}$ for the mass fraction $Y$ (a) and $\mathscr{K}_{Y^2}$ for the mass fraction energy $Y^2/2$ (b). \label{app10} }
\end{figure}
As for DGAD, its sign is solely determined by the coefficient $\mathscr{K}_{1,Y}$ and $\mathscr{K}_{1,Y^2}$ since dissipation rate $\nabla Y\cdot\nabla Y$ is strictly positive.
Due to the non-zero DGRD term, it is hard to predict the sign of $\mathrm{D}\left<Y\right>/\mathrm{D}t$ and $\mathrm{D}\left<Y^2/2\right>/\mathrm{D}t$ from Eqs.~(\ref{eq: dYdt-SDR-com}) and (\ref{eq: dY2dt-SDR-com}) directly. However, it is remarkable that by using the generalized Green's theorem $\iiint_{\mathcal{V}}(F\nabla^2G+\nabla F\cdot \nabla G)d\mathcal{V}=\oiint_\textrm{S}F(\partial G/\partial\vec{\textbf{n}})\cdot\vec{\textbf{n}}d\textrm{S}$, we can expand the non-zero DGRD term integration in the form of a strictly positive dissipation term. Starting from Eq.~(\ref{eq: dYdt-SDR-com}), the DGRD term can be expressed as:
\begin{eqnarray}\label{eq: YI_DGRD}
  \left<\frac{1}{\mathrm{Pe}}\mathscr{K}_{2,Y}(\sigma,Y)\nabla^2 Y\right> & = &
  -\left<\frac{1}{\mathrm{Pe}}\frac{\sigma}{(1-\sigma)Y+\sigma}\nabla^2 Y\right> \\
 & = &-\left<\frac{1}{\mathrm{Pe}} \frac{\sigma(1-\sigma)}{ \left((1-\sigma)Y+\sigma\right)^2 }\nabla Y\cdot\nabla Y\right>.
\end{eqnarray}
Then, the coefficient on growth rate for the mean mass fraction is defined as:
\begin{equation}\label{eq: YI_DGRD-2}
  \frac{\mathrm{D}\left<Y\right>}{\mathrm{D}t}=\left<-\frac{1}{\mathrm{Pe}}\mathscr{K}_Y(\sigma,Y)\nabla Y\cdot\nabla Y\right>,
\end{equation}
where,
\begin{equation}\label{eq: YI_DGRD-3}
  \mathscr{K}_Y(\sigma,Y)=\Psi\cdot\left(1+\Psi Y\right)+\frac{\sigma(1-\sigma)}{ \left((1-\sigma)Y+\sigma\right)^2 } \quad \left(\Psi=\frac{1-\sigma}{(1-\sigma)Y+\sigma}\right).
\end{equation}
As for DGRD term in Eq.~(\ref{eq: dY2dt-SDR-com}), it can be expressed as:
\begin{eqnarray}\label{eq: YI2_DGRD}
  \left<\frac{1}{\mathrm{Pe}}\mathscr{K}_{2,Y^2}(\sigma,Y)\nabla^2 Y\right> & = &
  \frac{1}{2}\left<\frac{1}{\mathrm{Pe}} \frac{(1-\sigma)Y^2}{(1-\sigma)Y+\sigma} \nabla^2 Y\right> \\
 & = & \frac{1}{2} \left<\frac{1}{\mathrm{Pe}}  \left[\left(\frac{\sigma}{(1-\sigma)Y+\sigma}\right)^2-1\right]
      \nabla Y\cdot\nabla Y\right>.
\end{eqnarray}
In the same way, the coefficient on growth rate for the mean mass fraction energy is defined as:
\begin{equation}\label{eq: YI2_DGRD-2}
  \frac{\mathrm{D}\left<Y^2/2\right>}{\mathrm{D}t}=\left<-\frac{1}{\mathrm{Pe}}\mathscr{K}_{Y^2}(\sigma,Y)\nabla Y\cdot\nabla Y\right>,
\end{equation}
where,
\begin{equation}\label{eq: YI2_DGRD-3}
  \mathscr{K}_{Y^2}(\sigma,Y)=\frac{(1+\Psi Y)^2}{2}-\frac{1}{2}\left(\frac{\sigma}{(1-\sigma)Y+\sigma}\right)^2+1.
\end{equation}
Then, we can determine the sign of $\mathrm{D}\left<Y\right>/\mathrm{D}t$ and $\mathrm{D}\left<Y^2/2\right>/\mathrm{D}t$ from the coefficients $\mathscr{K}_Y$ and $\mathscr{K}_{Y^2}$.
Figure~\ref{app10} shows the variation of coefficients $\mathscr{K}_{Y}$ and $\mathscr{K}_{Y^2}$ with mass fraction $Y$ and post-shock density ratio $\sigma$. It can be found that when $\sigma>1$, $\mathscr{K}_Y$ is strictly negative, and the mean mass fraction will increase due to the positive source in Eq.~(\ref{eq: YI_DGRD-2}).
As for mean mass fraction energy, it is noteworthy that the coefficient $\mathscr{K}_{Y^2}$ will become negative at a large density ratio and large mass fraction. Thus, the boundary of the strictly positive coefficient will be reached when $\mathscr{K}_{Y^2}(\sigma,Y=1)=0\Rightarrow\sigma=3/2$. Positive coefficient $\mathscr{K}_{Y^2}$ at the range of $\sigma<3/2$ identifies the conclusive decrease of mean mass fraction energy.

\bibliography{mybibfile}

\end{document}